\documentclass[aps, prx, twocolumn, superscriptaddress, 10pt]{revtex4-2}

\usepackage{amsmath}
\usepackage{amssymb}
\usepackage{amsfonts}
\usepackage{bm}
\usepackage{graphicx}
\usepackage{braket}
\usepackage{microtype}

\makeatletter

\DeclareMathOperator{\tr}{\rm{tr}}

\DeclareMathOperator{\M}{\mathcal{M}}
\newcommand{\raisedchi}{\protect\raisebox{2pt}{$\chi$}}

\newcommand{\jila}{\affiliation{JILA, NIST and Department of Physics, University of Colorado, Boulder, Colorado, USA}}

\newcommand{\ctqm}{\affiliation{Center for Theory of Quantum Matter, University of Colorado, Boulder, Colorado, USA}}

\newcommand{\nbi}{\affiliation{Niels Bohr Institute, University of Copenhagen, Blegdamsvej 17, DK-2100 Copenhagen, Denmark}}

\newcommand{\jqi}{\affiliation{Joint Quantum Institute, NIST/University of Maryland, College Park, MD, 20742, USA}}

\newcommand{\jcqics}{\affiliation{Joint Center for Quantum Information and Computer Science, NIST/University of Maryland, College Park, MD, 20742, USA}}

\newcommand{\harvard}{\affiliation{Department of Physics, Harvard University, Cambridge, MA 02138 USA}}

\newcommand{\iit}{\affiliation{Department of Physics, Indian Institute of Technology Madras, Chennai 600036, India}}

\newcommand{\cqicc}{\affiliation{ Center for Quantum Information, Communication and Computing, Indian Institute of Technology Madras, Chennai 600036, India}}

\usepackage[unicode=true,
 bookmarks=false,
 breaklinks=false, colorlinks=true]
 {hyperref}
\hypersetup{
  linkcolor = blue,
  citecolor = blue,
  urlcolor = blue
}

\begin{abstract}
    We explore a two-node, entanglement-enhanced sensor network for differential phase sensing that exploits decoherence-free subspaces to suppress common-mode noise, a primary limitation of many state-of-the-art quantum sensors. We identify a class of entangled states that, while not strictly optimal, achieve the same asymptotic sensitivity scaling as optimal states and can be prepared efficiently from initially unentangled atomic ensembles. Importantly, the preparation time decreases with increasing system size. This makes the states compatible with realistic noise processes in present-day quantum sensors that operate with large particle numbers but lack full error correction. We illustrate these ideas using two cavity-mediated preparation protocols: (i) coherent, unitary entanglement generation analogous to bosonic two-mode squeezing, yielding Heisenberg scaling; and (ii) dissipative preparation via collective emission into a shared cavity mode, providing a square-root improvement beyond the standard quantum limit. Numerical simulations show that both approaches remain effective at experimentally realistic cavity cooperativities, establishing a practical path toward scalable, quantum-enhanced differential phase sensing.
\end{abstract}

\begin{document}

\title{Lieb-Mattis states for robust entangled differential phase sensing}

\author{Raphael Kaubruegger}\jila
\author{Diego Fallas Padilla}\jila
\author{Athreya Shankar}\iit\cqicc
\author{Christoph Hotter}\nbi
\author{Sean R.~Muleady}\jqi\jcqics
\author{Jacob Bringewatt}\harvard
\author{Youcef Baamara}\jila
\author{Erfan Abbasgholinejad}\jqi\jcqics
\author{Alexey V.~Gorshkov}\jqi\jcqics
\author{Klaus Mølmer}\nbi
\author{James K.~Thompson}\jila
\author{Ana Maria Rey}\jila\ctqm

\maketitle

\section{Introduction}

Recent advancements in quantum sensors have enabled atomic interferometers \cite{rosi2014precision, parker2018measurement, overstreet2022observation} and optical clocks \cite{aeppli2024clock} to reach the fundamental sensitivity limit set by quantum projection noise in ensembles of uncorrelated atoms. Nevertheless, the performance of many state-of-the-art quantum sensors is limited by technical noise sources common to all the atoms, such as the  finite coherence time of the electromagnetic fields used for manipulation of the atoms, fluctuations in ambient magnetic fields, or vibrations of the optical path of atomic interferometers. A promising strategy to overcome this limitation is the use of quantum sensor networks \cite{eldredge2018optimal,proctor2018multiparameter, ge2018distributed, gross2020one, zhang2021distributed, bringewatt2021protocols, malitesta2023distributed}. These networks can be operated in regimes that are insensitive to noise common to all sensor nodes, while preserving sensitivity to spatially varying signals—such as differential frequency shifts between nodes.

Differential phase measurement protocols play a central role in precision metrology, enabling high-accuracy determinations of fundamental constants \cite{fixler2007atom, rosi2014precision}. They are employed in inertial sensing applications \cite{durfee2006long, stray2022quantum} and in experimental tests of the equivalence principle \cite{schlippert2014quantum, asenbaum2020atom, barrett2022testing}. In optical magnetometers operated in a gradiometric configuration, differential sensing can also suppress common-mode magnetic field noise \cite{jiang2019magnetic, lucivero2021femtotesla, wu2023quantum}. Moreover, networks of optical clocks have recently enabled the observation of the gravitational redshift over millimeter-scale height differences \cite{bothwell2022resolving, zheng2023lab, aeppli2024clock}, as well as differential phase estimation using entangled states \cite{robinson2024direct, yang2025clock}.

\begin{figure}[t!] 
   \centering
   \includegraphics[]{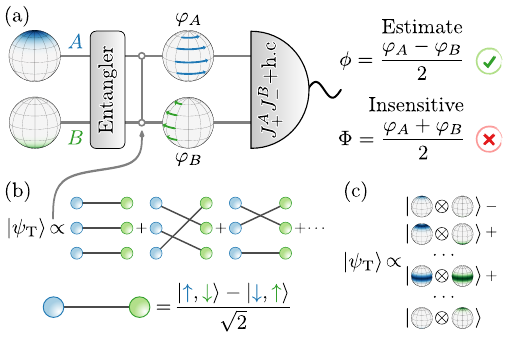} 
   \caption{(a) Schematic illustration of the differential phase sensing protocol. The atoms in ensembles $A (B)$ are initially prepared in their respective excited (ground) states, after which an entangling operation is applied. Subsequently, each ensemble acquires a distinct phase $\varphi_A$ and $\varphi_B$. Finally, a joint measurement of both ensembles is carried out to estimate the differential phase $\phi$ with high precision, while remaining insensitive to fluctuations of the common phase  $\Phi$. (b) The target state after the entangling process, $\ket{\psi_T}$, is an entangled Lieb-Mattis state (see Appendix~\ref{app:parentHamiltonian}) that can be understood as a permutation-symmetric superposition over all states where each atom in $A$ forms a singlet with an atom in $B$. (c) In the Dicke basis of the individual ensembles, represented by their Wigner distributions, this state can be expressed as an equal superposition, with alternating signs, of all Dicke-state combinations where the number of atoms in the excited state in ensemble $A$  equals the number of ground-state atoms in ensemble $B$. 
 }
   \label{fig:sequence}
\end{figure}

These quantum sensors typically operate with ensembles of $N$ unentangled atoms, which imposes a fundamental limit on their sensitivity. Specifically, the variance in estimating a phase encoded in such an unentangled system is bounded from below by $1/N$, a limit known as the standard quantum limit of quantum metrology. Quantum entanglement offers a path to surpass this limitation, enabling measurement precision to approach the Heisenberg limit of $1/N^2$, the ultimate bound permitted by quantum mechanics \cite{pezze2018quantum, huang2024entanglement}.

A central challenge in entanglement-enhanced frequency estimation arises from the fact that the attainable uncertainty decreases with increasing interrogation time. At long interrogation times, common-mode noise often emerges as the dominant source of decoherence. While highly entangled states are particularly vulnerable to such noise, unentangled atomic ensembles tend to be more robust. Moreover, the performance of unentangled probes can often be effectively restored through classical post-processing techniques—such as ellipse fitting \cite{foster2002method}—which are routinely employed in differential atom interferometry and optical atomic clocks \cite{fixler2007atom, rosi2015measurement, barrett2016dual, langlois2017differential, parker2018measurement, marti2018imaging, elliott2023quantum, bothwell2022resolving, zheng2023lab}.
Even though such ellipse fitting strategies can also be extended to a class of entangled states known as spin-squeezed states \cite{wineland1992spin, kitagawa1993squeezed, ma2011quantum}, which have recently been realized experimentally \cite{eckner2023realizing}, in scenarios where sensor performance is constrained by common-phase noise, the degree of useful spin squeezing—and hence the attainable sensitivity enhancement—is reduced. Consequently, the scaling of sensitivity with $N$ achieved using two spin-squeezed ensembles falls short of the ideal scaling expected in the absence of noise \cite{corgier2025optimized}. The primary goal of this manuscript is therefore to identify strategies capable of achieving  a more favorable sensitivity scaling in the common-mode-noise–limited regime while  remaining compatible with currently available sensing hardware.

A viable strategy to restore Heisenberg-limited scaling while preserving robustness against laser noise is to operate the sensor network within a decoherence free subspace (DFS) \cite{roos2006designer, monz201114, dorner2012quantum, landini2014phase, jeske2014quantum, altenburg2017estimation, eldredge2018optimal, sekatski2020optimal, hamann2022approximate, hamann2024optimal, hainzer2024correlation}. This is achieved by preparing entangled states and performing measurements that commute with the noise-generating operators, thereby rendering the system insensitive to common-mode noise by construction. DFS-based techniques have recently been experimentally demonstrated in a system with three and two trapped ions, respectively \cite{bate2025experimental, dietze2026entanglement}. 

The optimal states for differential phase measurements within a DFS are typically GHZ-type states \cite{roos2006designer, dorner2012quantum, landini2014phase}. These states can, in principle, achieve Heisenberg limit precision, reducing the estimator variance to $1/N^2$ \cite{bollinger1996optimal}. However, GHZ states are notoriously fragile. Their susceptibility to local noise renders their preparation challenging in practice, particularly for sensor architectures that involve many atoms per node.

In this work, we investigate an alternative family of DFS states that retains Heisenberg scaling and, importantly, preserves it even in the presence of local noise during state preparation; as a representative example, we consider free-space emission from the excited state. This robustness makes the states compatible with large-scale quantum systems. We further propose practical schemes to generate these states by exploiting cavity-mediated interactions between sensor nodes.

To extract the encoded phase information for the identified states, we propose to measure two-body operators accessible for example through photon counting of the light emitted from a cavity. Such measurements are technically more demanding than conventional single-body readout, and have a restricted dynamic range, the interval of differential phase values for which high-precision estimation is possible scales as $\mathcal{O}(1/N)$. The small dynamical range limitation can be mitigated by repeating the measurement for the same coded phase sufficiently often and applying a maximum likelihood estimator.

The remainder of the manuscript is organized as follows. In Sec.~\ref{sec:DifferentialSensing}, we introduce the formalism for differential phase sensing with two nodes, each comprising an ensemble of spin-1/2 atoms. Sec.~\ref{sec:DFS} reviews the concept of a DFS and explains how it can eliminate the effects of common laser noise, followed by a discussion in Sec.~\ref{Sec:Precision} of the fundamental limits on sensor precision. Section~\ref{sec:tow-body} identifies the two-body observable that is optimal within the decoherence-free subspace. Building on this result, Sec.~\ref{sec:TargetState} introduces a target entangled state that achieves Heisenberg scaling. Sec.~\ref{sec:Cavity} details two practical methods for generating entanglement between sensor nodes via a shared cavity \cite{periwal2021programmable}. In Sec.~\ref{sec:unitary_preparation}, we propose a unitary preparation scheme based on a bosonic two-mode squeezing interaction realizable with multilevel alkaline-earth atoms, while Sec.~\ref{sec:stochastic_preparation} discusses a stochastic preparation approach that exploits collective emission. Although this latter method yields only a $\sqrt{N}$ improvement beyond the standard quantum limit (rather than full Heisenberg scaling), it can be directly implemented on present-day experiments and is robust to particle-number fluctuations, a significant concern in cavity-based experiments. Finally, Sec.~\ref{sec:Outlook} concludes with a summary and outlook for future applications.

\section{Differential phase sensing}
\label{sec:DifferentialSensing}
We consider a two-node quantum sensor network, as illustrated in Fig.~\ref{fig:sequence}(a), in which a total of $N$ atoms are divided into two ensembles, $A$ and $B$, each containing $N_A = N_B = N/2$ atoms. The atoms feature a long-lived optical transition between the states $\ket{\downarrow}$ and $\ket{\uparrow}$. The dynamics of this effective spin-1/2 system are described by the Pauli operators $\sigma_{x,y,z}$, and the whole system can be described by the collective spin operators $J_{\alpha}^{K}=\sum_{k=1}^{N_{K}}\sigma^{(k)}_{\alpha}/2$ of the respective ensembles, where $\alpha = x,y,z$ and $K=A,B$.

In the interferometric sequence that we will study in this manuscript, the quantum system is first prepared in a potentially entangled quantum state $\ket{\psi}$. Thereafter, the phases $\varphi_A$ and $\varphi_B$ are encoded onto the quantum system according to 
\begin{align}
    \ket{\psi_{\varphi_A, \varphi_B}} =  e^{-i\left(\varphi_AJ_z^A + \varphi_B J_z^B\right)}\ket{\psi}=U({\varphi_A, \varphi_B})\ket{\psi}.
\end{align}
The final step is to perform a von Neumann measurement $\mathcal{M}= \sum_{\mu}\mu\, \ket{\mu}\bra{\mu}$ on the quantum system and convert the measurement outcomes $\mu$ into an estimate of the encoded  phase.

The objective is to optimize the precision in estimating the differential phase $\phi=\tfrac{\varphi_A-\varphi_B}{2}$, encoded by the operator $J_z^-=J_z^A-J_z^B$, while ensuring that the interferometer remains insensitive to the common phase $\Phi=\tfrac{\varphi_A+\varphi_B}{2}$, encoded by the operator $J_z^+=J_z^A+J_z^B$. 

The common phase fluctuations can be mitigated by preparing a product state between the two subensembles and then performing repeated measurements of the same differential phase while the common phase varies randomly between measurement repetitions. In each repetition, independent measurements on the two subensembles yield estimates of the phases $\varphi_A$ and $\varphi_B$, which collectively trace out an ellipse whose geometry depends on the differential phase $\phi$ \cite{foster2002method}. This approach has been experimentally demonstrated using ensembles of unentangled atoms \cite{bothwell2022resolving, zheng2023lab} as well as spin-squeezed states \cite{eckner2023realizing}.

However, the usefulness of spin-squeezed states in a differential phase-sensing setting, where only single-body observables are measured and the common phase is fully randomized, is more limited than in conventional phase estimation with a single spin ensemble. In contrast to unentangled ensembles, spin-squeezed ensembles exhibit phase-dependent measurement noise: while the noise can be reduced for certain phase values, it simultaneously increases for others. When the common phase fluctuates randomly across measurement repetitions, the sensor effectively samples the full range of phase-dependent noise. As a result, spin-squeezed states that achieve a reduction of the estimator variance by a factor of $N^{-2/3}$ when estimating the phase common to all the spins are already too strongly squeezed to operate effectively in such a differential measurement scenario. The largest achievable reduction in estimator variance in the differential phase sensing scenario scales as $N^{-1/3}$ \cite{corgier2025optimized}. See Appendix~\ref{app:seperable} for further discussion. 

This contrasts with the regime in which the common-phase noise is sufficiently small, i.e., decreases in proportion to the amount of spin squeezing generated. Only under these conditions can spin-squeezed protocols recover the usual sensitivity scalings associated with spin squeezing in a single ensemble \cite{corgier2023quantum,corgier2025optimized}.

\subsection{Decoherence free subspace}
\label{sec:DFS}

An alternative approach is to render the sensor inherently insensitive to the common phase by operating the quantum sensor within a DFS with respect to the operator $J_z^+$.

Operating in a DFS is achieved by preparing an eigenstate of $J_z^+$, ensuring that any random fluctuations in the common phase result only in an irrelevant  global phase factor. Additionally, measurements must be performed using an operator that commutes with $J_z^+$, thereby making the measurement outcomes independent of the averaging over $\Phi$.

From this point onward, we restrict our analysis to unitary operations and measurements that commute with $J_z^+$, and thus do not couple different DFSs. Although we account for noise processes that lead to mixed states with populations distributed across multiple DFSs, these processes do not induce coherences between them. Consequently, the quantum state remains independent of $\Phi$ at all times, and we therefore omit the dependence on $\Phi$ in all subsequent expressions.

We emphasize that in practice common-mode noise is never the sole noise source in a quantum sensor.  Noise processes with other spatial correlations are not suppressed by operating in a DFS. Nevertheless, in state-of-the-art sensors, common-mode noise remains the dominant contribution. For sensors operated in a differential configuration, the next most relevant class of noise processes are differential in nature, which are indistinguishable from the signal itself and therefore degrade the performance of any differential sensor equally. In addition, we analyze free-space emission as a local noise process that induces leakage into other DFSs, and we show that a realistic level of such noise remains tolerable during the state preparation without significantly compromising the overall sensitivity of the sensor.

\subsection{Sensor precision}
\label{Sec:Precision}

The encoded phase is estimated by performing repeated measurements of the observable $\M$ over $r$ repetitions and using the sample mean of the outcomes to invert the functional dependence of the expectation value $\braket{\mathcal{M}_{\phi}} \equiv \braket{\psi|U^{\dagger}(\phi)\, \mathcal{M}\, U(\phi)|\psi}$ on $\phi$, where $U(\phi) = e^{-i\phi J_z^-}$. The variance of this estimator is determined via error propagation 
\begin{gather}
\Delta^2_{\phi} = \frac{\braket{\M^2_{\phi}} - \braket{\M_{\phi}}^2}{r\left|\frac{\partial}{\partial \phi} \braket{\M_{\phi}}\right|^2}.
\label{eq:EstVar}
\end{gather}

In the limit of many repeated measurements $r$, the sensor becomes unbiased, meaning that the average value of the estimator for a given phase converges to the true encoded phase. In this regime, the estimator’s performance can be directly compared to fundamental bounds established for unbiased estimators. The variance of an unbiased estimator is lower bounded by the quantum Cramér–Rao bound (QCRB) \cite{braunstein1994statistical, paris2009quantum, demkowicz2015quantum, pezze2018quantum, huang2024entanglement},
\begin{align}
    \Delta_{\phi}^2 \geq \frac{1}{r\, F^{\rm Q}_{\ket{\psi}}},
\end{align}
where $F^{\mathrm{Q}}_{\ket{\psi}}$ represents the quantum Fisher information (QFI) of the state $\ket{\psi}$. Throughout the remainder of this work, we evaluate the performance in terms of the r-independent contribution to the estimator variance, and thus set $r = 1$.

The QFI quantifies the best precision that can be achieved in parameter estimation for a given quantum state across all physically realizable measurements and estimators. Quantitatively, for a pure state and a unitary parameter encoding, the QFI can be expressed in terms of the variance of the operator responsible for phase encoding:
\begin{align}
    F^{\rm Q}_{\ket{\psi}}= 4\left(\braket{\psi|J_z^-J_z^-|\psi}-\braket{\psi|J_z^-|\psi}^2\right).
    \label{eq:QFI}
\end{align}
The fundamental quantum limit on the estimator variance for a sensor comprising $N$ unentangled atoms is given by $\Delta_{\phi}^2 \geq 1/N$ , known as the standard quantum limit. This bound can be surpassed through the use of entanglement, but cannot be improved beyond the Heisenberg limit, $\Delta_{\phi}^2 \geq 1/N^2$.

The state that maximizes the QFI and attains the Heisenberg limit is the coherent superposition of the eigenstates of $J_z^z$ corresponding to the largest and smallest eigenvalues \cite{roos2006designer, dorner2012quantum, landini2014phase},
\begin{align}
    \ket{\psi_{\rm HL}} = \frac{\ket{\uparrow,\dots,\uparrow}\ket{\downarrow,\dots,\downarrow}+\ket{\downarrow,\dots,\downarrow}\ket{\uparrow,\dots,\uparrow}}{\sqrt{2}},
    \label{eq:GHZ}
\end{align}
which can be identified as a GHZ state combined with a local $\pi$-rotation on ensemble $B$. Importantly, $\ket{\psi_{\rm HL}}$ is an eigenstate of the collective operator $J_z^+$ and thus resides within a DFS and thus is inherently insensitive to fluctuations of the common phase. These characteristics make this state particularly appealing for preparation in quantum sensors, as it offers enhanced sensitivity to the desired signal. However, this amplified sensitivity also extends to most noise sources present in current quantum sensing platforms, making the preparation of large-scale GHZ states exceptionally challenging. Consequently, the generation of GHZ states has become a benchmark for evaluating the capabilities of quantum computing platforms \cite{omran2019generation, pogorelov2021compact, mooney2021generation}, with current efforts achieving GHZ-state sizes of up to 32 qubits \cite{moses2023race}.

To date, GHZ-like states have been the workhorse of discussions on decoherence-free subspaces. Here, by contrast, we introduce a new class of states that addresses a key limitation hindering the experimental utility of GHZ states in scalable arrays—namely, their pronounced fragility to noise, which, together with the unfavorable scaling of their preparation time with system size, restricts both their experimental realization to intermediate system sizes and their metrological utility even further.

\section{Optimal Two-body observable}
\label{sec:tow-body}

In our DFS, the expectation values of the single-particle observables $\sigma_{x, y}^{(k)}$ vanish. This is because these operators measure coherences between states that differ in the number of excited atoms by 1. Consequently, to obtain a nontrivial measurement outcome, it is necessary to consider an observable composed of at least two-body operators. The only two-body operators that do not commute with the phase encoding and measure coherences within a given DFS---without coupling DFSs with different number of atoms in the excited state---are of the form $\sigma_{+}^{(k)}\sigma_{-}^{(l)}$, where $2\sigma_{\pm}^{(k)}=\sigma_{x}^{(k)}\pm i\sigma_y^{(k)}$, and the indices $k$ and $l$ refer to atoms from different ensembles. 

One might be tempted to employ conventional two-body observables that include terms coupling different DFSs, such as $J_x^A J_x^B$, based on the assumption that these couplings do not influence the expectation value of the measurement. However, while the mean signal remains unaffected, such terms lead to an increased measurement variance due to second-order processes involving coupling to other DFSs and back.

The only two-body observable that satisfies the conditions of not containing any terms that commute with the phase encoding, not coupling different DFSs, and being permutationally invariant within the subensembles is
\begin{align}
    \M=J_{+}^AJ_{-}^B + J_-^AJ_+^B, 
    \label{eq:Measurement}
\end{align}
where $J^{K}_{\pm}=J^{K}_x\pm i J^{K}_y$. Note the distinction between these collective spin raising and lowering operators and the common and differential phase operators $J_z^{\pm}$. Expressing $\M$ in the transformed frame yields $\M_{\phi} = U^{\dagger}(\phi) \mathcal{M} U(\phi) =\cos(2\phi)\left(J_{+}^AJ_{-}^B + \rm{h.c.}\right)+i\sin(2\phi)\left(J_{+}^AJ_{-}^B - \rm{h.c}\right)$. Therefore, the measured expectation value of $\M$ exhibits interference fringes. However, unlike conventional Ramsey fringes, the response to the phase $\phi$ is doubled, reflecting the fact that a two-body observable is being measured rather than a single-body observable. See Appendix~\ref{app:Measurement} for a discussion on the implementation of a measurement scheme that mimics the optimal measurement in a cavity-based setup, realized by detecting photons emitted from the cavity, following the phase encoding. An alternative approach \cite{kaubruegger2021quantum} to implementing the measurement is to employ a variationally optimized circuit that approximates the unitary transformation mapping the eigenbasis of the operator $J_z^+$ onto that of $\M$. 

For a practical implementation of the quantum sensor, it is desirable to identify initial states that simultaneously exhibit both a large interference fringe contrast and a low measurement variance. However, achieving this balance inherently involves a trade-off: increasing entanglement enhances the QFI and thereby reduces the estimator variance, but at the cost of a reduced fringe amplitude, and vice versa.

In the following section, we present a quantum state characterized by QFI and fringe amplitude that both exhibit optimal scaling as $N^2$.  Furthermore, we demonstrate that the measurement of $\mathcal{M}$ achieves the QCRB for this initial state. 

\section{Entangled target state}
\label{sec:TargetState}

Since the phase encoding operator $J_z^-$ and the measurement $\M$ are invariant under permutations within each subensemble, a highly sensitive state should exhibit the same symmetry. Consequently, we restrict our analysis to states that are eigenstates of $\bm{J}^{K}\cdot\bm{J}^{K}$ with the maximum eigenvalue $\tfrac{N}{4}(\tfrac{N}{4}+1)$, thereby preserving the subensemble permutation symmetry for $K=A,B$. Here, $\bm{J}^{K}=(J_x^{K}, J_y^{K}, J_z^{K})^{T}$ represents the vector of the three collective spin operators for the respective subensemble $K=A,B$ and $\bm{J} = \bm{J}^A+\bm{J^B}$. States respecting this symmetry can be expressed in terms of the basis states $\ket{\tfrac{N}{4}, M^A, \tfrac{N}{4}, M^B}$, which satisfy $J_z^{K}\ket{\tfrac{N}{4},M^A,\tfrac{N}{4},M^B} =  M^K\ket{\tfrac{N}{4},M^A,\tfrac{N}{4},M^B}$. This basis is related to the $\ket{J,M}$ basis, which are eigenstates of $\bm{J}\cdot \bm{J}$ and $J_z^+$, via the Clebsch-Gordan coefficients $\braket{J, M|\tfrac{N}{4},M^A,\tfrac{N}{4},M^B}$.  Note that $\ket{J,M}$ of an ensemble of atoms are generally degenerate for $J<N/2$; however, imposing permutation symmetry within the subensembles ensures that the $\ket{J,M}$ states are unique.

Furthermore, restricting the analysis to states within the DFS in which $N/2$ atoms occupy the excited state constrains the states to the form
\begin{align}
    \ket{\psi_{\rm DFS}} = \sum_{M=-\tfrac{N}{4}}^{\tfrac{N}{4}}c_M\, \ket{\tfrac{N}{4},M,\tfrac{N}{4},-M}.
\end{align}
When the coefficients $c_M$ are chosen to be $\propto \left[-\tanh(\alpha)\right]^{N/4-M}$, these states form a family of states that mimics bosonic two-mode squeezed states within a spin system \cite{cable2010parameter,sundar2023bosonic, mamaev2025non}, where $\alpha$ is the squeezing strength. The maximally two-mode squeezed state corresponds to the limit  $\alpha\rightarrow \infty$, where the state coincides with the ground state of the Lieb-Mattis Hamiltonian (see Appendix.~\ref{app:parentHamiltonian})
\begin{align}
    \ket{\psi_{\rm T}} &  = \frac{1}{\sqrt{\tfrac{N}{2} + 1}}\sum_M (-1)^{\tfrac{N}{4}-M} \ket{\tfrac{N}{4},M,\tfrac{N}{4},-M} \notag \\
    & = \ket{J=0,M=0}.
    \label{eq:target_state}
\end{align} 
The Lieb-Mattis state, illustrated in  Fig.~\ref{fig:sequence}(b), is the state symmetrized over all permutations within each subsystem in which each atom in one ensemble forms a singlet with an atom in the other ensemble. A related mixed state—an equal probabilistic mixture over all such singlet coverings—has previously been considered for field-gradient estimation\cite{urizar2013macroscopic}. Although this mixed state is intrinsically insensitive to common-mode noise, it does not provide entanglement-enhanced sensitivity for gradient sensing~\cite{altenburg2017estimation}.

The state $\ket{\psi_{\rm T}}$ is uniquely characterized as the eigenstate of $\bm{J}\cdot\bm{J}$ with eigenvalue zero, and therefore corresponds to $\ket{J=0,M=0}$ in the non-degenerate $\ket{J,M}$ basis introduced earlier. It combines a large interference-fringe contrast with a low measurement variance, making it a natural “target state” for the state-preparation protocols explored in this work.

Specifically, the measurement expectation value as a function of the encoded phase is
\begin{align}
    \braket{\psi_{\rm T}|\M_{\phi}|\psi_{\rm T}}&=-\frac{N^2+4N}{12}\cos(2\phi),\\
    \braket{\psi_{\rm T}|\M^2_{\phi}|\psi_{\rm T}}&=\frac{N^2+4N}{240}\bigg(N^2+4N+8 +\notag\\
    &\qquad\qquad \ \, (N^2 + 4N - 12) \cos(4 \phi)\bigg),
\end{align}
and thus the fringe amplitude $(N^2+4N)/12$ scales asymptotically as $N^2$,  which is the maximum scaling achievable for a two-body observable with $N^2$ terms. Moreover, the estimator variance is minimized at $\phi=\pi/4$, corresponding to the zero crossings of the interference fringes, achieving the QCRB. The bound is characterized by the QFI
\begin{align}
    F^{\rm Q}_{\ket{\psi_{\rm T}}}=\frac{4N+N^2}{3},
\end{align}
which is derived in Appendix \ref{app:ang_mom_states}. The QFI, like the fringe amplitude, asymptotically exhibits the best possible scaling, deviating only by a factor of three from the fundamental Heisenberg limit. Having established that the Lieb-Mattis state is a desirable state for entanglement enhanced differential sensing we will proceed in the next sections by describing methods for preparing this state and proxies of this state in a cavity system. 

\section{State preparation}
\label{sec:Cavity}

\begin{figure}[t] 
   \centering
   \includegraphics[]{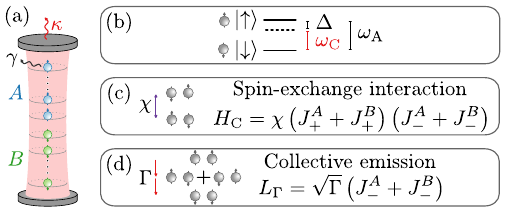} 
   \caption{(a) Schematic of the cavity setup. Two ensembles of atoms, labeled $A$ and $B$, trapped in a magic-wavelength optical lattice (gray ellipses) inside a cavity and are initially prepared in the excited and ground state, respectively. Photons leak out of the cavity at a rate $\kappa$, and atoms in the excited state can emit photons into free space at a rate $\gamma$. (b)  The cavity mode frequency $\omega_{\mathrm{C}}$ is detuned by $\Delta$ from the atomic transition frequency $\omega_{\mathrm{A}}$, which quantifies the energy difference between $\ket{\downarrow}$ and $\ket{\uparrow}$. (c)  Spin-exchange interactions, described by the Hamiltonian $H_{\mathrm{C}}$, are mediated by the exchange of virtual photons through the cavity mode. (d) The atoms can collectively emit into the cavity mode. This process is described by the jump operator $L_{\Gamma}$. 
 }
   \label{fig:cavity}
\end{figure}

\begin{figure}[t]
    \centering
    \includegraphics[]{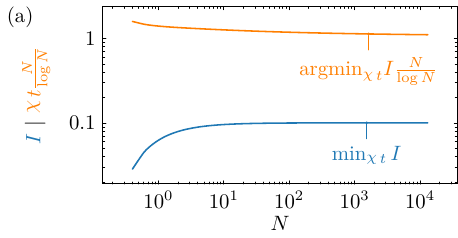}\\
    \includegraphics[]{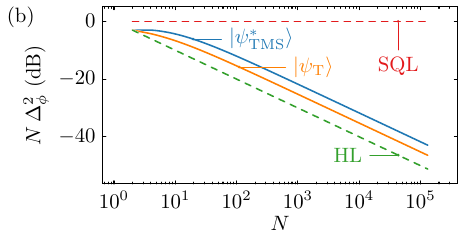}
   \caption{(a) The minimized infidelity $ I = 1 - \left| \braket{\psi_{\rm T}|\psi_{\rm TMS}(t)} \right|^2 $, between the target state  $\ket{\psi_{\rm T}} $ and a state that is generated by quenching the product state  $ \ket{\psi_0} $ under the two-mode squeezing Hamiltonian  $H_{\rm TMS}$  [Eq.~\eqref{eq:H_TMS}] (blue). The optimal time required to reach the minimized infidelity for different atom numbers (orange). (b) Scaling of the estimator variance of the quenched state $\ket{\psi_{\rm TMS}^*}$ at the optimal time  and phase $\phi_0=\pi/4$ in comparison to the target state $\ket{\psi_{\rm T}}$ and the standard quantum limit (SQL) and Heisenberg limit (HL).}
   \label{fig:UnitaryTMS}
\end{figure}

We consider a cavity setup as the one depicted in Fig.~\ref{fig:cavity}, where atoms are confined in a deep one-dimensional magic-wavelength optical lattice within an optical cavity, effectively suppressing their motion along the lattice. A single cavity mode, characterized by an angular frequency $\omega_{\rm C}$ and power decay rate $\kappa$, couples to a long-lived transition between an excited state $\ket{\uparrow}$ and a ground state $\ket{\downarrow}$ with single-photon Rabi frequency of $2g$. The atomic transition has an angular frequency $\omega_{\rm A}$ and a natural decay rate $\gamma \ll \kappa$. 

In the far-detuned limit, $|\Delta|=|\omega_{\rm A}-\omega_{\rm C}|\gg \kappa, g\sqrt{N}$, the cavity field can be adiabatically eliminated, while virtual photons  mediate effective  unitary interactions between the atoms, which can be described by the effective spin Hamiltonian \cite{muniz2020exploring}
\begin{align}
    H_{\rm C} = \raisedchi \left(J_+^A+J_+^B\right) \left(J_-^A+J_-^B\right). 
    \label{eq:H_cavity}
\end{align}
The interaction strength $\raisedchi =4g^2 \Delta/\left(4\Delta^2+\kappa^2\right)$ can be tuned by changing the detuning $\Delta$.  In order to realize a spin Hamiltonian of this particular form, the atoms need to couple uniformly to a single cavity mode, which can be achieved by trapping the atoms in a lattice with the right spacing or in a ring cavity. 

Besides unitary interactions, photons leaking out of the cavity also lead to dissipation in the form of collective emission, described by the collective jump operator $L_{\Gamma}=\sqrt{\Gamma/2}\left(J_-^A+J_-^B\right)$, where $\Gamma = 4 g^2 \kappa /\left(4\Delta^2 +\kappa^2\right)$.  By increasing the detuning $\Delta$, the ratio $\Gamma/\raisedchi$ can, in principle, be made arbitrarily small, thus supressing collective decay. However, this also reduces the overall interaction strength $\raisedchi$, eventually making emission into free space, at rate $\gamma$, the dominant source of decoherence. The jump operator describing free-space emission from the excited state of atom $k$  is given by $L_{\gamma}^{(k)}=\sqrt{\gamma}\sigma_-^{(k)}$, where $k$ indexes the atoms in both ensembles. In practice, one must therefore choose a detuning that balances the detrimental effects of collective and free-space emission to minimize decoherence. The so-called collective cooperativity parameter, $NC=4 N g^2/(\kappa \gamma)$, determines the effectiveness of state preparation in the presence of these effects.

In Section \ref{sec:unitary_preparation}, we discuss two methods for preparing or approximating the target state by considering only the unitary part of the cavity Hamiltonian in Eq.~\eqref{eq:H_cavity}. In contrast, in Section \ref{sec:stochastic_preparation}, we propose an alternative route to generate a proxy of the target entangled state by starting from a suitably chosen initial product state and directly employing collective dissipation.

\subsection{Unitary generation of entanglement}
\label{sec:unitary_preparation}

\begin{figure}[t] 
   \centering
   \includegraphics[]{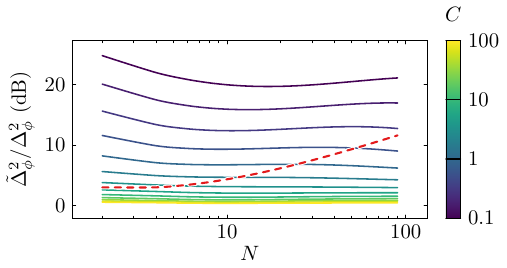}
 \caption{The optimal estimator variance, denoted as $\tilde{\Delta}_{\phi}^2$, attainable under a quench with the two-mode squeezing Hamiltonian while simultaneously being subject to collective and free-space emission. It is evaluated relative to the ideal variance achievable in the absence of collective and free-space emission, denoted as $\Delta_{\phi}^2$. $\tilde{\Delta}_{\phi}^2$ is optimized with respect to the quench duration and the detuning between the cavity and the atomic transition frequency. The red dashed line marks the particle number $N$ at which the standard quantum limit is exceeded for a given single-particle cooperativity $C$.
}
   \label{fig:TMS_Collective_Spontaneous_Emission}
\end{figure}

An approach to preparing the target state is to initialize the system in the state  $\ket{\psi_0}=\ket{\tfrac{N}{4},+\tfrac{N}{4},\tfrac{N}{4},-\tfrac{N}{4}}$, where all atoms in ensemble $A$  are in the excited state, and all atoms in ensemble $B$  are in the ground state. By adiabatically sweeping a field gradient $H_{\delta}(t)=-\delta_t(t) \left(J_z^A-J_z^B\right)$ in addition to the cavity Hamiltonian, from $\delta_t (0)/\raisedchi \gg 1$ to $\delta_t (t_{\rm final}) / \raisedchi\ll 1$, the target state can be prepared with arbitrarily high fidelity, provided the sweep is sufficiently slow. This is because the target state corresponds to the ground state of the cavity Hamiltonian when permutation symmetry within each sub-ensemble is imposed, whereas the initial state is an eigenstate of the field gradient. 
The speed at which the adiabatic sweep can be performed depends on the instantaneous energy gap of the Hamiltonian to the state closest in energy that shares the same symmetries  as the cavity Hamiltonian and the initial state. In the two limits $\delta_t / \raisedchi\gg1$ and $\delta_t/\raisedchi \ll 1$, the gap between the ground and first excited state is independent of $N$ and scales with $2\delta_t$ and $2\raisedchi$ respectively, (see Appendix \ref{app:adiabatic}). 
The non-collective gap, impose time scales that are challenging. With decoherence rates competing with the unitary dynamics, achieving a sweep that is adiabatic and at the same time fast enough compared to the decoherence rates requires $C\gg1$. This condition poses a significant challenge for currently available cavity systems, which typically operate in the regime $C\approx 1$. 

A more practical approach in the presence of decoherence is to generate a proxy of the target state by performing a quench using a Hamiltonian of the form
\begin{align}
    H_{\rm TMS} = \raisedchi\, i\, \left(J_+^AJ_-^B-J_-^AJ_+^B\right), 
    \label{eq:H_TMS}
\end{align}
applied to the same initial state $\ket{\psi_0}$. This Hamiltonian can be viewed as the atomic analog of the bosonic Hamiltonian responsible for generating two-mode squeezing \cite{sundar2023bosonic}. Such a Hamiltonian can be implemented by introducing additional one-axis twisting interactions of the form $H_{\rm OAT}=\raisedchi\left( J_z^AJ_z^A+J_z^BJ_z^B\right)$  to a spin-exchange Hamiltonian. This addition effectively cancels the interactions within each subensemble, since $\bm{J}^K \cdot \bm{J}^K = J_+^K J_-^K + J_z^K J_z^K$ does not generate dynamics when the instantaneous state is an eigenstate of this operator. In Appendix \ref{app:TMS_H}, we detail a potential method for realizing this interaction with multi-level atoms in a cavity. Note that the Hamiltonian derived in Appendix \ref{app:TMS_H} is equivalent to the expression in Eq.~\eqref{eq:H_TMS} up to a rotation generated by $J_z^-$. Specifically,
$i \left(J_+^A J_-^B - J_-^A J_+^B\right) = e^{+i\frac{\pi}{4} J_z^-} \left(J_+^A J_-^B + J_-^A J_+^B\right) e^{-i\frac{\pi}{4} J_z^-}$.

Figure~\ref{fig:UnitaryTMS} (a) illustrates the outcome of a quench designed to minimize the infidelity $I= 1 - \left|\braket{\psi_{\rm TMS}(t)|\psi_{\rm T}}\right|^2$ between the quenched state $\ket{\psi_{\rm TMS}(t)} = e^{-i H_{\rm TMS}\,t} \ket{\psi_0}$ and the target state. Notably, the infidelity asymptotically approaches a finite value of approximately 0.1. It is worth emphasizing that the quench dynamics under purely unitary evolution is constrained to happen in a DFS with $N/2$ atoms in the excited state, whose dimension grows linearly with $N$. Therefore, an exponential increase in infidelity, as observed in genuine many-body systems, is not expected. Nevertheless, it is remarkable that the infidelity asymptotically approaches a finite value significantly below 1.

An additional encouraging result is that the value of $\raisedchi\,t$ that minimizes the infidelity, despite lying beyond the regime of validity of the standard Holstein–Primakoff approximation~\cite{kurucz2010multilevel, sundar2023bosonic, khan2025generating} still exhibits a linear decrease with system size, up to a logarithmic correction. Time scales that decrease as $\log N/N$ are beneficial if the unitary interaction typically competes with spatially uncorrelated noise such as free-space emission. The optimal sensitivity of the quenched state is shown in Fig.~\ref{fig:UnitaryTMS} (b), demonstrating that it asymptotically has the same scaling as the target state, differing only by a constant prefactor of approximately 2.2.

\begin{figure}[t] 
   \centering
   \includegraphics[]{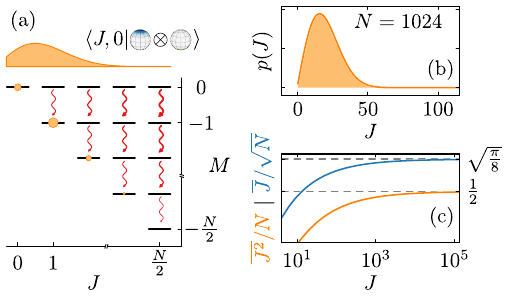} 
   \caption{(a) Collective emission (red curly arrows) evolves the initial state into a steady state, corresponding to a mixture of Dicke states located on the lower diagonal of the Dicke state ladder, characterized by quantum numbers $J$ and $M$. The steady-state distribution (orange circles) corresponds to the probability of projecting the initial state onto a specific $\ket{J,0}$ state (orange distribution). (b) The probability of projecting onto a given $\ket{J, 0}$ state, is determined for the initial state where the atoms in ensemble A are in the excited state and those in ensemble B are in the ground state. (c) The first moment, $\overline{J}=\sum_{J}J\, p(J)$, and the second moment, $\overline{J^2}=\sum_{J}J^2\, p(J)$,  of the distribution presented in panel (b) are shown for different atom numbers $N$, rescaled to highlight their respective asymptotic scaling. The dashed lines indicate the prefactors corresponding to the asymptotic scaling.}
   \label{fig:DickeLadder}
\end{figure}

In Fig.~\ref{fig:TMS_Collective_Spontaneous_Emission}, we analyze the interplay between the unitary generation of entanglement and the detrimental effects of noise in the system. This is achieved by optimizing over the detuning $\Delta$, which governs the ratios $\Gamma/\raisedchi$ and $\gamma/\raisedchi$, as well as the quench duration. 

For system sizes up to $N \leq 90$, the dynamics of two interacting atomic ensembles subject to both collective and free-space emission can be simulated by exact diagonalization of the Lindblad master equation. This is made computationally tractable by restricting the evolution to a Liouville space whose dimension scales as $\propto N^6$, leveraging the permutation symmetry of the free-space decay jump operators \cite{chase2008collective, gong2016steady, shammah2018open, nadolny2025nonreciprocal}. For larger particle numbers, a viable approach is to employ a Monte Carlo wavefunction simulation \cite{zhang2018monte}.

By optimizing the detuning $\Delta$, a trade-off is achieved that balances the detrimental effects of both collective and free-space emission. Our results indicate that, once the cooperativity-enhanced collective coupling $N\, C$ is sufficiently large to surpass the standard quantum limit, the presence of noise does not alter the scaling behavior of the estimator variance compared to the noiseless Heisenberg scaling. Notably, the deviations manifest solely in the prefactor, which increases as $C$ decreases, while the asymptotic scaling remains seemingly unaffected. 

Extrapolating this behavior to larger values of $N\, C$, we expect that a sensitivity enhancement consistent with Heisenberg scaling should be attainable in existing cavity-based experimental platforms. This stands in contrast to cavity-mediated protocols aimed at preparing GHZ states using standard one-axis twisting or quantum non-demolition schemes. When free-space emission errors are taken into account, captured by a finite single-particle cooperativity, such approaches cannot prepare GHZ states and can yield at most a $1/\sqrt{NC}$ improvement over the standard quantum limit \cite{chu2021quantum, koppenhofer2023revisiting}. This limitation is tied to the GHZ-state preparation time, which even in the fastest case (one-axis twisting) is independent of system size and therefore cannot compete with the rate of a single free-space emission error, which scales as $\gamma N$.

Furthermore, the performance of this scheme could be enhanced by dynamically varying the detuning during the quench given that as the instantaneous state approaches the target state during the evolution, the detrimental effect of collective emission diminishes. Consequently, the detuning can be gradually reduced, thereby also decreasing the ratio $\gamma/\chi$.

In the next section, we introduce an alternative approach that circumvents the need to engineer the desired unitary interactions directly. Instead, we exploit collective superradiance decay as the mechanism for generating metrologically useful entanglement.

\subsection{Stochastic preparation of entanglement}
\label{sec:stochastic_preparation}
Collective emission becomes the dominant dynamical process when the cavity is tuned close to resonance with the atomic transition frequency. This strong dissipative mechanism can be exploited to generate entanglement. Conceptually, this is akin to the approach in Ref.~\cite{masson2019rapid}, although that work instead focused on three-level atoms and identified states sensitive to a signal encoded by an operator resembling the two-mode squeezing Hamiltonian of Eq.~\eqref{eq:H_TMS}. Ref.~\cite{mamaev2025non} has proposed a dissipative approach for generating a state closely related to our target state, emphasizing its potential for multiparameter estimation. Their implementation extends dissipative spin-squeezing protocols to two subsystems by employing two cavity modes, four internal atomic levels, and several laser drives. While this scheme provides an instructive demonstration of dissipative entanglement generation, our aim is to realize comparable steady-state properties with reduced experimental complexity.

Specifically, by relying solely on collective emission, we now show how to prepare a statistical mixture of Lieb–Mattis states with different excited-to-ground state imbalances. While this simplified approach does not achieve Heisenberg scaling, it nevertheless provides a scalable sensitivity enhancement that exceeds what is attainable with spin-squeezed states in the common-mode noise–limited regime. Moreover, the protocol is directly compatible with the cooperativities achievable in current experimental platforms.

The basic  idea is to use collective emission acting  on the right initial state to generate an entangled density matrix . Here we will consider the same initial state $\ket{\psi_0}=\ket{\tfrac{N}{4},+\tfrac{N}{4},\tfrac{N}{4},-\tfrac{N}{4}}$ as previously. Projecting this state onto the $\ket{J,0}$ states, which satisfy $\bm{J}\cdot\bm{J}\ket{J,M}=J(J+1)\ket{J, M}$ and $J_z^+\ket{J,M}=M\ket{J,M}$,  results in the probability distribution shown in Fig.~\ref{fig:DickeLadder}(b,c),
\begin{align}
    p(J) &= \left|\braket{J, 0|\tfrac{N}{4},+\tfrac{N}{4},\tfrac{N}{4},-\tfrac{N}{4}}\right|^2\notag \\
    & \approx \frac{2J+1}{N / 2 + 1}\left(\frac{N  + 2 - J}{N  + 2 + J}\right)^{J+1},
\end{align} where the approximation is valid in the limit of asymptotically large $N$. Figures~\ref{fig:DickeLadder}(b,c) further illustrate that, for large $N$, the distribution becomes asymptotically centered at a value scaling $\propto \sqrt{N}$, with a standard deviation that also scales $\propto \sqrt{N}$. This means that the distribution has a small but finite probability to be in the target state but generally skewed towards small but non-zero values of $J$. 

\begin{figure}[t] 
   \centering
   \includegraphics[]{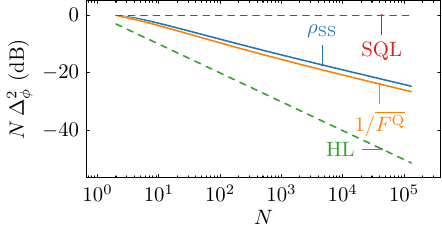} 
   \caption{The estimator variance of the steady state $\rho_{\rm SS}$ [Eq.~\eqref{eq:steady_state}], prepared via collective emission, is evaluated at the optimal phase $\phi=\pi/4$ for various atom numbers $N$ and the measurement $\mathcal{M}$. The resulting performance is benchmarked against the QCRB associated with the inverse of the average QFI [Eq.~\eqref{eq:QCR_SS}] of the pure states appearing in the spectral decomposition of $\rho_{\rm SS}$, as well as against the standard quantum limit and the Heisenberg limit.
 }
   \label{fig:EstVarScaling}
\end{figure}

This probability distribution contains all the necessary information to describe the steady state under collective emission sketched in Fig.~\ref{fig:DickeLadder}(a). This follows from the fact that the jump operator $L_{\Gamma}=\sqrt{\Gamma}J_-$ commutes with $\bm{J} \cdot \bm{J}$, which ensures that the $J$ quantum number distribution remains invariant under the evolution and merely reduces $M$, since $J_-\ket{J,M}\propto \ket{J, M - 1}$. Consequently, each $\ket{J,M}$ state decays into the corresponding state with the minimal number of excited state atoms, $\ket{J, -J}$. The resulting steady state is therefore given by
\begin{align}
    \rho_{\rm SS} = \sum_{J= 0}^{N/2}p(J)\, \ket{J, -J}\bra{J, -J}. 
    \label{eq:steady_state}
\end{align}
The steady state corresponds to a mixture of $\ket{J, -J}$ states, which all possess a large QFI for differential phase sensing: 
\begin{align}
    F^{\rm Q}_{\ket{J, -J}}= \frac{4N + N^2 - 8J - 4J^2}{3+2J} 
    \label{eq:FQ_JJ}
\end{align}
at $\phi = \pi / 4$ if $J\ll N$,  see Appendix~\ref{app:ang_mom_states}. Furthermore, the states $\ket{J, -J}$ correspond to the Lieb-Mattis ground state when the Lieb-Mattis Hamiltonian is projected onto a DFS with $N/2 - J$ atoms in the excited state.

Figure~\ref{fig:EstVarScaling} shows that, in the vicinity of the optimal phase $\phi = \pi/4$, the estimator variance for the steady state exhibits a scaling behavior closely matching the QCRB for the corresponding mixed state (see Appendix~\ref{app:avgQFI}),
\begin{align}
    \frac{1}{\Delta_{\phi}^2}\leq \overline{F^{\rm Q}}= \sum_{J} p(J)\, F^{\rm Q}_{\ket{J, -J}}.
    \label{eq:QCR_SS}
\end{align}
Numerical calculations reveal asymptotic scaling exponents of $- 0.50$ for the variance and $-0.51$ for the bound, both corresponding to the same scaling as the state  $\ket{\overline{J}, -\overline{J}}$ where $\overline{J }$ is the $J$ that is closest to the mean value  $\sum_J J\, p(J)$. 

We draw attention to another opportunity that becomes particularly relevant when the time required to prepare the initial state is short compared to the phase accumulation time. In this regime, measuring the photons leaking out of the cavity is advantageous, because these photons reveal which state $\ket{J,-J}$ has been prepared. With this information, one can restart the state preparation process if the detected photon count is too high (i.e., the $J$ quantum number of the stochastically prepared state is too large). While this procedure can reduce the estimator variance, its effectiveness strongly depends on the relative timescales of the state preparation and the phase accumulation, as well as on the photon count threshold chosen to discard a prepared state.

\begin{figure}[t] 
   \centering
   \includegraphics[]{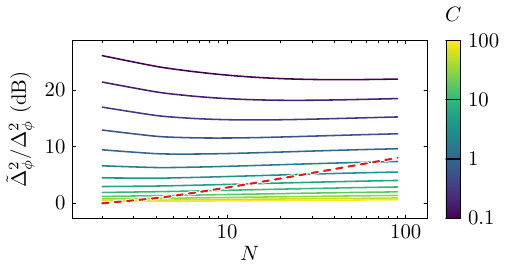}
 \caption{The optimal estimator variance attainable under stochastic entanglement generation, while being subject to free-space emission, denoted as $\tilde{\Delta}_{\phi}^2$, is evaluated relative to the variance achievable in the absence of free-space emission, denoted as $\Delta_{\phi}^2$. $\tilde{\Delta}_{\phi}^2$ is optimized with respect to the duration for which the initial state $\ket{\psi_0}$ undergoes collective and free-space emission. The red dashed line marks the particle number $N$ at which the standard quantum limit is exceeded for a given single-particle cooperativity $C$.
}
   \label{fig:Collective_Spontaneous_Emission}
\end{figure}

In Fig.~\ref{fig:Collective_Spontaneous_Emission}, we analyze the influence of free-space emission on the stochastic state preparation process. In the near-resonant bad cavity regime $N\Gamma\approx NC\gamma $, an increase in collective cooperativity enhances the favorable collective emission relative to the detrimental free-space emission rate. Notably, the scaling of the estimator variance remains largely unaffected even as cooperativity decreases. We emphasize that this is only feasible because the timescales required to prepare a state sufficiently close to the steady state decrease with $N$, thereby compensating for the $N$-fold increased probability of free-space emission events.

If this trend persists for larger systems, it suggests that significant improvements beyond the standard quantum limit could be realized for ensembles of a few hundred atoms. Such improvements appear feasible for cooperativity values around
 $C\approx 0.4$, which are within reach of current experimental setups \cite{norcia2018cavity, muniz2020exploring}.

Another notable advantage of the stochastic state preparation process is its robustness to fluctuations in the number of atoms within the subensembles. Assuming a fixed total number of atoms with an imbalance of $N^{\rm I}=|N^A-N^B|$, the probability distribution satisfies $p(J')=0$ for $J'<N^{\rm I}$, while the proportions of the distribution remain similar for  $J'>N^{\rm I}$. This can be understood as a consequence of the reduced number of atoms available in one of the ensembles to form singlet pairs. As a result, the unpaired atoms relax to the ground state, leading to a reduction in the total number of excited atoms in the steady state.

Consequently, as long as the imbalance satisfies  $N^{\rm I}\ll \sqrt{N}$, no significant increase in the estimator variance is expected. Additionally, this scheme benefits from the fact that the time required to reach the steady state is not a fine-tuned parameter dependent on specific particle numbers.

\section{Outlook}
\label{sec:Outlook}

In this work, we have investigated entanglement-enhanced sensing of a phase difference encoded in two atomic ensembles. By leveraging the concept of a DFS, we have analyzed an interferometric scheme that is intrinsically robust against common-mode noise. Moreover, we identified the ground state of the Lieb–Mattis Hamiltonian and showed that, when combined with an appropriate measurement strategy, it yields an estimator variance that scales at the Heisenberg limit. Importantly, compared to other protocols that achieve Heisenberg scaling, this approach exhibits more favorable scaling properties in the presence of preparation noise. 

Having established this state as the target for entanglement-enhanced sensing, we have proposed and analyzed two state preparation protocols that exploit cavity-mediated entanglement to approximate the target state. The first protocol relies on an interaction that can be interpreted as a bosonic two-mode squeezing interaction, which for example can be realized using multilevel alkaline-earth atoms, as described in Appendix~\ref{app:TMS_H}.

The second protocol stochastically prepares an entangled density matrix through collective emission into a cavity mode. Although it does not achieve Heisenberg-limited scaling of the estimator variance, it provides a sensitivity enhancement over the standard quantum limit by a factor scaling as $1/\sqrt{N}$. This surpasses what is achievable with two independently spin-squeezed ensembles under dominant common-mode noise. Importantly, the protocol is directly compatible with current cavity-based experimental platforms and is expected to be robust against atom number fluctuations on the order of $\sqrt{N}$.

Additionally, we have evaluated the impact of collective and free-space emission on both preparation schemes. Our analysis indicates that, while these decoherence mechanisms increase the prefactor of the estimator variance, they do not affect its fundamental scaling. For collective cooperativity values achievable in state-of-the-art experiments, a substantial enhancement beyond the standard quantum limit remains feasible.

While our discussion has focused  on  cavity  mediated infinite-range interactions,  ---  currently the most relevant experimental architecture for quantum enhanced  sensing as it holds the record level of  metrologically relevant entangled states \cite{cox2016deterministic, hosten2016measurement}---,  a natural extension of this work is the consideration of systems with finite-range
interactions. Such interactions naturally arise in a variety of quantum sensing platforms, including optical lattice \cite{milner2025coherent} or tweezer \cite{eckner2023realizing} clocks, trapped ions \cite{franke2023quantum}, and Rydberg atoms \cite{bornet2023scalable, hines2023spin}. In  Appendix \ref{app:parentHamiltonian}, we discuss a finite-range Hamiltonian that serves as a parent Hamiltonian for the target state, raising the question of whether such a Hamiltonian can be experimentally realized and whether its ground state can be prepared, for instance by an adiabatic ramp or a quench. Other directions to explore are  variational methods using parameterized quantum circuits to prepare the ground state, thereby minimizing the achievable estimator variance \cite{kaubruegger2019variational, koczor2020variational, kaubruegger2021quantum} or methods that use mid-circuit measurements and feedback \cite{yu2026efficient}.

In many practical applications, the quantity of interest is not the phase itself, but the frequency that accumulates as a phase over the interrogation time. The precision of frequency estimation improves with longer interrogation durations, during which the atoms interact with the signal. However, a fundamental limitation to indefinitely extending the interrogation time arises from the finite lifetime of the excited state.  To overcome this constraint, one may utilize pairs of fermionic atoms prepared in dark states, in which emission into free space is strongly suppressed \cite{pineiro2019dark}. By transferring the atoms into these dark states during the phase encoding, this approach effectively lifts the fundamental limitation on the interrogation time.

Another promising avenue for future extensions involves applying continuous measurement protocols \cite{gammelmark2014fisher, shankar2019continuous,yang2023efficient, duan2025concurrent}, in which for example the photons leaking from the cavity are monitored in real time while the differential phase is simultaneously encoded onto the atomic system. This approach could offer additional pathways to surpass classical sensing limits by continuously tracking the phase evolution. 

\section{Acknowledgments}
We thank Marcus Bintz for insightful discussions and Maya Miklos, Theodor Lukin Yelin, and Adam Kaufman for useful feedback on the manuscript.  This work is  supported by the Vannevar Bush Faculty Fellowship, AFOSR FA9550-24-1-0179, the NSF JILA-PFC PHY-2317149 and NSF QLCI awards OMA-2016244 and  OMA-2120757, the U.S. Department of Energy, Office of Science, National Quantum Information Science Research Centers, Quantum Systems Accelerator the Heising-Simons foundation and  NIST. RK acknowledges funding by the German National Academy of Sciences Leopoldina under grant LPDS 2024-08. KM acknowledges the JILA visiting fellows program and support from the Danish National Research Foundation (Center of Excellence “Hy-Q,” grant number DNRF139). CH is supported by the  Carlsberg Foundation through the “Semper Ardens” Research Project QCooL. S.R.M. is supported by the
NSF QLCI award OMA-2120757. E.A.~and A.V.G.~were supported in part by ARL (W911NF-24-2-0107), DARPA SAVaNT ADVENT, AFOSR MURI, the DoE ASCR Quantum Testbed Pathfinder program (awards No.~DE-SC0019040 and No.~DE-SC0024220), NSF STAQ program, and NQVL:QSTD:Pilot:FTL. E.A.~and A.V.G.~also acknowledge support from the U.S.~Department of Energy, Office of Science, Accelerated Research in Quantum Computing, Fundamental Algorithmic Research toward Quantum Utility (FAR-Qu). A.S. acknowledges support by the Department of Science and Technology, Govt. of India through the INSPIRE Faculty Award (DST/INSPIRE/04/2023/001486), by the Anusandhan National Research Foundation (ANRF), Govt. of India through the Prime Minister’s Early Career Research Grant (PMECRG) (ANRF/ECRG/2024/001160/PMS) and by IIT Madras through the New Faculty Initiation Grant (NFIG). 

\textit{Note added.}—After submission of this work, Ref.~\cite{chu2025reconfigurable} reported an alternative scheme for the dissipative preparation of the target and related states.

\appendix 
\section{Separable strategies}
\label{app:seperable}

In this section, we establish the limits on sensing the differential phase using two spin ensembles, with no entanglement shared between the ensembles but allowing for spin squeezing within each ensemble. To achieve this, we employ the classical Fisher information, which provides a lower bound on the sensitivity of any unbiased estimator. This bound, in turn, applies to specific estimation strategies, such as ellipse fitting \cite{foster2002method, fixler2007atom, rosi2015measurement, barrett2016dual, langlois2017differential, parker2018measurement, marti2018imaging, elliott2023quantum, bothwell2022resolving, zheng2023lab}.

\begin{figure}[t] 
   \centering
   \includegraphics[]{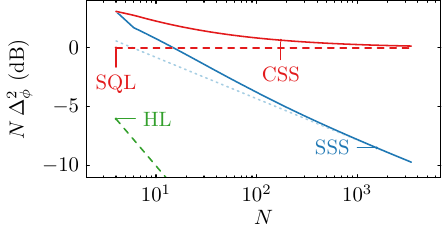} 
   \caption{The Fisher information in Eq.~\eqref{eq:CFI} is evaluated at the phase $\phi$ that maximizes the Fisher information for a sensor network comprising two coherent spin states (CSS) and two spin-squeezed states (SSS). In the case of spin-squeezed states, the Fisher information is further optimized over the degree of spin squeezing. The dotted lines represent the asymptotic scaling of the Fisher information in the limit of large atom numbers $N$. The dashed green line corresponds to the fundamental Heisenberg limit, given by  $F\leq N^2$. 
 }
   \label{fig:SeperableStates}
\end{figure}

We examine two scenarios. In the first scenario, both ensembles are initialized in a coherent spin state, denoted as
$\ket{\psi_{\rm CSS}}=e^{-i \tfrac{\pi}{2}J_y}\ket{\tfrac{N}{4},-\tfrac{N}{4},\tfrac{N}{4},-\tfrac{N}{4}}$, where all spins are aligned along the $x$-axis.  In the second scenario, the ensembles are prepared in spin-squeezed states using the one-axis-twisting interaction \cite{kitagawa1993squeezed}, represented as $\ket{\psi_{\rm SSS}}=e^{-i \nu J_x}e^{-i \mu \left(J_z^AJ_z^A+J_z^BJ_z^B\right)}\ket{\psi_{\rm CSS}}$, where the parameter $\nu$ is chosen such that the direction of the squeezed variance is along the $y$-axis.

Next, the phase $\phi$ is encoded onto the respective states according to  $\ket{\psi_{\phi}}=e^{-i\phi J_z^-}\ket{\psi}$. Since both states are not confined to a DFS, it is necessary to account explicitly for random fluctuations of the common phase. This is achieved by considering a density matrix of the form
\begin{align}
    \rho_{\phi}=\frac{1}{2\pi}\int_0^{2\pi} d\Phi\, e^{-i \Phi J_z^+}\ket{\psi_{\phi}}\bra{\psi_{\phi}}e^{+i \Phi J_z^+}.
\end{align}
Here, $J_z^-$ and $J_z^+$ are the operators associated with the differential and common phases, respectively. Since these operators commute, their effects can be treated consecutively.

The final step of the interferometric sequence involves applying a $\pi/2$-pulse around the $x$-axis and independently measuring the population imbalance,  $M^A$ and $M^B$, of each ensemble independently.  The conditional probability of observing a specific population imbalance, given the encoded differential phase $\phi$, is expressed as
\begin{align}
    p(M^A, M^B|\phi ) = \tr\Big[\Pi_{M^A, M^B}\, e^{-i\tfrac{\pi}{2}J_x}\rho_{\phi}e^{+i\tfrac{\pi}{2}J_x}\Big],
    \label{eq:CFI}
\end{align}
where $\Pi_{M^A, M^B}=\ket{\tfrac{N}{4},M^A,\tfrac{N}{4},M^B}\bra{\tfrac{N}{4},M^A,\tfrac{N}{4},M^B}$ is the projection operator onto the state characterized by the specified population imbalances.

The classical Fisher information, derived from the conditional probability distribution, is given by
\begin{align}
    F=\sum_{M^A,M^B}\frac{\left(\frac{\partial}{\partial \phi}p(M^A, M^B|{ \phi})\right)^2}{p(M^A, M^B|{\phi})} 
\end{align}
and serves as a lower bound for the estimator variance in Eq.~\eqref{eq:EstVar}, as established by the Cramér-Rao bound (CRB), $\Delta_{\phi}^2\geq 1 /F$, which is a tighter version of the QCRB that explicitly depends on the measurement. It is important to emphasize that the CRB derived here represents a fundamental lower bound for any unbiased ellipse-fitting method. However, achieving this bound with a maximum likelihood estimator would require precise knowledge of the probability distribution $p(M^A, M^B|\phi )$, which may be challenging to obtain experimentally. Nonetheless, the Fisher information provides a meaningful benchmark for evaluating the performance of the measurement protocols discussed in the main text.

In Fig.~\ref{fig:SeperableStates}, the inverse Fisher information for two coherent spin states initially exceeds the standard quantum limit but asymptotically approaches the standard quantum limit as $N$ becomes large. Whereas, independently squeezing the individual ensembles enables surpassing the standard quantum limit; however, the presence of common phase fluctuations diminishes the potential benefits of spin squeezing. In the absence of common phase noise, spin squeezing generated via the one-axis twisting Hamiltonian provides an improvement over the standard quantum limit, with an asymptotic scaling of $N^{-2/3}$ \cite{kitagawa1993squeezed}. 
To determine the asymptotic scaling behavior in the presence of phase noise, we fit the inverse Fisher information optimized over the squeezing strength. The resulting scaling exponent, represented by the dashed blue line in Fig.~\ref{fig:SeperableStates}, is found to be $-0.35$, which is in agreement with the previously reported exponent of $-1/3$ obtained from ellipse fitting \cite{corgier2025optimized}.
This notably smaller scaling exponent arises from the fact that the measurement variance of a spin-squeezed state is reduced only for specific phase values, while it increases for others. The presence of random common phase fluctuations leads to a situation where measurement variance is averaged over all phase values. In order to compensate, the optimal amount of squeezing is smaller than in the noiseless case. 

This underscores the advantages of a sensor operating within a DFS, where scaling as favorable as $N^{-1}$ can be achieved. Even in the most experimentally realistic scenario, where entanglement is generated via collective emission, the achieved scaling of  $N^{-1/2}$ significantly outperforms the scaling of two independently squeezed ensembles.

We also remark on an alternative separable scheme that, in principle, enables Heisenberg-limited sensitivities, but remains demanding to realize with current large-scale quantum-sensor hardware. This approach relies on preparing a sine state \cite{buvzek1999optimal} as the initial state and performing a measurement of the phase operator \cite{pegg1988unitary}. Such a protocol allows phase estimation with an estimator variance of $\pi^2/N^2$, independent of the encoded phase. Implementing this strategy on both ensembles would enable ellipse fitting or maximum-likelihood estimation of the differential phase with Heisenberg scaling. However, neither the preparation of sine states nor the direct measurement of the phase operator are straightforward with available platforms. Possible workarounds include approximations based on variational circuits \cite{kaubruegger2021quantum} or protocols employing ensembles of GHZ states of different sizes in conjunction with a quantum Fourier transform \cite{direkci2026heisenberg}. Both of these approaches, or variants thereof, have thus far only been demonstrated in small-scale quantum systems \cite{marciniak2022optimal, cao2024multi, finkelstein2024universal}. This highlights the importance of developing hardware-efficient schemes capable of achieving comparable performance in larger, experimentally relevant platforms.

\section{Photon Measurement}
\label{app:Measurement}

\begin{figure}[t] 
   \centering
   \includegraphics[]{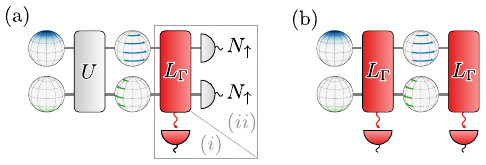} \\
   \caption{(a) After phase encoding, a measurement quench is applied by tuning the cavity into resonance, such that the subsequent dynamics are dominated by collective emission through the cavity mode, described by the jump operator $L_{\Gamma}$. (i) Detecting all photons leaking from the cavity until the steady state is reached implements a measurement of the diagonal operator in Eq.~\eqref{eq:photon_measurement}. (ii) Equivalent information can be obtained from a post-quench measurement of the remaining excited-state population $N_{\uparrow}$, e.g., via excitation-number-resolved vacuum--Rabi spectroscopy, referenced to the initial excitation number. (b) For a stochastically prepared initial state, achieving the full dynamic range additionally requires collecting the photons emitted during state preparation, which reveals which $\ket{J,-J}$ state was prepared in that particular run.}
   \label{fig:Measurement_schematic}
\end{figure}

Since directly implementing the optimal measurement in Eq.~\ref{eq:Measurement} is challenging, we instead consider an experimentally accessible readout scheme that can still saturate the QCRB. Specifically, we use the protocol sketched in Fig.~\ref{fig:Measurement_schematic}(a): after phase encoding, the cavity is abruptly tuned (quenched) into resonance with the relevant atomic transition. In the following, we refer to this step as the \emph{measurement quench}. In this regime, the dynamics are dominated by collective emission described by the jump operator $L_{\Gamma}$.

If all photons leaking from the cavity are detected until the atoms relax into the steady state of $L_{\Gamma}$, then the measurement record (the total number of detected photons) is equivalent to a projective measurement of the operator
\begin{align}
    O = \left(J + M\right)\ket{J,M}\bra{J,M}, 
    \label{eq:photon_measurement}
\end{align}
which is diagonal in the $\ket{J,M}$ basis. This interpretation arises from the fact that a state $\ket{J,M}$ decays to $\ket{J,-J}$ via the emission of $J + M$ photons, which, in the bad cavity limit, are rapidly lost and can thus be detected as they exit the cavity.

Rather than resolving the number of photons during the measurement quench, the same information can be obtained by measuring the remaining excited-state population after the quench, as sketched in Fig.~\ref{fig:Measurement_schematic}(a)(ii). Comparing the post-quench excitation number to the initially prepared excitation number (equal to the number of atoms in ensemble $A$) yields a measurement record equivalent to photon counting, up to additional imperfections from noise processes during the protocol, for example free-space spontaneous emission.

\begin{figure}[t] 
   \centering
   \includegraphics[]{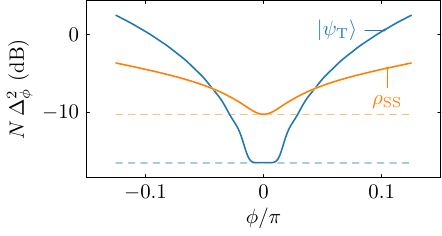} \\
   \caption{Estimator variance associated with measuring the number of photons emitted from the cavity after the phase has been encoded, evaluated for both the target state $\ket{\psi_{\rm T}}$ and the steady state resulting from collective emission $\rho_{\rm SS}$ of a system of $N=128$ atoms. The dashed lines indicate the corresponding QCRBs.}
   \label{fig:PhotonMeasurement}
\end{figure}

An auxiliary cavity mode can strongly couple atoms in $\ket{\uparrow}$ to a readout-only ancilla state, producing a vacuum–Rabi splitting whose size depends on the number of atoms in the excited state. For single-particle cooperativity $>1$, the splitting can be resolved well enough to enable single-excitation resolution \cite{chen2011conditional, chen2014cavity}.

To benchmark the performance of an explicit photon-counting implementation, analyze the corresponding estimator variance, shown in Fig.~\ref{fig:PhotonMeasurement}. For both the target state and the steady state resulting from collective emission, the QCRB is saturated at $\phi = 0$. A limitation of this measurement scheme, however, is that at $\phi = 0$, the target state is an eigenstate of the measurement operator and the steady state is a mixture of eigenstates. Therefore in  this limit, both the numerator and denominator of the estimator variance vanish, making the sensor particularly vulnerable to additional noise in the system which overwrites the vanishing numerator and results in a large estimator variance. 

\subsection{Dynamic range}

If we restrict attention to an estimator whose variance is inferred solely from the mean detected photon number, an apparent limitation is the small dynamic range: there is only a narrow interval of phases over which the estimator variance remains low. For the target state, the variance stays approximately constant only within a window of width $\sim 1/N$ around $\phi=0$, and it rises as $\phi$ moves away from zero. By contrast, for the stochastically prepared state there is no comparably flat plateau; instead, the variance grows more gradually with $|\phi|$. 

\begin{figure}[t] 
   \centering
   \includegraphics[]{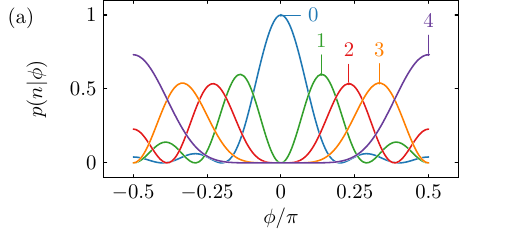} \\
   \includegraphics[]{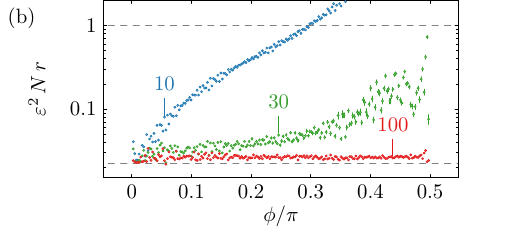} \\
   \caption{(a) Conditional probabilities of a $N=8$ sensor consisting the target state and a measurement of the number $n$ of photons leaking outside of the cavity  as function of the encoded phase $\phi$. (b) Mean-squared error $\varepsilon^2$ of a maximum-likelihood phase estimation protocol, defined as the squared difference between the encoded phase and the phase inferred by maximizing the likelihood function. The error is obtained from $r = 10, 30,$ and $100$ simulated measurement records of photons leaking from the cavity for an $N = 128$ atom system prepared in the target state $\ket{\psi_{\rm T}}$. The likelihood is optimized over the interval $[0,\pi/2]$, over which the photon-number probability distributions as the one shown in panel (a) are unique. The upper horizontal dashed line indicate the standard quantum limit and the lower the quantum Cramér-Rao bound of the target state.}
   \label{fig:dynamic_range}
\end{figure}

To put this in perspective, consider GHZ interferometry, where the dynamic range is fundamentally limited to $\sim \pi/N$. This restriction follows from the $2\pi/N$ periodicity of the measurement statistics, which makes phases outside $(-\pi/(2N),+\pi/(2N))$ ambiguous unless one combines data from multiple effective ensemble sizes \cite{cao2024multi,finkelstein2024universal}. By contrast, our full photon-counting distributions are mirrored about $0$ and $\pi/2$ and do not repeat within $(0,\pi/2)$ [Fig.\ref{fig:dynamic_range}(a)], enabling unambiguous estimation over an order-unity interval that is independent of $N$. Moreover, a maximum-likelihood estimator that uses the entire photon-counting record can approach the QCRB across this full range given sufficient repetitions (e.g., $r=100$ in Fig.\ref{fig:dynamic_range}(b)), providing a practical route to overcoming the limited dynamic range suggested by mean-based estimators.

In addition, the stochastic nature of the state-preparation stage can further constrain the dynamic range. If the phase is inferred only from the photons detected during the measurement quench, the dynamic range is reduced to a window of order $1/N$, as observed in simulations (not shown). By contrast, if one also records the photons emitted during the stochastic preparation stage as sketched in Fig.~\ref{fig:Measurement_schematic}, the measurement record retains run-by-run information about which $\ket{J,-J}$ Lieb--Mattis state was prepared prior to phase encoding. This additional pre-measurement information restores the dynamic range to the full interval $(0,\pi/2)$.

\subsection{Detection efficiency}

\begin{figure}[t] 
   \centering
   \includegraphics[]{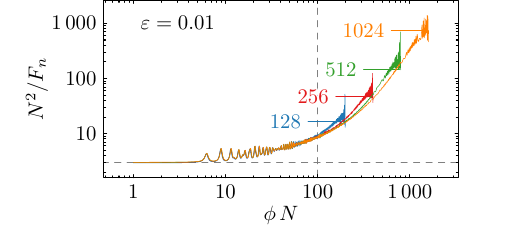} 
   \caption{ Finite detection errors for the target-state protocol. Rescaled inverse Fisher information $N^{2}/F_{N}$ (the Cramér–Rao bound for unbiased phase estimation) as a function of the rescaled phase $\phi N$, shown for $N=128,256,512,1024$ at a fixed detection-error probability $\varepsilon=0.01$. For each N, the phases span the full dynamic range $\phi\in(0,\pi/2)$. The horizontal dashed line marks the QCRB scaling threshold $N^{2}/F_{N}>3$.  The vertical dashed line indicates $1/\epsilon$. 
}
   \label{fig:Detection_Noise}
\end{figure}

A practical concern for the photon-counting measurement is how finite detection fidelity impacts the achievable phase sensitivity. To quantify this effect, we model imperfect detection by assigning an independent probability $\varepsilon$ for each photon reaching the detector to be missed. In practice, error mechanisms during the readout, such as free-space emission events occurring throughout the detection window, reduce the number of photons emerging from the cavity and therefore contribute to an effective detection infidelity beyond the intrinsic detector inefficiency. The appropriate value of $\varepsilon$ is thus an effective parameter that depends among other things on the detector performance and on the timescale of the collective-emission pulse.

If $p_n(\phi)$ denotes the ideal probability to detect $n$ photons for an encoded phase $\phi$, then the experimentally observed distribution is obtained by binomial thinning,
\begin{align}
\tilde{p}_{n,\varepsilon}(\phi)
=(1-\varepsilon)^n\sum_{n’\ge n}\binom{n’}{n}\varepsilon^{n’-n}p_{n’}(\phi).
\end{align}
From $\tilde{p}_{n,\varepsilon}(\phi)$ we compute the classical Fisher information $F_n = \sum_{n=0}^{N / 2} \frac{\left(\frac{\partial}{\partial_{\phi}}\tilde{p}_{n,\varepsilon}(\phi)\right)^2}{\tilde{p}_{n,\varepsilon}(\phi)}$ and the corresponding Cramér–Rao bound $ 1 / F_n$, shown in Fig.~\ref{fig:Detection_Noise}.

The collapse of the CRB for $\phi \lesssim 1/(\varepsilon N)$ shows that, in this regime, detection noise preserves the scaling of the sensitivity: it enters only as a constant prefactor relative to the detection-noise–free, Heisenberg-scaling performance of the target state. Equivalently, finite detection fidelity restricts the phase interval over which Heisenberg-scaling sensitivities are achievable to a dynamic range of order $1/(\varepsilon N)$, which remains larger than the $\sim 1/N$ range of a GHZ interferometer with spin-parity readout. If the detection error could be reduced further to scale as $\varepsilon \sim 1/N$—the stringent requirement for recovering Heisenberg scaling with GHZ-state and parity measurement protocol—then Heisenberg scaling would be restored over $(0, \pi/2)$ dynamic range in our protocol as well.

We further note that the curves in Fig.~\ref{fig:Detection_Noise} span the entire interval $(0,\pi/2)$ and, for the error rate shown, remain below the corresponding standard-quantum-limit sensitivity throughout the full dynamic range. Recent advances in single-photon detector technology report detection fidelities approaching $\sim 0.98$ \cite{reddy2020superconducting, ding2025photon}, placing the proposed photon-counting readout within reach of current experimental capabilities.

\section{Lieb-Mattis Hamiltonian}
\label{app:parentHamiltonian}
In this section, we  discuss the parent Hamiltonian of the target state Eq.~\eqref{eq:target_state} which is commonly referred to as the Lieb-Mattis Hamiltonian \cite{beekman2019introduction}
\begin{align}
    H_{\rm LM}&=2\raisedchi\, \bm{J}^A\cdot \bm{J}^B \notag \\
    &= \raisedchi\left(\bm{J}\cdot\bm{J} - \bm{J}^A\cdot\bm{J}^A-\bm{J}^B\cdot\bm{J}^B\right),
    \label{eq:LiebMattis}
\end{align}
The Lieb-Mattis Hamiltonian serves as a compelling and exactly solvable toy model for investigating ferromagnetism on a square lattice and exploring the mechanism of spontaneous symmetry breaking. As shown in the second line of Eq.~\eqref{eq:LiebMattis}, its ground state is characterized by maximizing the expectation values $\bm{J}^K\cdot\bm{J}^K$ within each ensemble while minimizing the expectation value of $\bm{J}\cdot \bm{J}$, ultimately yielding a unique state. This ground state can also be interpreted as a superposition of a continuum of Néel states, each pointing with equal probability in all possible directions on the Bloch sphere \cite{rademaker2019exact}. In the thermodynamic limit, this symmetry is spontaneously broken \cite{beekman2019introduction}. However, in finite systems, the symmetry can be explicitly broken by a perturbation, such as the phase-encoding Hamiltonian $J_z^-$, which selects a single Néel state from the continuum of  Néel states pointing in different directions.

If the system is restricted to a different DFS, or equivalently to an eigenspace of $J_z^+$ with eigenvalue $M\neq0$, the Lieb-Mattis ground state within this subspace corresponds to the simultaneous eigenstate of $\bm{J}\cdot \bm{J}$ and $J_z^+$, $\ket{|M|,M}$.  These states are uniquely defined under the condition that the expectation value of $\bm{J}^K\cdot\bm{J}^K$ is maximized. Notably, the estimator variance Eq.~\eqref{eq:EstVar} for any of these states, in conjunction with a measurement of the optimal two-body observable Eq.~\eqref{eq:Measurement}, achieves the QCRB for $\phi=\pi/4$. The QFI for these states is given by
\begin{align}
    F^{\rm Q}_{\ket{|M|, M}}= \frac{4N + N^2 - 8M - 4M^2}{3+2M},
    \label{eq:FQ_LM}
\end{align}
which exhibits Heisenberg scaling when $M\ll N$. A derivation of this expression for the QFI can be found in Appendix \ref{app:ang_mom_states}.

To find a parent Hamiltonian for the target state, it is not strictly necessary to have infinite-range interactions. Instead, it suffices to consider a bipartite lattice model in any dimension, in which nearest-neighbor sites belong to different sublattices, while next-nearest neighbors belong to the same sublattice. The $J_1$-$J_2$ Hamiltonian 
\begin{align}
    H_{J_1-J_2}=J_1\sum_{\braket{i, j}}\bm{\sigma}^{(i)}\cdot \bm{\sigma}^{(j)}-J_{2} \sum_{\braket{\braket{i, j}}}\bm{\sigma}^{(i)}\cdot \bm{\sigma}^{(j)}
\end{align}
is a parent Hamiltonian of the target state, provided that the ratio of nearest-neighbor to next-nearest-neighbor interaction strengths satisfies $J_1/J_2\ll 1$. Here the summation indices $\braket{i, j}$ and $\braket{\braket{i, j}}$ denote sums over all nearest-neighbor and next-nearest-neighbor pairs, respectively.

Each term in the $J_1$-$J_2$ Hamiltonian commutes with $\bm{J}\cdot \bm{J}$, ensuring that the ground state is an eigenstate of $\bm{J}\cdot \bm{J}$. However, the nearest-neighbor interactions do not commute with $\bm{J}^{A}\cdot\bm{J}^{A}$ or $\bm{J}^{B}\cdot\bm{J}^{B}$, implying that the fully permutation-symmetric state within each subensemble is not the ground state unless $J_1/J_2 \ll 1$. In that regime, the $J_1$-$J_2$ Hamiltonian approximately commutes with $\bm{J}^{A}\cdot\bm{J}^{A}$ and $\bm{J}^{B}\cdot\bm{J}^{B}$, causing the ground state to converge to the target state.

\section{Adiabatic preparation}
\label{app:adiabatic}

\begin{figure}[t] 
   \centering
   \includegraphics[]{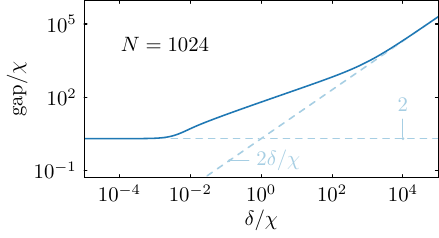} \\
   \caption{Energy gap between the ground and first excited state of the Hamiltonian $H=\raisedchi\left(J_+^A+J_+^B\right)\left(J_-^A+J_-^B\right)+\delta\left(J_z^A-J_z^B\right)$ for $N=1024$ atoms. }
   \label{fig:AdiabaticGap}
\end{figure}

In this section, we provide additional intuition for the adiabatic sweep, which is implemented using the cavity Hamiltonian in Eq.~\eqref{eq:H_cavity} together with a time-dependent relative detuning $\delta_t$ between the two subensembles. The detuning is gradually ramped to zero as a function of the ramp time t. The resulting time-dependent Hamiltonian takes the form 
\begin{align}
    H_{\rm ad}(t)=& \raisedchi \Big(J_+^A+J_+^B\Big)\Big(J_-^A+J_-^B\Big)\notag \\
    & -\delta_t\Big(J_z^A-J_z^B\Big).
\end{align}
and enables the adiabatic preparation of the target state. This state corresponds to the unique ground state of the Lieb-Mattis Hamiltonian introduced in Eq.~\eqref{eq:LiebMattis}) $H_{\rm LM}={J_+J_-+J_z^2-\bm{J}^A\cdot\bm{J}^A-\bm{J}^B\cdot\bm{J}^B}$.
The terms $J_z^2, \bm{J}^A\cdot\bm{J}^A, \bm{J}^B\cdot\bm{J}^B$, which are required to transform the cavity Hamiltonian into the parent Hamiltonian, commute with all components of the cavity Hamiltonian. Moreover, both the initial state and the final state are eigenstates of these operators with identical eigenvalues. Consequently, these terms do not need to be explicitly included during the adiabatic sweep to achieve the desired target state.

The speed at which the adiabatic sweep, from $t=0$ to  $t_{\rm{final}}$, can be performed depends on the instantaneous energy gap of the Hamiltonian to the state closest in energy that shares the same symmetries  as the cavity Hamiltonian and the initial state. In the two limits $\delta_t(t=0) / \raisedchi\gg1$ and $\delta_t(t=t_{\rm{final}})/\raisedchi \ll 1$, the gap between the ground and first excited state is independent of $N$ and scales with $2\delta_t$ and $2\raisedchi$ respectively, see Fig.~\ref{fig:AdiabaticGap}. To ensure the success of the sweep, its speed must be small compared to these energy scales while remaining sufficiently fast to avoid perturbations such as free-space or collective emissions  that could take the system out of the symmetry subspace. However, achieving this balance poses challenges under typical cavity parameters that rely  on the collective cooperativity $N C $ and not just $C$.

\section{Realizing a two-mode squeezing interaction with multi-level alkaline earth atoms}
\label{app:TMS_H}

\begin{figure}[t] 
   \centering
   \includegraphics[]{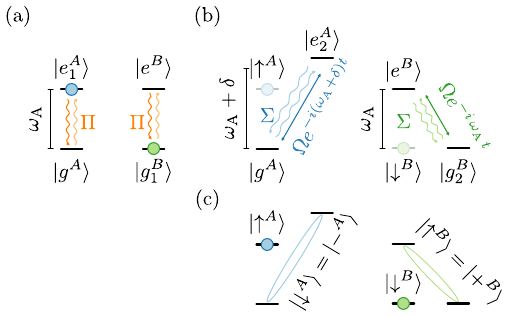} 
   \caption{Level scheme for generating the two-mode squeezing interaction between two atomic ensembles, $A$ and $B$, mediated by a cavity. (a) $\Pi$-polarized light facilitates spin-exchange interactions among all atoms, as described by Eq.~\eqref{eq:H_cavity}. The atoms in ensemble $A$ are initially prepared in the $\ket{e^A_1}$ state, while those in ensemble $B$ are initialized in the $\ket{g_1^B}$ state.
(b) Resonant driving to auxiliary levels, whose energies are shifted relative to one another, combined with $\Sigma$-polarized light, induces effective interactions within each ensemble. This intra-ensemble interaction modifies the global spin-exchange dynamics and gives rise to the desired two-mode squeezing interaction described by Eq.~\eqref{eq:H_TMS}. (c) The effective spin-1/2 states of ensemble $A$ are $\ket{\uparrow^A}=\ket{e^A_1}$ and the hybridized state $\ket{\downarrow^A}=\ket{-^A}=\frac{\ket{g^A}-\ket{e^A_2}}{\sqrt{2}}$ and $
\ket{\downarrow^B}=\ket{g^B_1}$ and $\ket{\uparrow^B}=\ket{-^B}=\frac{\ket{g^B_2}+\ket{e^B}}{\sqrt{2}}$.
 }
   \label{fig:Interaction}
\end{figure}

In this section, we present a potential implementation of the two-mode squeezing interaction described by Eq.~\eqref{eq:H_TMS}. Our focus is on alkaline-earth atoms with large nuclear spin, such as $^{87}\rm Sr$. We select the nuclear spin quantization axis to be perpendicular to the cavity axis,  so that the linearly polarized light ($\Pi$) and horizontally polarized light ($\Sigma$) drive different transitions between the ground and excited-state manifolds. 
To realize the desired interaction, we isolate three hyperfine states from the ground-state manifold, denoted as $\ket{g^A}, \ket{g_1^B}$, and $\ket{g_2^B}$, along with three states from the excited-state manifold, labeled $\ket{e_1^A}$, $\ket{e_2^A}$ and $\ket{e^B}$, as illustrated in Fig.~\ref{fig:Interaction}. The subscripts $A$ and $B$ indicate that atoms in ensemble $A$ exclusively occupy states labeled with $A$, while atoms in ensemble $B$ are restricted to states labeled with $B$. The isolation of these energy levels is achieved through appropriate Zeeman and AC Stark shifts. The energy splitting between the isolated ground and excited states is given by $\omega_{\rm A}$, except for the state $\ket{e_2^A}$, which experiences an additional shift
$\delta$ relative to $\ket{e_1^A}$ and $\ket{e^B}$. 

As sketched in Fig.~\ref{fig:Interaction}(a), the atoms in ensemble $A$ are prepared in state $\ket{e_1^A}$, while those in ensemble $B$ are initialized in state $\ket{g_1^B}$. The linearly polarized cavity mode drives transitions  $\ket{g^A}\leftrightarrow\ket{e^A_1}$ and $\ket{g^B_1}\leftrightarrow\ket{e^B}$. When the cavity is sufficiently far detuned $\Delta=\omega_{\rm C}-\omega_{\rm A}\gg g\sqrt{N}$, the cavity mode can be adiabatically eliminated and the unitary evolution of the spins is described by the Hamiltonian in Eq.~\eqref{eq:H_cavity}. The underlying process is sketched in Fig.~\ref{fig:Interaction}(a) where an excited state atom decays to the ground state and emits a linearly polarized photon which can be absorber by a ground-state atom in the same or the other ensemble leading to the desired spin-exchange between ensembles but also the undesired spin-exchange between atoms in the same ensemble. 

To mitigate the latter effect, we drive the transitions $\ket{g^A} \leftrightarrow \ket{e^A_2}$ and $\ket{g^B} \leftrightarrow \ket{e^B}$ using two lasers with equal Rabi frequency $\Omega$. Additionally, the state $\ket{e^A_2}$ is shifted by an energy $\delta$ relative to the other excited states. This additional detuning ensures that vertically polarized photons that drive the same transitions are reabsorbed exclusively by atoms within the same ensemble, thereby inducing effective interactions of the form $J_z^A J_z^A$ and $J_z^B J_z^B$, while suppressing any unwanted inter-ensemble interactions.

To derive this formally we consider the combined system of atoms and the cavity, which is described by an atomic Hamiltonian $H_{\rm A}$, light Hamiltonian $H_{\rm L}$, and atom-light Hamiltonian $H_{\rm AL}$. In a rotating frame where $\ket{e_1^A}\rightarrow e^{-i\omega_{\rm A}t}\ket{e_1^A}$, $\ket{e_2^A}\rightarrow e^{-i(\omega_{\rm A}+\delta)t}\ket{e_2^A}$, $\ket{e^B}\rightarrow e^{-i\omega_{\rm A}t}\ket{e^B}$, and $a^{}_{P}\rightarrow e^{i\omega_{\rm A}t} a^{}_P$, the Hamiltonians are
\begin{align}
    H_{\rm A} = 
    &\Omega\bigg(\ket{g^A}\bra{e^A_2}  +\ket{g_2^B}\bra{e^B}+{\rm h.c}\bigg), 
\end{align}
\begin{align}
    H_{\rm L}=\Delta\sum_{P=\Sigma, \Pi}a_P^{\dagger}a^{}_P,
\end{align}
and 
\begin{align}
    H_{\rm AL}=g\bigg(& e^{i\delta t}a^{\dagger}_{\Sigma}c_{g^A}^{e^A_2}\ket{g^A}\bra{e^A_2}+a^{\dagger}_{\Sigma}c_{g^B_2}^{e^B}\ket{g^B_2}\bra{e^B}\notag \\
    & +a^{\dagger}_{\Pi}c_{g^A}^{e_2^A}\ket{g^A}\bra{e_1^A}+a^{\dagger}_{\Pi}c_{g^B_1}^{e^B}\ket{g^B_1}\bra{e^B}\notag \\
    & + \rm {h.c.}\bigg). 
\end{align}
Here $\Delta = \omega_{\rm C}-\omega_{\rm A}$,  $a^{\dagger}_{P}$ is the creation operator of a photon in the respective polarization mode, and $c_{\alpha}^{\beta}$ are the Clebsh-Gordan coefficients between the states $\ket{\alpha}$ and $\ket{\beta}$. 

In the following, we consider the dressed-state basis defined by $\ket{\pm^A}=(\ket{e_2^A}\pm \ket{g^A})/\sqrt{2}$ and $\ket{\pm^B}=(\ket{e^B}\pm \ket{g_2^B})/\sqrt{2}$. In this basis, the atomic Hamiltonian takes the form
\begin{align}
    H_{\rm A} = \Omega\Big(&\ket{+^A}\bra{+^A}-\ket{-^A}\bra{-^A}\notag \\&+\ket{+^B}\bra{+^B}-\ket{-^B}\bra{-^B}\Big).
\end{align}
Finally, we express the atom-light interaction Hamiltonian in an interaction picture with respect to $H_{\rm A}$ and $H_{\rm L}$, which yields 
\begin{align}
    \tfrac{\sqrt{2}}{g}H_{\rm AL}^{\rm I}=
    & a^{\dagger}_{\Pi}c_{g^A}^{e_1^A}\ket{-^A}\bra{e_1^A}e^{-i(\Delta-\Omega)t}  + \notag \\
    &a^{\dagger}_{\Pi}c_{g_1^B}^{e^B}\ket{g_1^B}\bra{+^B}e^{-i(\Delta-\Omega)t} + \notag\\
    & \frac{a^{\dagger}_{\Sigma}c_{g^A}^{e^A_2}e^{-i(\Delta + \delta)t}}{\sqrt{2}}\bigg(\ket{+^A}
    \bra{+^A}-\notag \\
    & \qquad \quad\qquad \qquad\  \ket{-^A}\bra{-^A}\bigg) + \notag \\
    & \frac{a^{\dagger}_{\Sigma}c_{g_2^B}^{e^B}e^{-i\Delta t}}{\sqrt{2}}\bigg(\ket{+^B}
    \bra{+^B}-\notag \\ 
    & \qquad \quad\qquad \qquad\ \ket{-^B}\bra{-^B}\bigg) + \notag \\
    & +\rm {h.c},
\end{align}
where off-resonant terms that oscillate at frequencies $\Omega + \Delta$, $\Delta-2\Omega$, and $\Delta + 2\Omega$ are omitted. 

Identifying the collective spin operators discussed in the main text as $J_-^A=\sum_{i}\ket{-^A}_i\bra{e^A}$, $J_-^B=\sum_{i}\ket{g_1^B}\bra{+^B}$, $J_z^A=\sum_{i}\frac{\ket{e^A}_i\bra{e^A}-\ket{-^A}_i\bra{-^A}}{2}$, and $J_z^A=\sum_{i}\frac{\ket{+^B}_i\bra{+^B}-\ket{g^B}_i\bra{g^B}}{2}$ and adiabatically eliminating the two cavity modes yields an effective many-body spin Hamiltonian of the form 
\begin{align}
    H_{\rm eff} = &\tfrac{g^2}{2(\Delta-\Omega)}|c_{g^A}^{e_1^A}|^2 \left(J_-^A+J_-^B\right)\left(J_+^A+J_+^B\right)  + \notag \\
    & \tfrac{g^2}{\Delta}|c_{g^A}^{e^A_2}|^2\left(J_z^A+\frac{N}{4}\right)^2 + \notag \\
    & \tfrac{g^2}{\Delta + \delta}|c_{g^B_2}^{e^B}|^2\left(J_z^B+\frac{N}{4}\right)^2, 
\end{align}
where we have assumed that $|c_{g^A}^{e_1^A}|^2 = | c_{g_1^B}^{e^B}|^2$ and that the number of atoms in both ensembles is exactly the same. The final step is to identify hyperfine states and laser parameters for which $\frac{g^2}{2(\Delta-\Omega)}|c_{g^A}^{e_1^A}|^2  = \tfrac{g^2}{\Delta}|c_{g^A}^{e^A_2}|^2 = \tfrac{g^2}{\Delta + \delta}|c_{g^B_2}^{e^B}|^2$, such that the effective spin Hamiltonian simplifies to 
\begin{align}
    H_{\rm eff} = \tfrac{g^2}{2(\Delta-\Omega)}|c_{g^A}^{e_1^A}|^2 \Big(J_-^AJ_+^B+J_+^AJ_-^B +\bm{J}^A\cdot \bm{J}^A + \notag \\ \bm{J}^B\cdot \bm{J}^B + \frac{N}{2}\left(J_z^A-J_z^B\right)\Big). 
\end{align}
By adding an additional field gradient between the two ensembles and assuring that permutation invariance within the subensembles is maintained during the dynamics this yields the Hamiltonian of Eq.~\eqref{eq:H_TMS} up to a $\pi/4$-rotation under the phase encoding generator. 

As a concrete example, we consider the states $\ket{g^A}, \ket{g^B_1}$, and $\ket{g^B_2}$ to correspond to the $9/2, -9/2$, and $-7/2$ levels of the ${}^1S_0$ manifold of $^{87} \rm{Sr}$. Similarly, we assign the excited states $\ket{e}^A_1, \ket{e}^A_2$, and $\ket{e^B}$ to  the $9/2, 11/2$, and $- 9/2$ levels of the ${}^3P_1$ manifold, respectively. Furthermore, by choosing $\Delta=(9/8)\Omega$ and $\delta = (1/4)\Omega$, the condition $\frac{g^2}{2(\Delta-\Omega)}|c_{g^A}^{e_1^A}|^2  = \tfrac{g^2}{\Delta}|c_{g^A}^{e^A_2}|^2 = \tfrac{g^2}{\Delta + \delta}|c_{g^B_2}^{e^B}|^2$ is satisfied. 

\section{Analytic expressions for $\ket{J,M}$ states}
\label{app:ang_mom_states}

In this section, we explicitly derive the QFI and other expectation values for the $\ket{J, M}$ states, which exhibit permutational symmetry within the two subensembles. To achieve this, it is convenient to represent the generator of phase encoding in the $\ket{J, M}$ basis as follows:
\begin{align}
    J_z^-=\sum_{J=1}^{N/2}\sum_{M=-J+1}^{J} &\sqrt{\tfrac{\left(J^2 - M^2\right)\left(\left(N / 2 + 1\right)^2-J^2\right)}{4J^2-1}}\notag\\
    &\times\big( \ket{J-1,M} \langle J, M|+{\rm h.c.}\big). 
\end{align}
 In this basis, the generator of phase encoding establishes couplings between states with the same $M$ but $J$ differing by one.

From this expression, it is straightforward to verify that the QFI for the state $\ket{J, M}$ is given by
\begin{align}
    F_{\ket{J,M}}^{\rm Q} =& 4\Big(\braket{J,M|J_z^-J_z^-|J,M} - \braket{J,M|J_z^-|J,M}^2\Big) \notag \\
    =& \frac{12 M^2 + 8 J (1 + J) (J + J^2 - M^2-1) }{3 - 4 J (1 + J)} \notag \\
    &+ \frac{(1 - 2 J (1 + J) + 2 M^2)}{3 - 4 J (1 + J)}(N^2 + 4 N). 
 \label{eq:FQI_JM}
\end{align}
This result simplifies to the expression in Eq.~\eqref{eq:FQ_LM} when $J$ is replaced by $|M|$.

The expectation values of the measurement operator, given by Eq.~\eqref{eq:Measurement}, can be computed in the transformed frame with respect to the phase encoding. This is achieved using identities such as $J_+J_-= \bm{J}\cdot \bm{J} - J_zJ_z$ and leveraging the fact that the states $\ket{J, M}$ are eigenstates of the operators $\bm{J}\cdot\bm{J}, J_z, \bm{J}^K\cdot \bm{J}^K$ for $K=A,B$. The final expression for the measurement expectation value is
\begin{align}
    \bra{J,M}\M_{\phi}&\ket{J,M}/\cos(2\phi) \left(3-4 (4 J (J+1))\right)\notag \\
    =&J (J+1) \left(2+M^2-3 J (J+1)\right) \notag \\
    &+\left(J^2+J+M^2-1\right)\big(N^2/4 + N). 
\end{align}

Notably, the measurement expectation value for any values of $J$ and $M$ is proportional solely to $\cos(2\phi)$. 

For the Lieb-Mattis ground state within a given $J_z^+$ eigenspace, this expression further simplifies to 
\begin{align}
    \bra{|M|,M}\M_{\phi}&\ket{|M|,M}/\cos(2\phi) \notag \\
    & = \frac{(M+1) (2 M-N) (2 M+N+4)}{8 M+12}.
    \label{eq:M_exp_LM}
\end{align}
This result highlights that, eigenspaces where $|M|\ll N$, the fringe amplitudes of the measurement operator $\M$ scale quadratically with $N$. This scaling makes these states particularly promising candidates for sensing applications.

Similarly, the expression for $\braket{J, M|\M^2_{\phi}|J, M}$ can be derived following the same procedure. However, the resulting expression is too extensive to be presented here. Instead, we focus on its dependence on $\phi$, which takes the form $\braket{J,M|\M^2_{\rm \phi}|J,M}=\alpha(N,J,M)+\beta(N, J, M)\cos(4\phi)$. The general expressions for the real coefficients $\alpha(N, J, M)$ and $\beta(N, J, M)$ simplify to
\begin{align}
    \alpha(N,|M|&,M)\frac{32 (3 + 2 M) (5 + 2 M)}{(1 + M) (2 M - N) (4 + 
   2 M + N) }\notag \\
    =& 4 (1 + M) (-4 + M + M^2)\notag \\
    & - 4 (2 + M) N - (2 + M) N^2,
    \label{eq:MM_exp_LM1}
\end{align}
\begin{align}
    \beta(N,|M|&,M)\frac{32 (3 + 2 M) (5 + 2 M)}{(1 + M) (2 M - N) (4 + 
   2 M + N) } \notag \\
    =& (2 + M) (2 + 2 M - N) (6 + 2 M + N), 
    \label{eq:MM_exp_LM2}
\end{align}
for the Lieb-Mattis ground states in different eigenspaces of $J_z^+$. 

Substituting the expressions from Eqs.~(\ref{eq:M_exp_LM}, \ref{eq:MM_exp_LM1}, \ref{eq:MM_exp_LM2}) into the estimator variance defined in Eq.~\eqref{eq:EstVar}, it can be shown that the estimator variance is minimized at $\phi = \pi / 4$. Furthermore, under this condition, the estimator variance saturates the QCRB, which is defined in terms of the QFI given in Eq.~\eqref{eq:FQ_LM}.

\section{Average quantum Fisher information }
\label{app:avgQFI}
In this section, we discuss a condition under which the convex sum of the QFI of the pure states appearing in the spectral decomposition of a density matrix coincides with the QFI of the density matrix itself.
The QFI of a mixed state with spectral decomposition $\rho=\sum_{i}p_i \ket{\psi_i}\bra{\psi_i}$ which is not full rank is given by  \cite{liu2014quantum} 
\begin{align}
    F^{\rm Q}_{\rho}= \sum_{i}p_i F^{\rm Q}_{\ket{\psi_i} } - \sum_{i\neq j}\frac{8 p_i p_j}{p_i + p_j}\left|\bra{\psi_i}J_z^-\ket{\psi_j}\right|^2.
\end{align}
For the density matrix in Eq.~\eqref{eq:steady_state}, the states $\ket{\psi_J}=\ket{J, -J}$ which represent the spectral decomposition satisfy $\bra{\psi_i}J_z^-\ket{\psi_j} = 0$ for all $i, j$. Thus the QFI coincides with the average QFI with respect to the probability $p(J)$, see Eq.~\eqref{eq:QCR_SS}.  In general the average QFI is just an upper bound to the actual QFI of a density matrix \cite{pezze2014quantum}. 


\begin{thebibliography}{105}%
\makeatletter
\providecommand \@ifxundefined [1]{%
 \@ifx{#1\undefined}
}%
\providecommand \@ifnum [1]{%
 \ifnum #1\expandafter \@firstoftwo
 \else \expandafter \@secondoftwo
 \fi
}%
\providecommand \@ifx [1]{%
 \ifx #1\expandafter \@firstoftwo
 \else \expandafter \@secondoftwo
 \fi
}%
\providecommand \natexlab [1]{#1}%
\providecommand \enquote  [1]{``#1''}%
\providecommand \bibnamefont  [1]{#1}%
\providecommand \bibfnamefont [1]{#1}%
\providecommand \citenamefont [1]{#1}%
\providecommand \href@noop [0]{\@secondoftwo}%
\providecommand \href [0]{\begingroup \@sanitize@url \@href}%
\providecommand \@href[1]{\@@startlink{#1}\@@href}%
\providecommand \@@href[1]{\endgroup#1\@@endlink}%
\providecommand \@sanitize@url [0]{\catcode `\\12\catcode `\$12\catcode
  `\&12\catcode `\#12\catcode `\^12\catcode `\_12\catcode `\%12\relax}%
\providecommand \@@startlink[1]{}%
\providecommand \@@endlink[0]{}%
\providecommand \url  [0]{\begingroup\@sanitize@url \@url }%
\providecommand \@url [1]{\endgroup\@href {#1}{\urlprefix }}%
\providecommand \urlprefix  [0]{URL }%
\providecommand \Eprint [0]{\href }%
\providecommand \doibase [0]{https://doi.org/}%
\providecommand \selectlanguage [0]{\@gobble}%
\providecommand \bibinfo  [0]{\@secondoftwo}%
\providecommand \bibfield  [0]{\@secondoftwo}%
\providecommand \translation [1]{[#1]}%
\providecommand \BibitemOpen [0]{}%
\providecommand \bibitemStop [0]{}%
\providecommand \bibitemNoStop [0]{.\EOS\space}%
\providecommand \EOS [0]{\spacefactor3000\relax}%
\providecommand \BibitemShut  [1]{\csname bibitem#1\endcsname}%
\let\auto@bib@innerbib\@empty
\bibitem [{\citenamefont {Rosi}\ \emph {et~al.}(2014)\citenamefont {Rosi},
  \citenamefont {Sorrentino}, \citenamefont {Cacciapuoti}, \citenamefont
  {Prevedelli},\ and\ \citenamefont {Tino}}]{rosi2014precision}%
  \BibitemOpen
  \bibfield  {author} {\bibinfo {author} {\bibfnamefont {G.}~\bibnamefont
  {Rosi}}, \bibinfo {author} {\bibfnamefont {F.}~\bibnamefont {Sorrentino}},
  \bibinfo {author} {\bibfnamefont {L.}~\bibnamefont {Cacciapuoti}}, \bibinfo
  {author} {\bibfnamefont {M.}~\bibnamefont {Prevedelli}},\ and\ \bibinfo
  {author} {\bibfnamefont {G.}~\bibnamefont {Tino}},\ }\bibfield  {title}
  {\bibinfo {title} {Precision measurement of the newtonian gravitational
  constant using cold atoms},\ }\href@noop {} {\bibfield  {journal} {\bibinfo
  {journal} {Nature}\ }\textbf {\bibinfo {volume} {510}},\ \bibinfo {pages}
  {518} (\bibinfo {year} {2014})}\BibitemShut {NoStop}%
\bibitem [{\citenamefont {Parker}\ \emph {et~al.}(2018)\citenamefont {Parker},
  \citenamefont {Yu}, \citenamefont {Zhong}, \citenamefont {Estey},\ and\
  \citenamefont {M{\"u}ller}}]{parker2018measurement}%
  \BibitemOpen
  \bibfield  {author} {\bibinfo {author} {\bibfnamefont {R.~H.}\ \bibnamefont
  {Parker}}, \bibinfo {author} {\bibfnamefont {C.}~\bibnamefont {Yu}}, \bibinfo
  {author} {\bibfnamefont {W.}~\bibnamefont {Zhong}}, \bibinfo {author}
  {\bibfnamefont {B.}~\bibnamefont {Estey}},\ and\ \bibinfo {author}
  {\bibfnamefont {H.}~\bibnamefont {M{\"u}ller}},\ }\bibfield  {title}
  {\bibinfo {title} {Measurement of the fine-structure constant as a test of
  the standard model},\ }\href@noop {} {\bibfield  {journal} {\bibinfo
  {journal} {Science}\ }\textbf {\bibinfo {volume} {360}},\ \bibinfo {pages}
  {191} (\bibinfo {year} {2018})}\BibitemShut {NoStop}%
\bibitem [{\citenamefont {Overstreet}\ \emph {et~al.}(2022)\citenamefont
  {Overstreet}, \citenamefont {Asenbaum}, \citenamefont {Curti}, \citenamefont
  {Kim},\ and\ \citenamefont {Kasevich}}]{overstreet2022observation}%
  \BibitemOpen
  \bibfield  {author} {\bibinfo {author} {\bibfnamefont {C.}~\bibnamefont
  {Overstreet}}, \bibinfo {author} {\bibfnamefont {P.}~\bibnamefont
  {Asenbaum}}, \bibinfo {author} {\bibfnamefont {J.}~\bibnamefont {Curti}},
  \bibinfo {author} {\bibfnamefont {M.}~\bibnamefont {Kim}},\ and\ \bibinfo
  {author} {\bibfnamefont {M.~A.}\ \bibnamefont {Kasevich}},\ }\bibfield
  {title} {\bibinfo {title} {Observation of a gravitational aharonov-bohm
  effect},\ }\href@noop {} {\bibfield  {journal} {\bibinfo  {journal}
  {Science}\ }\textbf {\bibinfo {volume} {375}},\ \bibinfo {pages} {226}
  (\bibinfo {year} {2022})}\BibitemShut {NoStop}%
\bibitem [{\citenamefont {Aeppli}\ \emph {et~al.}(2024)\citenamefont {Aeppli},
  \citenamefont {Kim}, \citenamefont {Warfield}, \citenamefont {Safronova},\
  and\ \citenamefont {Ye}}]{aeppli2024clock}%
  \BibitemOpen
  \bibfield  {author} {\bibinfo {author} {\bibfnamefont {A.}~\bibnamefont
  {Aeppli}}, \bibinfo {author} {\bibfnamefont {K.}~\bibnamefont {Kim}},
  \bibinfo {author} {\bibfnamefont {W.}~\bibnamefont {Warfield}}, \bibinfo
  {author} {\bibfnamefont {M.~S.}\ \bibnamefont {Safronova}},\ and\ \bibinfo
  {author} {\bibfnamefont {J.}~\bibnamefont {Ye}},\ }\bibfield  {title}
  {\bibinfo {title} {Clock with 8$\times$ 10-19 systematic uncertainty},\
  }\href@noop {} {\bibfield  {journal} {\bibinfo  {journal} {Physical Review
  Letters}\ }\textbf {\bibinfo {volume} {133}},\ \bibinfo {pages} {023401}
  (\bibinfo {year} {2024})}\BibitemShut {NoStop}%
\bibitem [{\citenamefont {Eldredge}\ \emph {et~al.}(2018)\citenamefont
  {Eldredge}, \citenamefont {Foss-Feig}, \citenamefont {Gross}, \citenamefont
  {Rolston},\ and\ \citenamefont {Gorshkov}}]{eldredge2018optimal}%
  \BibitemOpen
  \bibfield  {author} {\bibinfo {author} {\bibfnamefont {Z.}~\bibnamefont
  {Eldredge}}, \bibinfo {author} {\bibfnamefont {M.}~\bibnamefont {Foss-Feig}},
  \bibinfo {author} {\bibfnamefont {J.~A.}\ \bibnamefont {Gross}}, \bibinfo
  {author} {\bibfnamefont {S.~L.}\ \bibnamefont {Rolston}},\ and\ \bibinfo
  {author} {\bibfnamefont {A.~V.}\ \bibnamefont {Gorshkov}},\ }\bibfield
  {title} {\bibinfo {title} {Optimal and secure measurement protocols for
  quantum sensor networks},\ }\href@noop {} {\bibfield  {journal} {\bibinfo
  {journal} {Physical Review A}\ }\textbf {\bibinfo {volume} {97}},\ \bibinfo
  {pages} {042337} (\bibinfo {year} {2018})}\BibitemShut {NoStop}%
\bibitem [{\citenamefont {Proctor}\ \emph {et~al.}(2018)\citenamefont
  {Proctor}, \citenamefont {Knott},\ and\ \citenamefont
  {Dunningham}}]{proctor2018multiparameter}%
  \BibitemOpen
  \bibfield  {author} {\bibinfo {author} {\bibfnamefont {T.~J.}\ \bibnamefont
  {Proctor}}, \bibinfo {author} {\bibfnamefont {P.~A.}\ \bibnamefont {Knott}},\
  and\ \bibinfo {author} {\bibfnamefont {J.~A.}\ \bibnamefont {Dunningham}},\
  }\bibfield  {title} {\bibinfo {title} {Multiparameter estimation in networked
  quantum sensors},\ }\href@noop {} {\bibfield  {journal} {\bibinfo  {journal}
  {Physical review letters}\ }\textbf {\bibinfo {volume} {120}},\ \bibinfo
  {pages} {080501} (\bibinfo {year} {2018})}\BibitemShut {NoStop}%
\bibitem [{\citenamefont {Ge}\ \emph {et~al.}(2018)\citenamefont {Ge},
  \citenamefont {Jacobs}, \citenamefont {Eldredge}, \citenamefont {Gorshkov},\
  and\ \citenamefont {Foss-Feig}}]{ge2018distributed}%
  \BibitemOpen
  \bibfield  {author} {\bibinfo {author} {\bibfnamefont {W.}~\bibnamefont
  {Ge}}, \bibinfo {author} {\bibfnamefont {K.}~\bibnamefont {Jacobs}}, \bibinfo
  {author} {\bibfnamefont {Z.}~\bibnamefont {Eldredge}}, \bibinfo {author}
  {\bibfnamefont {A.~V.}\ \bibnamefont {Gorshkov}},\ and\ \bibinfo {author}
  {\bibfnamefont {M.}~\bibnamefont {Foss-Feig}},\ }\bibfield  {title} {\bibinfo
  {title} {Distributed quantum metrology with linear networks and separable
  inputs},\ }\href@noop {} {\bibfield  {journal} {\bibinfo  {journal} {Physical
  review letters}\ }\textbf {\bibinfo {volume} {121}},\ \bibinfo {pages}
  {043604} (\bibinfo {year} {2018})}\BibitemShut {NoStop}%
\bibitem [{\citenamefont {Gross}\ and\ \citenamefont
  {Caves}(2020)}]{gross2020one}%
  \BibitemOpen
  \bibfield  {author} {\bibinfo {author} {\bibfnamefont {J.~A.}\ \bibnamefont
  {Gross}}\ and\ \bibinfo {author} {\bibfnamefont {C.~M.}\ \bibnamefont
  {Caves}},\ }\bibfield  {title} {\bibinfo {title} {One from many: Estimating a
  function of many parameters},\ }\href@noop {} {\bibfield  {journal} {\bibinfo
   {journal} {Journal of Physics A: Mathematical and Theoretical}\ }\textbf
  {\bibinfo {volume} {54}},\ \bibinfo {pages} {014001} (\bibinfo {year}
  {2020})}\BibitemShut {NoStop}%
\bibitem [{\citenamefont {Zhang}\ and\ \citenamefont
  {Zhuang}(2021)}]{zhang2021distributed}%
  \BibitemOpen
  \bibfield  {author} {\bibinfo {author} {\bibfnamefont {Z.}~\bibnamefont
  {Zhang}}\ and\ \bibinfo {author} {\bibfnamefont {Q.}~\bibnamefont {Zhuang}},\
  }\bibfield  {title} {\bibinfo {title} {Distributed quantum sensing},\
  }\href@noop {} {\bibfield  {journal} {\bibinfo  {journal} {Quantum Science
  and Technology}\ }\textbf {\bibinfo {volume} {6}},\ \bibinfo {pages} {043001}
  (\bibinfo {year} {2021})}\BibitemShut {NoStop}%
\bibitem [{\citenamefont {Bringewatt}\ \emph {et~al.}(2021)\citenamefont
  {Bringewatt}, \citenamefont {Boettcher}, \citenamefont {Niroula},
  \citenamefont {Bienias},\ and\ \citenamefont
  {Gorshkov}}]{bringewatt2021protocols}%
  \BibitemOpen
  \bibfield  {author} {\bibinfo {author} {\bibfnamefont {J.}~\bibnamefont
  {Bringewatt}}, \bibinfo {author} {\bibfnamefont {I.}~\bibnamefont
  {Boettcher}}, \bibinfo {author} {\bibfnamefont {P.}~\bibnamefont {Niroula}},
  \bibinfo {author} {\bibfnamefont {P.}~\bibnamefont {Bienias}},\ and\ \bibinfo
  {author} {\bibfnamefont {A.~V.}\ \bibnamefont {Gorshkov}},\ }\bibfield
  {title} {\bibinfo {title} {Protocols for estimating multiple functions with
  quantum sensor networks: Geometry and performance},\ }\href@noop {}
  {\bibfield  {journal} {\bibinfo  {journal} {Physical Review Research}\
  }\textbf {\bibinfo {volume} {3}},\ \bibinfo {pages} {033011} (\bibinfo {year}
  {2021})}\BibitemShut {NoStop}%
\bibitem [{\citenamefont {Malitesta}\ \emph {et~al.}(2023)\citenamefont
  {Malitesta}, \citenamefont {Smerzi},\ and\ \citenamefont
  {Pezz{\`e}}}]{malitesta2023distributed}%
  \BibitemOpen
  \bibfield  {author} {\bibinfo {author} {\bibfnamefont {M.}~\bibnamefont
  {Malitesta}}, \bibinfo {author} {\bibfnamefont {A.}~\bibnamefont {Smerzi}},\
  and\ \bibinfo {author} {\bibfnamefont {L.}~\bibnamefont {Pezz{\`e}}},\
  }\bibfield  {title} {\bibinfo {title} {Distributed quantum sensing with
  squeezed-vacuum light in a configurable array of mach-zehnder
  interferometers},\ }\href@noop {} {\bibfield  {journal} {\bibinfo  {journal}
  {Physical Review A}\ }\textbf {\bibinfo {volume} {108}},\ \bibinfo {pages}
  {032621} (\bibinfo {year} {2023})}\BibitemShut {NoStop}%
\bibitem [{\citenamefont {Fixler}\ \emph {et~al.}(2007)\citenamefont {Fixler},
  \citenamefont {Foster}, \citenamefont {McGuirk},\ and\ \citenamefont
  {Kasevich}}]{fixler2007atom}%
  \BibitemOpen
  \bibfield  {author} {\bibinfo {author} {\bibfnamefont {J.~B.}\ \bibnamefont
  {Fixler}}, \bibinfo {author} {\bibfnamefont {G.}~\bibnamefont {Foster}},
  \bibinfo {author} {\bibfnamefont {J.}~\bibnamefont {McGuirk}},\ and\ \bibinfo
  {author} {\bibfnamefont {M.}~\bibnamefont {Kasevich}},\ }\bibfield  {title}
  {\bibinfo {title} {Atom interferometer measurement of the newtonian constant
  of gravity},\ }\href@noop {} {\bibfield  {journal} {\bibinfo  {journal}
  {Science}\ }\textbf {\bibinfo {volume} {315}},\ \bibinfo {pages} {74}
  (\bibinfo {year} {2007})}\BibitemShut {NoStop}%
\bibitem [{\citenamefont {Durfee}\ \emph {et~al.}(2006)\citenamefont {Durfee},
  \citenamefont {Shaham},\ and\ \citenamefont {Kasevich}}]{durfee2006long}%
  \BibitemOpen
  \bibfield  {author} {\bibinfo {author} {\bibfnamefont {D.}~\bibnamefont
  {Durfee}}, \bibinfo {author} {\bibfnamefont {Y.}~\bibnamefont {Shaham}},\
  and\ \bibinfo {author} {\bibfnamefont {M.}~\bibnamefont {Kasevich}},\
  }\bibfield  {title} {\bibinfo {title} {Long-term stability of an
  area-reversible atom-interferometer sagnac gyroscope},\ }\href@noop {}
  {\bibfield  {journal} {\bibinfo  {journal} {Physical review letters}\
  }\textbf {\bibinfo {volume} {97}},\ \bibinfo {pages} {240801} (\bibinfo
  {year} {2006})}\BibitemShut {NoStop}%
\bibitem [{\citenamefont {Stray}\ \emph {et~al.}(2022)\citenamefont {Stray},
  \citenamefont {Lamb}, \citenamefont {Kaushik}, \citenamefont {Vovrosh},
  \citenamefont {Rodgers}, \citenamefont {Winch}, \citenamefont {Hayati},
  \citenamefont {Boddice}, \citenamefont {Stabrawa}, \citenamefont {Niggebaum}
  \emph {et~al.}}]{stray2022quantum}%
  \BibitemOpen
  \bibfield  {author} {\bibinfo {author} {\bibfnamefont {B.}~\bibnamefont
  {Stray}}, \bibinfo {author} {\bibfnamefont {A.}~\bibnamefont {Lamb}},
  \bibinfo {author} {\bibfnamefont {A.}~\bibnamefont {Kaushik}}, \bibinfo
  {author} {\bibfnamefont {J.}~\bibnamefont {Vovrosh}}, \bibinfo {author}
  {\bibfnamefont {A.}~\bibnamefont {Rodgers}}, \bibinfo {author} {\bibfnamefont
  {J.}~\bibnamefont {Winch}}, \bibinfo {author} {\bibfnamefont
  {F.}~\bibnamefont {Hayati}}, \bibinfo {author} {\bibfnamefont
  {D.}~\bibnamefont {Boddice}}, \bibinfo {author} {\bibfnamefont
  {A.}~\bibnamefont {Stabrawa}}, \bibinfo {author} {\bibfnamefont
  {A.}~\bibnamefont {Niggebaum}}, \emph {et~al.},\ }\bibfield  {title}
  {\bibinfo {title} {Quantum sensing for gravity cartography},\ }\href@noop {}
  {\bibfield  {journal} {\bibinfo  {journal} {Nature}\ }\textbf {\bibinfo
  {volume} {602}},\ \bibinfo {pages} {590} (\bibinfo {year}
  {2022})}\BibitemShut {NoStop}%
\bibitem [{\citenamefont {Schlippert}\ \emph {et~al.}(2014)\citenamefont
  {Schlippert}, \citenamefont {Hartwig}, \citenamefont {Albers}, \citenamefont
  {Richardson}, \citenamefont {Schubert}, \citenamefont {Roura}, \citenamefont
  {Schleich}, \citenamefont {Ertmer},\ and\ \citenamefont
  {Rasel}}]{schlippert2014quantum}%
  \BibitemOpen
  \bibfield  {author} {\bibinfo {author} {\bibfnamefont {D.}~\bibnamefont
  {Schlippert}}, \bibinfo {author} {\bibfnamefont {J.}~\bibnamefont {Hartwig}},
  \bibinfo {author} {\bibfnamefont {H.}~\bibnamefont {Albers}}, \bibinfo
  {author} {\bibfnamefont {L.~L.}\ \bibnamefont {Richardson}}, \bibinfo
  {author} {\bibfnamefont {C.}~\bibnamefont {Schubert}}, \bibinfo {author}
  {\bibfnamefont {A.}~\bibnamefont {Roura}}, \bibinfo {author} {\bibfnamefont
  {W.~P.}\ \bibnamefont {Schleich}}, \bibinfo {author} {\bibfnamefont
  {W.}~\bibnamefont {Ertmer}},\ and\ \bibinfo {author} {\bibfnamefont {E.~M.}\
  \bibnamefont {Rasel}},\ }\bibfield  {title} {\bibinfo {title} {Quantum test
  of the universality of free fall},\ }\href@noop {} {\bibfield  {journal}
  {\bibinfo  {journal} {Physical Review Letters}\ }\textbf {\bibinfo {volume}
  {112}},\ \bibinfo {pages} {203002} (\bibinfo {year} {2014})}\BibitemShut
  {NoStop}%
\bibitem [{\citenamefont {Asenbaum}\ \emph {et~al.}(2020)\citenamefont
  {Asenbaum}, \citenamefont {Overstreet}, \citenamefont {Kim}, \citenamefont
  {Curti},\ and\ \citenamefont {Kasevich}}]{asenbaum2020atom}%
  \BibitemOpen
  \bibfield  {author} {\bibinfo {author} {\bibfnamefont {P.}~\bibnamefont
  {Asenbaum}}, \bibinfo {author} {\bibfnamefont {C.}~\bibnamefont
  {Overstreet}}, \bibinfo {author} {\bibfnamefont {M.}~\bibnamefont {Kim}},
  \bibinfo {author} {\bibfnamefont {J.}~\bibnamefont {Curti}},\ and\ \bibinfo
  {author} {\bibfnamefont {M.~A.}\ \bibnamefont {Kasevich}},\ }\bibfield
  {title} {\bibinfo {title} {Atom-interferometric test of the equivalence
  principle at the 10- 12 level},\ }\href@noop {} {\bibfield  {journal}
  {\bibinfo  {journal} {Physical Review Letters}\ }\textbf {\bibinfo {volume}
  {125}},\ \bibinfo {pages} {191101} (\bibinfo {year} {2020})}\BibitemShut
  {NoStop}%
\bibitem [{\citenamefont {Barrett}\ \emph {et~al.}(2022)\citenamefont
  {Barrett}, \citenamefont {Condon}, \citenamefont {Chichet}, \citenamefont
  {Antoni-Micollier}, \citenamefont {Arguel}, \citenamefont {Rabault},
  \citenamefont {Pelluet}, \citenamefont {Jarlaud}, \citenamefont {Landragin},
  \citenamefont {Bouyer} \emph {et~al.}}]{barrett2022testing}%
  \BibitemOpen
  \bibfield  {author} {\bibinfo {author} {\bibfnamefont {B.}~\bibnamefont
  {Barrett}}, \bibinfo {author} {\bibfnamefont {G.}~\bibnamefont {Condon}},
  \bibinfo {author} {\bibfnamefont {L.}~\bibnamefont {Chichet}}, \bibinfo
  {author} {\bibfnamefont {L.}~\bibnamefont {Antoni-Micollier}}, \bibinfo
  {author} {\bibfnamefont {R.}~\bibnamefont {Arguel}}, \bibinfo {author}
  {\bibfnamefont {M.}~\bibnamefont {Rabault}}, \bibinfo {author} {\bibfnamefont
  {C.}~\bibnamefont {Pelluet}}, \bibinfo {author} {\bibfnamefont
  {V.}~\bibnamefont {Jarlaud}}, \bibinfo {author} {\bibfnamefont
  {A.}~\bibnamefont {Landragin}}, \bibinfo {author} {\bibfnamefont
  {P.}~\bibnamefont {Bouyer}}, \emph {et~al.},\ }\bibfield  {title} {\bibinfo
  {title} {Testing the universality of free fall using correlated 39k--87rb
  atom interferometers},\ }\href@noop {} {\bibfield  {journal} {\bibinfo
  {journal} {AVS Quantum Science}\ }\textbf {\bibinfo {volume} {4}} (\bibinfo
  {year} {2022})}\BibitemShut {NoStop}%
\bibitem [{\citenamefont {Jiang}\ \emph {et~al.}(2019)\citenamefont {Jiang},
  \citenamefont {Frutos}, \citenamefont {Wu}, \citenamefont {Blanchard},
  \citenamefont {Peng},\ and\ \citenamefont {Budker}}]{jiang2019magnetic}%
  \BibitemOpen
  \bibfield  {author} {\bibinfo {author} {\bibfnamefont {M.}~\bibnamefont
  {Jiang}}, \bibinfo {author} {\bibfnamefont {R.~P.}\ \bibnamefont {Frutos}},
  \bibinfo {author} {\bibfnamefont {T.}~\bibnamefont {Wu}}, \bibinfo {author}
  {\bibfnamefont {J.~W.}\ \bibnamefont {Blanchard}}, \bibinfo {author}
  {\bibfnamefont {X.}~\bibnamefont {Peng}},\ and\ \bibinfo {author}
  {\bibfnamefont {D.}~\bibnamefont {Budker}},\ }\bibfield  {title} {\bibinfo
  {title} {Magnetic gradiometer for the detection of zero-to ultralow-field
  nuclear magnetic resonance},\ }\href@noop {} {\bibfield  {journal} {\bibinfo
  {journal} {Physical Review Applied}\ }\textbf {\bibinfo {volume} {11}},\
  \bibinfo {pages} {024005} (\bibinfo {year} {2019})}\BibitemShut {NoStop}%
\bibitem [{\citenamefont {Lucivero}\ \emph {et~al.}(2021)\citenamefont
  {Lucivero}, \citenamefont {Lee}, \citenamefont {Dural},\ and\ \citenamefont
  {Romalis}}]{lucivero2021femtotesla}%
  \BibitemOpen
  \bibfield  {author} {\bibinfo {author} {\bibfnamefont {V.}~\bibnamefont
  {Lucivero}}, \bibinfo {author} {\bibfnamefont {W.}~\bibnamefont {Lee}},
  \bibinfo {author} {\bibfnamefont {N.}~\bibnamefont {Dural}},\ and\ \bibinfo
  {author} {\bibfnamefont {M.}~\bibnamefont {Romalis}},\ }\bibfield  {title}
  {\bibinfo {title} {Femtotesla direct magnetic gradiometer using a single
  multipass cell},\ }\href@noop {} {\bibfield  {journal} {\bibinfo  {journal}
  {Physical Review Applied}\ }\textbf {\bibinfo {volume} {15}},\ \bibinfo
  {pages} {014004} (\bibinfo {year} {2021})}\BibitemShut {NoStop}%
\bibitem [{\citenamefont {Wu}\ \emph {et~al.}(2023)\citenamefont {Wu},
  \citenamefont {Bao}, \citenamefont {Guo}, \citenamefont {Chen}, \citenamefont
  {Du}, \citenamefont {Shi}, \citenamefont {Yang}, \citenamefont {Chen},\ and\
  \citenamefont {Zhang}}]{wu2023quantum}%
  \BibitemOpen
  \bibfield  {author} {\bibinfo {author} {\bibfnamefont {S.}~\bibnamefont
  {Wu}}, \bibinfo {author} {\bibfnamefont {G.}~\bibnamefont {Bao}}, \bibinfo
  {author} {\bibfnamefont {J.}~\bibnamefont {Guo}}, \bibinfo {author}
  {\bibfnamefont {J.}~\bibnamefont {Chen}}, \bibinfo {author} {\bibfnamefont
  {W.}~\bibnamefont {Du}}, \bibinfo {author} {\bibfnamefont {M.}~\bibnamefont
  {Shi}}, \bibinfo {author} {\bibfnamefont {P.}~\bibnamefont {Yang}}, \bibinfo
  {author} {\bibfnamefont {L.}~\bibnamefont {Chen}},\ and\ \bibinfo {author}
  {\bibfnamefont {W.}~\bibnamefont {Zhang}},\ }\bibfield  {title} {\bibinfo
  {title} {Quantum magnetic gradiometer with entangled twin light beams},\
  }\href@noop {} {\bibfield  {journal} {\bibinfo  {journal} {Science Advances}\
  }\textbf {\bibinfo {volume} {9}},\ \bibinfo {pages} {eadg1760} (\bibinfo
  {year} {2023})}\BibitemShut {NoStop}%
\bibitem [{\citenamefont {Bothwell}\ \emph {et~al.}(2022)\citenamefont
  {Bothwell}, \citenamefont {Kennedy}, \citenamefont {Aeppli}, \citenamefont
  {Kedar}, \citenamefont {Robinson}, \citenamefont {Oelker}, \citenamefont
  {Staron},\ and\ \citenamefont {Ye}}]{bothwell2022resolving}%
  \BibitemOpen
  \bibfield  {author} {\bibinfo {author} {\bibfnamefont {T.}~\bibnamefont
  {Bothwell}}, \bibinfo {author} {\bibfnamefont {C.~J.}\ \bibnamefont
  {Kennedy}}, \bibinfo {author} {\bibfnamefont {A.}~\bibnamefont {Aeppli}},
  \bibinfo {author} {\bibfnamefont {D.}~\bibnamefont {Kedar}}, \bibinfo
  {author} {\bibfnamefont {J.~M.}\ \bibnamefont {Robinson}}, \bibinfo {author}
  {\bibfnamefont {E.}~\bibnamefont {Oelker}}, \bibinfo {author} {\bibfnamefont
  {A.}~\bibnamefont {Staron}},\ and\ \bibinfo {author} {\bibfnamefont
  {J.}~\bibnamefont {Ye}},\ }\bibfield  {title} {\bibinfo {title} {Resolving
  the gravitational redshift across a millimetre-scale atomic sample},\
  }\href@noop {} {\bibfield  {journal} {\bibinfo  {journal} {Nature}\ }\textbf
  {\bibinfo {volume} {602}},\ \bibinfo {pages} {420} (\bibinfo {year}
  {2022})}\BibitemShut {NoStop}%
\bibitem [{\citenamefont {Zheng}\ \emph {et~al.}(2023)\citenamefont {Zheng},
  \citenamefont {Dolde}, \citenamefont {Cambria}, \citenamefont {Lim},\ and\
  \citenamefont {Kolkowitz}}]{zheng2023lab}%
  \BibitemOpen
  \bibfield  {author} {\bibinfo {author} {\bibfnamefont {X.}~\bibnamefont
  {Zheng}}, \bibinfo {author} {\bibfnamefont {J.}~\bibnamefont {Dolde}},
  \bibinfo {author} {\bibfnamefont {M.~C.}\ \bibnamefont {Cambria}}, \bibinfo
  {author} {\bibfnamefont {H.~M.}\ \bibnamefont {Lim}},\ and\ \bibinfo {author}
  {\bibfnamefont {S.}~\bibnamefont {Kolkowitz}},\ }\bibfield  {title} {\bibinfo
  {title} {A lab-based test of the gravitational redshift with a miniature
  clock network},\ }\href@noop {} {\bibfield  {journal} {\bibinfo  {journal}
  {Nature Communications}\ }\textbf {\bibinfo {volume} {14}},\ \bibinfo {pages}
  {4886} (\bibinfo {year} {2023})}\BibitemShut {NoStop}%
\bibitem [{\citenamefont {Robinson}\ \emph {et~al.}(2024)\citenamefont
  {Robinson}, \citenamefont {Miklos}, \citenamefont {Tso}, \citenamefont
  {Kennedy}, \citenamefont {Bothwell}, \citenamefont {Kedar}, \citenamefont
  {Thompson},\ and\ \citenamefont {Ye}}]{robinson2024direct}%
  \BibitemOpen
  \bibfield  {author} {\bibinfo {author} {\bibfnamefont {J.~M.}\ \bibnamefont
  {Robinson}}, \bibinfo {author} {\bibfnamefont {M.}~\bibnamefont {Miklos}},
  \bibinfo {author} {\bibfnamefont {Y.~M.}\ \bibnamefont {Tso}}, \bibinfo
  {author} {\bibfnamefont {C.~J.}\ \bibnamefont {Kennedy}}, \bibinfo {author}
  {\bibfnamefont {T.}~\bibnamefont {Bothwell}}, \bibinfo {author}
  {\bibfnamefont {D.}~\bibnamefont {Kedar}}, \bibinfo {author} {\bibfnamefont
  {J.~K.}\ \bibnamefont {Thompson}},\ and\ \bibinfo {author} {\bibfnamefont
  {J.}~\bibnamefont {Ye}},\ }\bibfield  {title} {\bibinfo {title} {Direct
  comparison of two spin-squeezed optical clock ensembles at the 10- 17
  level},\ }\href@noop {} {\bibfield  {journal} {\bibinfo  {journal} {Nature
  Physics}\ }\textbf {\bibinfo {volume} {20}},\ \bibinfo {pages} {208}
  (\bibinfo {year} {2024})}\BibitemShut {NoStop}%
\bibitem [{\citenamefont {Yang}\ \emph {et~al.}(2025)\citenamefont {Yang},
  \citenamefont {Miklos}, \citenamefont {Tso}, \citenamefont {Kraus},
  \citenamefont {Hur},\ and\ \citenamefont {Ye}}]{yang2025clock}%
  \BibitemOpen
  \bibfield  {author} {\bibinfo {author} {\bibfnamefont {Y.}~\bibnamefont
  {Yang}}, \bibinfo {author} {\bibfnamefont {M.}~\bibnamefont {Miklos}},
  \bibinfo {author} {\bibfnamefont {Y.~M.}\ \bibnamefont {Tso}}, \bibinfo
  {author} {\bibfnamefont {S.}~\bibnamefont {Kraus}}, \bibinfo {author}
  {\bibfnamefont {J.}~\bibnamefont {Hur}},\ and\ \bibinfo {author}
  {\bibfnamefont {J.}~\bibnamefont {Ye}},\ }\bibfield  {title} {\bibinfo
  {title} {Clock precision beyond the standard quantum limit at 10-18 level},\
  }\href@noop {} {\bibfield  {journal} {\bibinfo  {journal} {Physical Review
  Letters}\ }\textbf {\bibinfo {volume} {135}},\ \bibinfo {pages} {193202}
  (\bibinfo {year} {2025})}\BibitemShut {NoStop}%
\bibitem [{\citenamefont {Pezze}\ \emph {et~al.}(2018)\citenamefont {Pezze},
  \citenamefont {Smerzi}, \citenamefont {Oberthaler}, \citenamefont {Schmied},\
  and\ \citenamefont {Treutlein}}]{pezze2018quantum}%
  \BibitemOpen
  \bibfield  {author} {\bibinfo {author} {\bibfnamefont {L.}~\bibnamefont
  {Pezze}}, \bibinfo {author} {\bibfnamefont {A.}~\bibnamefont {Smerzi}},
  \bibinfo {author} {\bibfnamefont {M.~K.}\ \bibnamefont {Oberthaler}},
  \bibinfo {author} {\bibfnamefont {R.}~\bibnamefont {Schmied}},\ and\ \bibinfo
  {author} {\bibfnamefont {P.}~\bibnamefont {Treutlein}},\ }\bibfield  {title}
  {\bibinfo {title} {Quantum metrology with nonclassical states of atomic
  ensembles},\ }\href@noop {} {\bibfield  {journal} {\bibinfo  {journal}
  {Reviews of Modern Physics}\ }\textbf {\bibinfo {volume} {90}},\ \bibinfo
  {pages} {035005} (\bibinfo {year} {2018})}\BibitemShut {NoStop}%
\bibitem [{\citenamefont {Huang}\ \emph {et~al.}(2024)\citenamefont {Huang},
  \citenamefont {Zhuang},\ and\ \citenamefont {Lee}}]{huang2024entanglement}%
  \BibitemOpen
  \bibfield  {author} {\bibinfo {author} {\bibfnamefont {J.}~\bibnamefont
  {Huang}}, \bibinfo {author} {\bibfnamefont {M.}~\bibnamefont {Zhuang}},\ and\
  \bibinfo {author} {\bibfnamefont {C.}~\bibnamefont {Lee}},\ }\bibfield
  {title} {\bibinfo {title} {Entanglement-enhanced quantum metrology: from
  standard quantum limit to heisenberg limit},\ }\href@noop {} {\bibfield
  {journal} {\bibinfo  {journal} {Applied Physics Reviews}\ }\textbf {\bibinfo
  {volume} {11}} (\bibinfo {year} {2024})}\BibitemShut {NoStop}%
\bibitem [{\citenamefont {Foster}\ \emph {et~al.}(2002)\citenamefont {Foster},
  \citenamefont {Fixler}, \citenamefont {McGuirk},\ and\ \citenamefont
  {Kasevich}}]{foster2002method}%
  \BibitemOpen
  \bibfield  {author} {\bibinfo {author} {\bibfnamefont {G.}~\bibnamefont
  {Foster}}, \bibinfo {author} {\bibfnamefont {J.}~\bibnamefont {Fixler}},
  \bibinfo {author} {\bibfnamefont {J.}~\bibnamefont {McGuirk}},\ and\ \bibinfo
  {author} {\bibfnamefont {M.}~\bibnamefont {Kasevich}},\ }\bibfield  {title}
  {\bibinfo {title} {Method of phase extraction between coupled atom
  interferometers using ellipse-specific fitting},\ }\href@noop {} {\bibfield
  {journal} {\bibinfo  {journal} {Optics letters}\ }\textbf {\bibinfo {volume}
  {27}},\ \bibinfo {pages} {951} (\bibinfo {year} {2002})}\BibitemShut
  {NoStop}%
\bibitem [{\citenamefont {Rosi}\ \emph {et~al.}(2015)\citenamefont {Rosi},
  \citenamefont {Cacciapuoti}, \citenamefont {Sorrentino}, \citenamefont
  {Menchetti}, \citenamefont {Prevedelli},\ and\ \citenamefont
  {Tino}}]{rosi2015measurement}%
  \BibitemOpen
  \bibfield  {author} {\bibinfo {author} {\bibfnamefont {G.}~\bibnamefont
  {Rosi}}, \bibinfo {author} {\bibfnamefont {L.}~\bibnamefont {Cacciapuoti}},
  \bibinfo {author} {\bibfnamefont {F.}~\bibnamefont {Sorrentino}}, \bibinfo
  {author} {\bibfnamefont {M.}~\bibnamefont {Menchetti}}, \bibinfo {author}
  {\bibfnamefont {M.}~\bibnamefont {Prevedelli}},\ and\ \bibinfo {author}
  {\bibfnamefont {G.}~\bibnamefont {Tino}},\ }\bibfield  {title} {\bibinfo
  {title} {Measurement of the gravity-field curvature by atom interferometry},\
  }\href@noop {} {\bibfield  {journal} {\bibinfo  {journal} {Physical Review
  Letters}\ }\textbf {\bibinfo {volume} {114}},\ \bibinfo {pages} {013001}
  (\bibinfo {year} {2015})}\BibitemShut {NoStop}%
\bibitem [{\citenamefont {Barrett}\ \emph {et~al.}(2016)\citenamefont
  {Barrett}, \citenamefont {Antoni-Micollier}, \citenamefont {Chichet},
  \citenamefont {Battelier}, \citenamefont {L{\'e}veque}, \citenamefont
  {Landragin},\ and\ \citenamefont {Bouyer}}]{barrett2016dual}%
  \BibitemOpen
  \bibfield  {author} {\bibinfo {author} {\bibfnamefont {B.}~\bibnamefont
  {Barrett}}, \bibinfo {author} {\bibfnamefont {L.}~\bibnamefont
  {Antoni-Micollier}}, \bibinfo {author} {\bibfnamefont {L.}~\bibnamefont
  {Chichet}}, \bibinfo {author} {\bibfnamefont {B.}~\bibnamefont {Battelier}},
  \bibinfo {author} {\bibfnamefont {T.}~\bibnamefont {L{\'e}veque}}, \bibinfo
  {author} {\bibfnamefont {A.}~\bibnamefont {Landragin}},\ and\ \bibinfo
  {author} {\bibfnamefont {P.}~\bibnamefont {Bouyer}},\ }\bibfield  {title}
  {\bibinfo {title} {Dual matter-wave inertial sensors in weightlessness},\
  }\href@noop {} {\bibfield  {journal} {\bibinfo  {journal} {Nature
  communications}\ }\textbf {\bibinfo {volume} {7}},\ \bibinfo {pages} {13786}
  (\bibinfo {year} {2016})}\BibitemShut {NoStop}%
\bibitem [{\citenamefont {Langlois}\ \emph {et~al.}(2017)\citenamefont
  {Langlois}, \citenamefont {Caldani}, \citenamefont {Trimeche}, \citenamefont
  {Merlet},\ and\ \citenamefont {Pereira~dos
  Santos}}]{langlois2017differential}%
  \BibitemOpen
  \bibfield  {author} {\bibinfo {author} {\bibfnamefont {M.}~\bibnamefont
  {Langlois}}, \bibinfo {author} {\bibfnamefont {R.}~\bibnamefont {Caldani}},
  \bibinfo {author} {\bibfnamefont {A.}~\bibnamefont {Trimeche}}, \bibinfo
  {author} {\bibfnamefont {S.}~\bibnamefont {Merlet}},\ and\ \bibinfo {author}
  {\bibfnamefont {F.}~\bibnamefont {Pereira~dos Santos}},\ }\bibfield  {title}
  {\bibinfo {title} {Differential phase extraction in dual interferometers
  exploiting the correlation between classical and quantum sensors},\
  }\href@noop {} {\bibfield  {journal} {\bibinfo  {journal} {Physical Review
  A}\ }\textbf {\bibinfo {volume} {96}},\ \bibinfo {pages} {053624} (\bibinfo
  {year} {2017})}\BibitemShut {NoStop}%
\bibitem [{\citenamefont {Marti}\ \emph {et~al.}(2018)\citenamefont {Marti},
  \citenamefont {Hutson}, \citenamefont {Goban}, \citenamefont {Campbell},
  \citenamefont {Poli},\ and\ \citenamefont {Ye}}]{marti2018imaging}%
  \BibitemOpen
  \bibfield  {author} {\bibinfo {author} {\bibfnamefont {G.~E.}\ \bibnamefont
  {Marti}}, \bibinfo {author} {\bibfnamefont {R.~B.}\ \bibnamefont {Hutson}},
  \bibinfo {author} {\bibfnamefont {A.}~\bibnamefont {Goban}}, \bibinfo
  {author} {\bibfnamefont {S.~L.}\ \bibnamefont {Campbell}}, \bibinfo {author}
  {\bibfnamefont {N.}~\bibnamefont {Poli}},\ and\ \bibinfo {author}
  {\bibfnamefont {J.}~\bibnamefont {Ye}},\ }\bibfield  {title} {\bibinfo
  {title} {Imaging optical frequencies with 100 $\mu$ hz precision and 1.1
  $\mu$ m resolution},\ }\href@noop {} {\bibfield  {journal} {\bibinfo
  {journal} {Physical review letters}\ }\textbf {\bibinfo {volume} {120}},\
  \bibinfo {pages} {103201} (\bibinfo {year} {2018})}\BibitemShut {NoStop}%
\bibitem [{\citenamefont {Elliott}\ \emph {et~al.}(2023)\citenamefont
  {Elliott}, \citenamefont {Aveline}, \citenamefont {Bigelow}, \citenamefont
  {Boegel}, \citenamefont {Botsi}, \citenamefont {Charron}, \citenamefont
  {d’Incao}, \citenamefont {Engels}, \citenamefont {Estrampes}, \citenamefont
  {Gaaloul} \emph {et~al.}}]{elliott2023quantum}%
  \BibitemOpen
  \bibfield  {author} {\bibinfo {author} {\bibfnamefont {E.~R.}\ \bibnamefont
  {Elliott}}, \bibinfo {author} {\bibfnamefont {D.~C.}\ \bibnamefont
  {Aveline}}, \bibinfo {author} {\bibfnamefont {N.~P.}\ \bibnamefont
  {Bigelow}}, \bibinfo {author} {\bibfnamefont {P.}~\bibnamefont {Boegel}},
  \bibinfo {author} {\bibfnamefont {S.}~\bibnamefont {Botsi}}, \bibinfo
  {author} {\bibfnamefont {E.}~\bibnamefont {Charron}}, \bibinfo {author}
  {\bibfnamefont {J.~P.}\ \bibnamefont {d’Incao}}, \bibinfo {author}
  {\bibfnamefont {P.}~\bibnamefont {Engels}}, \bibinfo {author} {\bibfnamefont
  {T.}~\bibnamefont {Estrampes}}, \bibinfo {author} {\bibfnamefont
  {N.}~\bibnamefont {Gaaloul}}, \emph {et~al.},\ }\bibfield  {title} {\bibinfo
  {title} {Quantum gas mixtures and dual-species atom interferometry in
  space},\ }\href@noop {} {\bibfield  {journal} {\bibinfo  {journal} {Nature}\
  }\textbf {\bibinfo {volume} {623}},\ \bibinfo {pages} {502} (\bibinfo {year}
  {2023})}\BibitemShut {NoStop}%
\bibitem [{\citenamefont {Wineland}\ \emph {et~al.}(1992)\citenamefont
  {Wineland}, \citenamefont {Bollinger}, \citenamefont {Itano}, \citenamefont
  {Moore},\ and\ \citenamefont {Heinzen}}]{wineland1992spin}%
  \BibitemOpen
  \bibfield  {author} {\bibinfo {author} {\bibfnamefont {D.~J.}\ \bibnamefont
  {Wineland}}, \bibinfo {author} {\bibfnamefont {J.~J.}\ \bibnamefont
  {Bollinger}}, \bibinfo {author} {\bibfnamefont {W.~M.}\ \bibnamefont
  {Itano}}, \bibinfo {author} {\bibfnamefont {F.}~\bibnamefont {Moore}},\ and\
  \bibinfo {author} {\bibfnamefont {D.~J.}\ \bibnamefont {Heinzen}},\
  }\bibfield  {title} {\bibinfo {title} {Spin squeezing and reduced quantum
  noise in spectroscopy},\ }\href@noop {} {\bibfield  {journal} {\bibinfo
  {journal} {Physical Review A}\ }\textbf {\bibinfo {volume} {46}},\ \bibinfo
  {pages} {R6797} (\bibinfo {year} {1992})}\BibitemShut {NoStop}%
\bibitem [{\citenamefont {Kitagawa}\ and\ \citenamefont
  {Ueda}(1993)}]{kitagawa1993squeezed}%
  \BibitemOpen
  \bibfield  {author} {\bibinfo {author} {\bibfnamefont {M.}~\bibnamefont
  {Kitagawa}}\ and\ \bibinfo {author} {\bibfnamefont {M.}~\bibnamefont
  {Ueda}},\ }\bibfield  {title} {\bibinfo {title} {Squeezed spin states},\
  }\href@noop {} {\bibfield  {journal} {\bibinfo  {journal} {Physical Review
  A}\ }\textbf {\bibinfo {volume} {47}},\ \bibinfo {pages} {5138} (\bibinfo
  {year} {1993})}\BibitemShut {NoStop}%
\bibitem [{\citenamefont {Ma}\ \emph {et~al.}(2011)\citenamefont {Ma},
  \citenamefont {Wang}, \citenamefont {Sun},\ and\ \citenamefont
  {Nori}}]{ma2011quantum}%
  \BibitemOpen
  \bibfield  {author} {\bibinfo {author} {\bibfnamefont {J.}~\bibnamefont
  {Ma}}, \bibinfo {author} {\bibfnamefont {X.}~\bibnamefont {Wang}}, \bibinfo
  {author} {\bibfnamefont {C.-P.}\ \bibnamefont {Sun}},\ and\ \bibinfo {author}
  {\bibfnamefont {F.}~\bibnamefont {Nori}},\ }\bibfield  {title} {\bibinfo
  {title} {Quantum spin squeezing},\ }\href@noop {} {\bibfield  {journal}
  {\bibinfo  {journal} {Physics Reports}\ }\textbf {\bibinfo {volume} {509}},\
  \bibinfo {pages} {89} (\bibinfo {year} {2011})}\BibitemShut {NoStop}%
\bibitem [{\citenamefont {Eckner}\ \emph {et~al.}(2023)\citenamefont {Eckner},
  \citenamefont {Darkwah~Oppong}, \citenamefont {Cao}, \citenamefont {Young},
  \citenamefont {Milner}, \citenamefont {Robinson}, \citenamefont {Ye},\ and\
  \citenamefont {Kaufman}}]{eckner2023realizing}%
  \BibitemOpen
  \bibfield  {author} {\bibinfo {author} {\bibfnamefont {W.~J.}\ \bibnamefont
  {Eckner}}, \bibinfo {author} {\bibfnamefont {N.}~\bibnamefont
  {Darkwah~Oppong}}, \bibinfo {author} {\bibfnamefont {A.}~\bibnamefont {Cao}},
  \bibinfo {author} {\bibfnamefont {A.~W.}\ \bibnamefont {Young}}, \bibinfo
  {author} {\bibfnamefont {W.~R.}\ \bibnamefont {Milner}}, \bibinfo {author}
  {\bibfnamefont {J.~M.}\ \bibnamefont {Robinson}}, \bibinfo {author}
  {\bibfnamefont {J.}~\bibnamefont {Ye}},\ and\ \bibinfo {author}
  {\bibfnamefont {A.~M.}\ \bibnamefont {Kaufman}},\ }\bibfield  {title}
  {\bibinfo {title} {Realizing spin squeezing with rydberg interactions in an
  optical clock},\ }\href@noop {} {\bibfield  {journal} {\bibinfo  {journal}
  {Nature}\ }\textbf {\bibinfo {volume} {621}},\ \bibinfo {pages} {734}
  (\bibinfo {year} {2023})}\BibitemShut {NoStop}%
\bibitem [{\citenamefont {corgier}\ \emph {et~al.}(2025)\citenamefont
  {corgier}, \citenamefont {Malitesta}, \citenamefont {Sidorenkov},
  \citenamefont {Pereira~dos Santos}, \citenamefont {Rosi}, \citenamefont
  {Tino}, \citenamefont {Smerzi}, \citenamefont {Salvi},\ and\ \citenamefont
  {Pezz{\`e}}}]{corgier2025optimized}%
  \BibitemOpen
  \bibfield  {author} {\bibinfo {author} {\bibfnamefont {r.}~\bibnamefont
  {corgier}}, \bibinfo {author} {\bibfnamefont {M.}~\bibnamefont {Malitesta}},
  \bibinfo {author} {\bibfnamefont {L.}~\bibnamefont {Sidorenkov}}, \bibinfo
  {author} {\bibfnamefont {F.}~\bibnamefont {Pereira~dos Santos}}, \bibinfo
  {author} {\bibfnamefont {G.}~\bibnamefont {Rosi}}, \bibinfo {author}
  {\bibfnamefont {G.~M.}\ \bibnamefont {Tino}}, \bibinfo {author}
  {\bibfnamefont {A.}~\bibnamefont {Smerzi}}, \bibinfo {author} {\bibfnamefont
  {L.}~\bibnamefont {Salvi}},\ and\ \bibinfo {author} {\bibfnamefont
  {L.}~\bibnamefont {Pezz{\`e}}},\ }\bibfield  {title} {\bibinfo {title}
  {Optimized squeezing for accurate differential sensing under large phase
  noise},\ }\href@noop {} {\bibfield  {journal} {\bibinfo  {journal} {Quantum
  Science and Technology}\ } (\bibinfo {year} {2025})}\BibitemShut {NoStop}%
\bibitem [{\citenamefont {Roos}\ \emph {et~al.}(2006)\citenamefont {Roos},
  \citenamefont {Chwalla}, \citenamefont {Kim}, \citenamefont {Riebe},\ and\
  \citenamefont {Blatt}}]{roos2006designer}%
  \BibitemOpen
  \bibfield  {author} {\bibinfo {author} {\bibfnamefont {C.~F.}\ \bibnamefont
  {Roos}}, \bibinfo {author} {\bibfnamefont {M.}~\bibnamefont {Chwalla}},
  \bibinfo {author} {\bibfnamefont {K.}~\bibnamefont {Kim}}, \bibinfo {author}
  {\bibfnamefont {M.}~\bibnamefont {Riebe}},\ and\ \bibinfo {author}
  {\bibfnamefont {R.}~\bibnamefont {Blatt}},\ }\bibfield  {title} {\bibinfo
  {title} {‘designer atoms’ for quantum metrology},\ }\href@noop {}
  {\bibfield  {journal} {\bibinfo  {journal} {Nature}\ }\textbf {\bibinfo
  {volume} {443}},\ \bibinfo {pages} {316} (\bibinfo {year}
  {2006})}\BibitemShut {NoStop}%
\bibitem [{\citenamefont {Monz}\ \emph {et~al.}(2011)\citenamefont {Monz},
  \citenamefont {Schindler}, \citenamefont {Barreiro}, \citenamefont {Chwalla},
  \citenamefont {Nigg}, \citenamefont {Coish}, \citenamefont {Harlander},
  \citenamefont {H{\"a}nsel}, \citenamefont {Hennrich},\ and\ \citenamefont
  {Blatt}}]{monz201114}%
  \BibitemOpen
  \bibfield  {author} {\bibinfo {author} {\bibfnamefont {T.}~\bibnamefont
  {Monz}}, \bibinfo {author} {\bibfnamefont {P.}~\bibnamefont {Schindler}},
  \bibinfo {author} {\bibfnamefont {J.~T.}\ \bibnamefont {Barreiro}}, \bibinfo
  {author} {\bibfnamefont {M.}~\bibnamefont {Chwalla}}, \bibinfo {author}
  {\bibfnamefont {D.}~\bibnamefont {Nigg}}, \bibinfo {author} {\bibfnamefont
  {W.~A.}\ \bibnamefont {Coish}}, \bibinfo {author} {\bibfnamefont
  {M.}~\bibnamefont {Harlander}}, \bibinfo {author} {\bibfnamefont
  {W.}~\bibnamefont {H{\"a}nsel}}, \bibinfo {author} {\bibfnamefont
  {M.}~\bibnamefont {Hennrich}},\ and\ \bibinfo {author} {\bibfnamefont
  {R.}~\bibnamefont {Blatt}},\ }\bibfield  {title} {\bibinfo {title} {14-qubit
  entanglement: Creation and coherence},\ }\href@noop {} {\bibfield  {journal}
  {\bibinfo  {journal} {Physical Review Letters}\ }\textbf {\bibinfo {volume}
  {106}},\ \bibinfo {pages} {130506} (\bibinfo {year} {2011})}\BibitemShut
  {NoStop}%
\bibitem [{\citenamefont {Dorner}(2012)}]{dorner2012quantum}%
  \BibitemOpen
  \bibfield  {author} {\bibinfo {author} {\bibfnamefont {U.}~\bibnamefont
  {Dorner}},\ }\bibfield  {title} {\bibinfo {title} {Quantum frequency
  estimation with trapped ions and atoms},\ }\href@noop {} {\bibfield
  {journal} {\bibinfo  {journal} {New Journal of Physics}\ }\textbf {\bibinfo
  {volume} {14}},\ \bibinfo {pages} {043011} (\bibinfo {year}
  {2012})}\BibitemShut {NoStop}%
\bibitem [{\citenamefont {Landini}\ \emph {et~al.}(2014)\citenamefont
  {Landini}, \citenamefont {Fattori}, \citenamefont {Pezz{\`e}},\ and\
  \citenamefont {Smerzi}}]{landini2014phase}%
  \BibitemOpen
  \bibfield  {author} {\bibinfo {author} {\bibfnamefont {M.}~\bibnamefont
  {Landini}}, \bibinfo {author} {\bibfnamefont {M.}~\bibnamefont {Fattori}},
  \bibinfo {author} {\bibfnamefont {L.}~\bibnamefont {Pezz{\`e}}},\ and\
  \bibinfo {author} {\bibfnamefont {A.}~\bibnamefont {Smerzi}},\ }\bibfield
  {title} {\bibinfo {title} {Phase-noise protection in quantum-enhanced
  differential interferometry},\ }\href@noop {} {\bibfield  {journal} {\bibinfo
   {journal} {New Journal of Physics}\ }\textbf {\bibinfo {volume} {16}},\
  \bibinfo {pages} {113074} (\bibinfo {year} {2014})}\BibitemShut {NoStop}%
\bibitem [{\citenamefont {Jeske}\ \emph {et~al.}(2014)\citenamefont {Jeske},
  \citenamefont {Cole},\ and\ \citenamefont {Huelga}}]{jeske2014quantum}%
  \BibitemOpen
  \bibfield  {author} {\bibinfo {author} {\bibfnamefont {J.}~\bibnamefont
  {Jeske}}, \bibinfo {author} {\bibfnamefont {J.~H.}\ \bibnamefont {Cole}},\
  and\ \bibinfo {author} {\bibfnamefont {S.~F.}\ \bibnamefont {Huelga}},\
  }\bibfield  {title} {\bibinfo {title} {Quantum metrology subject to spatially
  correlated markovian noise: restoring the heisenberg limit},\ }\href@noop {}
  {\bibfield  {journal} {\bibinfo  {journal} {New Journal of Physics}\ }\textbf
  {\bibinfo {volume} {16}},\ \bibinfo {pages} {073039} (\bibinfo {year}
  {2014})}\BibitemShut {NoStop}%
\bibitem [{\citenamefont {Altenburg}\ \emph {et~al.}(2017)\citenamefont
  {Altenburg}, \citenamefont {Oszmaniec}, \citenamefont {W{\"o}lk},\ and\
  \citenamefont {G{\"u}hne}}]{altenburg2017estimation}%
  \BibitemOpen
  \bibfield  {author} {\bibinfo {author} {\bibfnamefont {S.}~\bibnamefont
  {Altenburg}}, \bibinfo {author} {\bibfnamefont {M.}~\bibnamefont
  {Oszmaniec}}, \bibinfo {author} {\bibfnamefont {S.}~\bibnamefont
  {W{\"o}lk}},\ and\ \bibinfo {author} {\bibfnamefont {O.}~\bibnamefont
  {G{\"u}hne}},\ }\bibfield  {title} {\bibinfo {title} {Estimation of gradients
  in quantum metrology},\ }\href@noop {} {\bibfield  {journal} {\bibinfo
  {journal} {Physical Review A}\ }\textbf {\bibinfo {volume} {96}},\ \bibinfo
  {pages} {042319} (\bibinfo {year} {2017})}\BibitemShut {NoStop}%
\bibitem [{\citenamefont {Sekatski}\ \emph {et~al.}(2020)\citenamefont
  {Sekatski}, \citenamefont {W{\"o}lk},\ and\ \citenamefont
  {D{\"u}r}}]{sekatski2020optimal}%
  \BibitemOpen
  \bibfield  {author} {\bibinfo {author} {\bibfnamefont {P.}~\bibnamefont
  {Sekatski}}, \bibinfo {author} {\bibfnamefont {S.}~\bibnamefont {W{\"o}lk}},\
  and\ \bibinfo {author} {\bibfnamefont {W.}~\bibnamefont {D{\"u}r}},\
  }\bibfield  {title} {\bibinfo {title} {Optimal distributed sensing in noisy
  environments},\ }\href@noop {} {\bibfield  {journal} {\bibinfo  {journal}
  {Physical Review Research}\ }\textbf {\bibinfo {volume} {2}},\ \bibinfo
  {pages} {023052} (\bibinfo {year} {2020})}\BibitemShut {NoStop}%
\bibitem [{\citenamefont {Hamann}\ \emph {et~al.}(2022)\citenamefont {Hamann},
  \citenamefont {Sekatski},\ and\ \citenamefont
  {D{\"u}r}}]{hamann2022approximate}%
  \BibitemOpen
  \bibfield  {author} {\bibinfo {author} {\bibfnamefont {A.}~\bibnamefont
  {Hamann}}, \bibinfo {author} {\bibfnamefont {P.}~\bibnamefont {Sekatski}},\
  and\ \bibinfo {author} {\bibfnamefont {W.}~\bibnamefont {D{\"u}r}},\
  }\bibfield  {title} {\bibinfo {title} {Approximate decoherence free subspaces
  for distributed sensing},\ }\href@noop {} {\bibfield  {journal} {\bibinfo
  {journal} {Quantum Science and Technology}\ }\textbf {\bibinfo {volume}
  {7}},\ \bibinfo {pages} {025003} (\bibinfo {year} {2022})}\BibitemShut
  {NoStop}%
\bibitem [{\citenamefont {Hamann}\ \emph {et~al.}(2024)\citenamefont {Hamann},
  \citenamefont {Sekatski},\ and\ \citenamefont {D{\"u}r}}]{hamann2024optimal}%
  \BibitemOpen
  \bibfield  {author} {\bibinfo {author} {\bibfnamefont {A.}~\bibnamefont
  {Hamann}}, \bibinfo {author} {\bibfnamefont {P.}~\bibnamefont {Sekatski}},\
  and\ \bibinfo {author} {\bibfnamefont {W.}~\bibnamefont {D{\"u}r}},\
  }\bibfield  {title} {\bibinfo {title} {Optimal distributed multi-parameter
  estimation in noisy environments},\ }\href@noop {} {\bibfield  {journal}
  {\bibinfo  {journal} {Quantum Science and Technology}\ }\textbf {\bibinfo
  {volume} {9}},\ \bibinfo {pages} {035005} (\bibinfo {year}
  {2024})}\BibitemShut {NoStop}%
\bibitem [{\citenamefont {Hainzer}\ \emph {et~al.}(2024)\citenamefont
  {Hainzer}, \citenamefont {Kiesenhofer}, \citenamefont {Ollikainen},
  \citenamefont {Bock}, \citenamefont {Kranzl}, \citenamefont {Joshi},
  \citenamefont {Yoeli}, \citenamefont {Blatt}, \citenamefont {Gefen},\ and\
  \citenamefont {Roos}}]{hainzer2024correlation}%
  \BibitemOpen
  \bibfield  {author} {\bibinfo {author} {\bibfnamefont {H.}~\bibnamefont
  {Hainzer}}, \bibinfo {author} {\bibfnamefont {D.}~\bibnamefont
  {Kiesenhofer}}, \bibinfo {author} {\bibfnamefont {T.}~\bibnamefont
  {Ollikainen}}, \bibinfo {author} {\bibfnamefont {M.}~\bibnamefont {Bock}},
  \bibinfo {author} {\bibfnamefont {F.}~\bibnamefont {Kranzl}}, \bibinfo
  {author} {\bibfnamefont {M.}~\bibnamefont {Joshi}}, \bibinfo {author}
  {\bibfnamefont {G.}~\bibnamefont {Yoeli}}, \bibinfo {author} {\bibfnamefont
  {R.}~\bibnamefont {Blatt}}, \bibinfo {author} {\bibfnamefont
  {T.}~\bibnamefont {Gefen}},\ and\ \bibinfo {author} {\bibfnamefont
  {C.}~\bibnamefont {Roos}},\ }\bibfield  {title} {\bibinfo {title}
  {Correlation spectroscopy with multiqubit-enhanced phase estimation},\
  }\href@noop {} {\bibfield  {journal} {\bibinfo  {journal} {Physical Review
  X}\ }\textbf {\bibinfo {volume} {14}},\ \bibinfo {pages} {011033} (\bibinfo
  {year} {2024})}\BibitemShut {NoStop}%
\bibitem [{\citenamefont {Bate}\ \emph {et~al.}(2025)\citenamefont {Bate},
  \citenamefont {Hamann}, \citenamefont {Canteri}, \citenamefont {Winkler},
  \citenamefont {Koong}, \citenamefont {Krutyanskiy}, \citenamefont {D{\"u}r},\
  and\ \citenamefont {Lanyon}}]{bate2025experimental}%
  \BibitemOpen
  \bibfield  {author} {\bibinfo {author} {\bibfnamefont {J.}~\bibnamefont
  {Bate}}, \bibinfo {author} {\bibfnamefont {A.}~\bibnamefont {Hamann}},
  \bibinfo {author} {\bibfnamefont {M.}~\bibnamefont {Canteri}}, \bibinfo
  {author} {\bibfnamefont {A.}~\bibnamefont {Winkler}}, \bibinfo {author}
  {\bibfnamefont {Z.~X.}\ \bibnamefont {Koong}}, \bibinfo {author}
  {\bibfnamefont {V.}~\bibnamefont {Krutyanskiy}}, \bibinfo {author}
  {\bibfnamefont {W.}~\bibnamefont {D{\"u}r}},\ and\ \bibinfo {author}
  {\bibfnamefont {B.~P.}\ \bibnamefont {Lanyon}},\ }\bibfield  {title}
  {\bibinfo {title} {Experimental distributed quantum sensing in a noisy
  environment},\ }\href@noop {} {\bibfield  {journal} {\bibinfo  {journal}
  {Physical Review Letters}\ }\textbf {\bibinfo {volume} {135}},\ \bibinfo
  {pages} {220801} (\bibinfo {year} {2025})}\BibitemShut {NoStop}%
\bibitem [{\citenamefont {Dietze}\ \emph {et~al.}(2026)\citenamefont {Dietze},
  \citenamefont {Pelzer}, \citenamefont {Krinner}, \citenamefont {Dawel},
  \citenamefont {Kramer}, \citenamefont {Spethmann}, \citenamefont {Kielinski},
  \citenamefont {Hammerer}, \citenamefont {Stahl}, \citenamefont {Klose} \emph
  {et~al.}}]{dietze2026entanglement}%
  \BibitemOpen
  \bibfield  {author} {\bibinfo {author} {\bibfnamefont {K.}~\bibnamefont
  {Dietze}}, \bibinfo {author} {\bibfnamefont {L.}~\bibnamefont {Pelzer}},
  \bibinfo {author} {\bibfnamefont {L.}~\bibnamefont {Krinner}}, \bibinfo
  {author} {\bibfnamefont {F.}~\bibnamefont {Dawel}}, \bibinfo {author}
  {\bibfnamefont {J.}~\bibnamefont {Kramer}}, \bibinfo {author} {\bibfnamefont
  {N.~C.}\ \bibnamefont {Spethmann}}, \bibinfo {author} {\bibfnamefont
  {T.}~\bibnamefont {Kielinski}}, \bibinfo {author} {\bibfnamefont
  {K.}~\bibnamefont {Hammerer}}, \bibinfo {author} {\bibfnamefont
  {K.}~\bibnamefont {Stahl}}, \bibinfo {author} {\bibfnamefont
  {J.}~\bibnamefont {Klose}}, \emph {et~al.},\ }\bibfield  {title} {\bibinfo
  {title} {Entanglement-enhanced optical ion clock},\ }\href@noop {} {\bibfield
   {journal} {\bibinfo  {journal} {Physical Review Letters}\ }\textbf {\bibinfo
  {volume} {136}},\ \bibinfo {pages} {073601} (\bibinfo {year}
  {2026})}\BibitemShut {NoStop}%
\bibitem [{\citenamefont {Bollinger}\ \emph {et~al.}(1996)\citenamefont
  {Bollinger}, \citenamefont {Itano}, \citenamefont {Wineland},\ and\
  \citenamefont {Heinzen}}]{bollinger1996optimal}%
  \BibitemOpen
  \bibfield  {author} {\bibinfo {author} {\bibfnamefont {J.~J.}\ \bibnamefont
  {Bollinger}}, \bibinfo {author} {\bibfnamefont {W.~M.}\ \bibnamefont
  {Itano}}, \bibinfo {author} {\bibfnamefont {D.~J.}\ \bibnamefont
  {Wineland}},\ and\ \bibinfo {author} {\bibfnamefont {D.~J.}\ \bibnamefont
  {Heinzen}},\ }\bibfield  {title} {\bibinfo {title} {Optimal frequency
  measurements with maximally correlated states},\ }\href@noop {} {\bibfield
  {journal} {\bibinfo  {journal} {Physical Review A}\ }\textbf {\bibinfo
  {volume} {54}},\ \bibinfo {pages} {R4649} (\bibinfo {year}
  {1996})}\BibitemShut {NoStop}%
\bibitem [{\citenamefont {Periwal}\ \emph {et~al.}(2021)\citenamefont
  {Periwal}, \citenamefont {Cooper}, \citenamefont {Kunkel}, \citenamefont
  {Wienand}, \citenamefont {Davis},\ and\ \citenamefont
  {Schleier-Smith}}]{periwal2021programmable}%
  \BibitemOpen
  \bibfield  {author} {\bibinfo {author} {\bibfnamefont {A.}~\bibnamefont
  {Periwal}}, \bibinfo {author} {\bibfnamefont {E.~S.}\ \bibnamefont {Cooper}},
  \bibinfo {author} {\bibfnamefont {P.}~\bibnamefont {Kunkel}}, \bibinfo
  {author} {\bibfnamefont {J.~F.}\ \bibnamefont {Wienand}}, \bibinfo {author}
  {\bibfnamefont {E.~J.}\ \bibnamefont {Davis}},\ and\ \bibinfo {author}
  {\bibfnamefont {M.}~\bibnamefont {Schleier-Smith}},\ }\bibfield  {title}
  {\bibinfo {title} {Programmable interactions and emergent geometry in an
  array of atom clouds},\ }\href@noop {} {\bibfield  {journal} {\bibinfo
  {journal} {Nature}\ }\textbf {\bibinfo {volume} {600}},\ \bibinfo {pages}
  {630} (\bibinfo {year} {2021})}\BibitemShut {NoStop}%
\bibitem [{\citenamefont {Corgier}\ \emph {et~al.}(2023)\citenamefont
  {Corgier}, \citenamefont {Malitesta}, \citenamefont {Smerzi},\ and\
  \citenamefont {Pezz{\`e}}}]{corgier2023quantum}%
  \BibitemOpen
  \bibfield  {author} {\bibinfo {author} {\bibfnamefont {R.}~\bibnamefont
  {Corgier}}, \bibinfo {author} {\bibfnamefont {M.}~\bibnamefont {Malitesta}},
  \bibinfo {author} {\bibfnamefont {A.}~\bibnamefont {Smerzi}},\ and\ \bibinfo
  {author} {\bibfnamefont {L.}~\bibnamefont {Pezz{\`e}}},\ }\bibfield  {title}
  {\bibinfo {title} {Quantum-enhanced differential atom interferometers and
  clocks with spin-squeezing swapping},\ }\href@noop {} {\bibfield  {journal}
  {\bibinfo  {journal} {Quantum}\ }\textbf {\bibinfo {volume} {7}},\ \bibinfo
  {pages} {965} (\bibinfo {year} {2023})}\BibitemShut {NoStop}%
\bibitem [{\citenamefont {Braunstein}\ and\ \citenamefont
  {Caves}(1994)}]{braunstein1994statistical}%
  \BibitemOpen
  \bibfield  {author} {\bibinfo {author} {\bibfnamefont {S.~L.}\ \bibnamefont
  {Braunstein}}\ and\ \bibinfo {author} {\bibfnamefont {C.~M.}\ \bibnamefont
  {Caves}},\ }\bibfield  {title} {\bibinfo {title} {Statistical distance and
  the geometry of quantum states},\ }\href@noop {} {\bibfield  {journal}
  {\bibinfo  {journal} {Physical Review Letters}\ }\textbf {\bibinfo {volume}
  {72}},\ \bibinfo {pages} {3439} (\bibinfo {year} {1994})}\BibitemShut
  {NoStop}%
\bibitem [{\citenamefont {Paris}(2009)}]{paris2009quantum}%
  \BibitemOpen
  \bibfield  {author} {\bibinfo {author} {\bibfnamefont {M.~G.}\ \bibnamefont
  {Paris}},\ }\bibfield  {title} {\bibinfo {title} {Quantum estimation for
  quantum technology},\ }\href@noop {} {\bibfield  {journal} {\bibinfo
  {journal} {International Journal of Quantum Information}\ }\textbf {\bibinfo
  {volume} {7}},\ \bibinfo {pages} {125} (\bibinfo {year} {2009})}\BibitemShut
  {NoStop}%
\bibitem [{\citenamefont {Demkowicz-Dobrza{\'n}ski}\ \emph
  {et~al.}(2015)\citenamefont {Demkowicz-Dobrza{\'n}ski}, \citenamefont
  {Jarzyna},\ and\ \citenamefont {Ko{\l}ody{\'n}ski}}]{demkowicz2015quantum}%
  \BibitemOpen
  \bibfield  {author} {\bibinfo {author} {\bibfnamefont {R.}~\bibnamefont
  {Demkowicz-Dobrza{\'n}ski}}, \bibinfo {author} {\bibfnamefont
  {M.}~\bibnamefont {Jarzyna}},\ and\ \bibinfo {author} {\bibfnamefont
  {J.}~\bibnamefont {Ko{\l}ody{\'n}ski}},\ }\bibfield  {title} {\bibinfo
  {title} {Quantum limits in optical interferometry},\ }\href@noop {}
  {\bibfield  {journal} {\bibinfo  {journal} {Progress in Optics}\ }\textbf
  {\bibinfo {volume} {60}},\ \bibinfo {pages} {345} (\bibinfo {year}
  {2015})}\BibitemShut {NoStop}%
\bibitem [{\citenamefont {Omran}\ \emph {et~al.}(2019)\citenamefont {Omran},
  \citenamefont {Levine}, \citenamefont {Keesling}, \citenamefont {Semeghini},
  \citenamefont {Wang}, \citenamefont {Ebadi}, \citenamefont {Bernien},
  \citenamefont {Zibrov}, \citenamefont {Pichler}, \citenamefont {Choi} \emph
  {et~al.}}]{omran2019generation}%
  \BibitemOpen
  \bibfield  {author} {\bibinfo {author} {\bibfnamefont {A.}~\bibnamefont
  {Omran}}, \bibinfo {author} {\bibfnamefont {H.}~\bibnamefont {Levine}},
  \bibinfo {author} {\bibfnamefont {A.}~\bibnamefont {Keesling}}, \bibinfo
  {author} {\bibfnamefont {G.}~\bibnamefont {Semeghini}}, \bibinfo {author}
  {\bibfnamefont {T.~T.}\ \bibnamefont {Wang}}, \bibinfo {author}
  {\bibfnamefont {S.}~\bibnamefont {Ebadi}}, \bibinfo {author} {\bibfnamefont
  {H.}~\bibnamefont {Bernien}}, \bibinfo {author} {\bibfnamefont {A.~S.}\
  \bibnamefont {Zibrov}}, \bibinfo {author} {\bibfnamefont {H.}~\bibnamefont
  {Pichler}}, \bibinfo {author} {\bibfnamefont {S.}~\bibnamefont {Choi}}, \emph
  {et~al.},\ }\bibfield  {title} {\bibinfo {title} {Generation and manipulation
  of schr{\"o}dinger cat states in rydberg atom arrays},\ }\href@noop {}
  {\bibfield  {journal} {\bibinfo  {journal} {Science}\ }\textbf {\bibinfo
  {volume} {365}},\ \bibinfo {pages} {570} (\bibinfo {year}
  {2019})}\BibitemShut {NoStop}%
\bibitem [{\citenamefont {Pogorelov}\ \emph {et~al.}(2021)\citenamefont
  {Pogorelov}, \citenamefont {Feldker}, \citenamefont {Marciniak},
  \citenamefont {Postler}, \citenamefont {Jacob}, \citenamefont
  {Krieglsteiner}, \citenamefont {Podlesnic}, \citenamefont {Meth},
  \citenamefont {Negnevitsky}, \citenamefont {Stadler} \emph
  {et~al.}}]{pogorelov2021compact}%
  \BibitemOpen
  \bibfield  {author} {\bibinfo {author} {\bibfnamefont {I.}~\bibnamefont
  {Pogorelov}}, \bibinfo {author} {\bibfnamefont {T.}~\bibnamefont {Feldker}},
  \bibinfo {author} {\bibfnamefont {C.~D.}\ \bibnamefont {Marciniak}}, \bibinfo
  {author} {\bibfnamefont {L.}~\bibnamefont {Postler}}, \bibinfo {author}
  {\bibfnamefont {G.}~\bibnamefont {Jacob}}, \bibinfo {author} {\bibfnamefont
  {O.}~\bibnamefont {Krieglsteiner}}, \bibinfo {author} {\bibfnamefont
  {V.}~\bibnamefont {Podlesnic}}, \bibinfo {author} {\bibfnamefont
  {M.}~\bibnamefont {Meth}}, \bibinfo {author} {\bibfnamefont {V.}~\bibnamefont
  {Negnevitsky}}, \bibinfo {author} {\bibfnamefont {M.}~\bibnamefont
  {Stadler}}, \emph {et~al.},\ }\bibfield  {title} {\bibinfo {title} {Compact
  ion-trap quantum computing demonstrator},\ }\href@noop {} {\bibfield
  {journal} {\bibinfo  {journal} {PRX quantum}\ }\textbf {\bibinfo {volume}
  {2}},\ \bibinfo {pages} {020343} (\bibinfo {year} {2021})}\BibitemShut
  {NoStop}%
\bibitem [{\citenamefont {Mooney}\ \emph {et~al.}(2021)\citenamefont {Mooney},
  \citenamefont {White}, \citenamefont {Hill},\ and\ \citenamefont
  {Hollenberg}}]{mooney2021generation}%
  \BibitemOpen
  \bibfield  {author} {\bibinfo {author} {\bibfnamefont {G.~J.}\ \bibnamefont
  {Mooney}}, \bibinfo {author} {\bibfnamefont {G.~A.}\ \bibnamefont {White}},
  \bibinfo {author} {\bibfnamefont {C.~D.}\ \bibnamefont {Hill}},\ and\
  \bibinfo {author} {\bibfnamefont {L.~C.}\ \bibnamefont {Hollenberg}},\
  }\bibfield  {title} {\bibinfo {title} {Generation and verification of
  27-qubit greenberger-horne-zeilinger states in a superconducting quantum
  computer},\ }\href@noop {} {\bibfield  {journal} {\bibinfo  {journal}
  {Journal of Physics Communications}\ }\textbf {\bibinfo {volume} {5}},\
  \bibinfo {pages} {095004} (\bibinfo {year} {2021})}\BibitemShut {NoStop}%
\bibitem [{\citenamefont {Moses}\ \emph {et~al.}(2023)\citenamefont {Moses},
  \citenamefont {Baldwin}, \citenamefont {Allman}, \citenamefont {Ancona},
  \citenamefont {Ascarrunz}, \citenamefont {Barnes}, \citenamefont
  {Bartolotta}, \citenamefont {Bjork}, \citenamefont {Blanchard}, \citenamefont
  {Bohn} \emph {et~al.}}]{moses2023race}%
  \BibitemOpen
  \bibfield  {author} {\bibinfo {author} {\bibfnamefont {S.~A.}\ \bibnamefont
  {Moses}}, \bibinfo {author} {\bibfnamefont {C.~H.}\ \bibnamefont {Baldwin}},
  \bibinfo {author} {\bibfnamefont {M.~S.}\ \bibnamefont {Allman}}, \bibinfo
  {author} {\bibfnamefont {R.}~\bibnamefont {Ancona}}, \bibinfo {author}
  {\bibfnamefont {L.}~\bibnamefont {Ascarrunz}}, \bibinfo {author}
  {\bibfnamefont {C.}~\bibnamefont {Barnes}}, \bibinfo {author} {\bibfnamefont
  {J.}~\bibnamefont {Bartolotta}}, \bibinfo {author} {\bibfnamefont
  {B.}~\bibnamefont {Bjork}}, \bibinfo {author} {\bibfnamefont
  {P.}~\bibnamefont {Blanchard}}, \bibinfo {author} {\bibfnamefont
  {M.}~\bibnamefont {Bohn}}, \emph {et~al.},\ }\bibfield  {title} {\bibinfo
  {title} {A race-track trapped-ion quantum processor},\ }\href@noop {}
  {\bibfield  {journal} {\bibinfo  {journal} {Physical Review X}\ }\textbf
  {\bibinfo {volume} {13}},\ \bibinfo {pages} {041052} (\bibinfo {year}
  {2023})}\BibitemShut {NoStop}%
\bibitem [{\citenamefont {Kaubruegger}\ \emph {et~al.}(2021)\citenamefont
  {Kaubruegger}, \citenamefont {Vasilyev}, \citenamefont {Schulte},
  \citenamefont {Hammerer},\ and\ \citenamefont
  {Zoller}}]{kaubruegger2021quantum}%
  \BibitemOpen
  \bibfield  {author} {\bibinfo {author} {\bibfnamefont {R.}~\bibnamefont
  {Kaubruegger}}, \bibinfo {author} {\bibfnamefont {D.~V.}\ \bibnamefont
  {Vasilyev}}, \bibinfo {author} {\bibfnamefont {M.}~\bibnamefont {Schulte}},
  \bibinfo {author} {\bibfnamefont {K.}~\bibnamefont {Hammerer}},\ and\
  \bibinfo {author} {\bibfnamefont {P.}~\bibnamefont {Zoller}},\ }\bibfield
  {title} {\bibinfo {title} {Quantum variational optimization of ramsey
  interferometry and atomic clocks},\ }\href@noop {} {\bibfield  {journal}
  {\bibinfo  {journal} {Physical review X}\ }\textbf {\bibinfo {volume} {11}},\
  \bibinfo {pages} {041045} (\bibinfo {year} {2021})}\BibitemShut {NoStop}%
\bibitem [{\citenamefont {Cable}\ and\ \citenamefont
  {Durkin}(2010)}]{cable2010parameter}%
  \BibitemOpen
  \bibfield  {author} {\bibinfo {author} {\bibfnamefont {H.}~\bibnamefont
  {Cable}}\ and\ \bibinfo {author} {\bibfnamefont {G.~A.}\ \bibnamefont
  {Durkin}},\ }\bibfield  {title} {\bibinfo {title} {Parameter estimation with
  entangled photons produced by parametric down-conversion},\ }\href@noop {}
  {\bibfield  {journal} {\bibinfo  {journal} {Physical review letters}\
  }\textbf {\bibinfo {volume} {105}},\ \bibinfo {pages} {013603} (\bibinfo
  {year} {2010})}\BibitemShut {NoStop}%
\bibitem [{\citenamefont {Sundar}\ \emph {et~al.}(2023)\citenamefont {Sundar},
  \citenamefont {Barberena}, \citenamefont {Orioli}, \citenamefont {Chu},
  \citenamefont {Thompson}, \citenamefont {Rey},\ and\ \citenamefont
  {Lewis-Swan}}]{sundar2023bosonic}%
  \BibitemOpen
  \bibfield  {author} {\bibinfo {author} {\bibfnamefont {B.}~\bibnamefont
  {Sundar}}, \bibinfo {author} {\bibfnamefont {D.}~\bibnamefont {Barberena}},
  \bibinfo {author} {\bibfnamefont {A.~P.}\ \bibnamefont {Orioli}}, \bibinfo
  {author} {\bibfnamefont {A.}~\bibnamefont {Chu}}, \bibinfo {author}
  {\bibfnamefont {J.~K.}\ \bibnamefont {Thompson}}, \bibinfo {author}
  {\bibfnamefont {A.~M.}\ \bibnamefont {Rey}},\ and\ \bibinfo {author}
  {\bibfnamefont {R.~J.}\ \bibnamefont {Lewis-Swan}},\ }\bibfield  {title}
  {\bibinfo {title} {Bosonic pair production and squeezing for optical phase
  measurements in long-lived dipoles coupled to a cavity},\ }\href@noop {}
  {\bibfield  {journal} {\bibinfo  {journal} {Physical Review Letters}\
  }\textbf {\bibinfo {volume} {130}},\ \bibinfo {pages} {113202} (\bibinfo
  {year} {2023})}\BibitemShut {NoStop}%
\bibitem [{\citenamefont {Mamaev}\ \emph {et~al.}(2025)\citenamefont {Mamaev},
  \citenamefont {Koppenh{\"o}fer}, \citenamefont {Pocklington},\ and\
  \citenamefont {Clerk}}]{mamaev2025non}%
  \BibitemOpen
  \bibfield  {author} {\bibinfo {author} {\bibfnamefont {M.}~\bibnamefont
  {Mamaev}}, \bibinfo {author} {\bibfnamefont {M.}~\bibnamefont
  {Koppenh{\"o}fer}}, \bibinfo {author} {\bibfnamefont {A.}~\bibnamefont
  {Pocklington}},\ and\ \bibinfo {author} {\bibfnamefont {A.~A.}\ \bibnamefont
  {Clerk}},\ }\bibfield  {title} {\bibinfo {title} {Non-gaussian generalized
  two-mode squeezing: Applications to two-ensemble spin squeezing and beyond},\
  }\href@noop {} {\bibfield  {journal} {\bibinfo  {journal} {Physical review
  letters}\ }\textbf {\bibinfo {volume} {134}},\ \bibinfo {pages} {073603}
  (\bibinfo {year} {2025})}\BibitemShut {NoStop}%
\bibitem [{\citenamefont {Urizar-Lanz}\ \emph {et~al.}(2013)\citenamefont
  {Urizar-Lanz}, \citenamefont {Hyllus}, \citenamefont {Egusquiza},
  \citenamefont {Mitchell},\ and\ \citenamefont
  {T{\'o}th}}]{urizar2013macroscopic}%
  \BibitemOpen
  \bibfield  {author} {\bibinfo {author} {\bibfnamefont {I.}~\bibnamefont
  {Urizar-Lanz}}, \bibinfo {author} {\bibfnamefont {P.}~\bibnamefont {Hyllus}},
  \bibinfo {author} {\bibfnamefont {I.~L.}\ \bibnamefont {Egusquiza}}, \bibinfo
  {author} {\bibfnamefont {M.~W.}\ \bibnamefont {Mitchell}},\ and\ \bibinfo
  {author} {\bibfnamefont {G.}~\bibnamefont {T{\'o}th}},\ }\bibfield  {title}
  {\bibinfo {title} {Macroscopic singlet states for gradient magnetometry},\
  }\href@noop {} {\bibfield  {journal} {\bibinfo  {journal} {Physical Review
  A—Atomic, Molecular, and Optical Physics}\ }\textbf {\bibinfo {volume}
  {88}},\ \bibinfo {pages} {013626} (\bibinfo {year} {2013})}\BibitemShut
  {NoStop}%
\bibitem [{\citenamefont {Muniz}\ \emph {et~al.}(2020)\citenamefont {Muniz},
  \citenamefont {Barberena}, \citenamefont {Lewis-Swan}, \citenamefont {Young},
  \citenamefont {Cline}, \citenamefont {Rey},\ and\ \citenamefont
  {Thompson}}]{muniz2020exploring}%
  \BibitemOpen
  \bibfield  {author} {\bibinfo {author} {\bibfnamefont {J.~A.}\ \bibnamefont
  {Muniz}}, \bibinfo {author} {\bibfnamefont {D.}~\bibnamefont {Barberena}},
  \bibinfo {author} {\bibfnamefont {R.~J.}\ \bibnamefont {Lewis-Swan}},
  \bibinfo {author} {\bibfnamefont {D.~J.}\ \bibnamefont {Young}}, \bibinfo
  {author} {\bibfnamefont {J.~R.}\ \bibnamefont {Cline}}, \bibinfo {author}
  {\bibfnamefont {A.~M.}\ \bibnamefont {Rey}},\ and\ \bibinfo {author}
  {\bibfnamefont {J.~K.}\ \bibnamefont {Thompson}},\ }\bibfield  {title}
  {\bibinfo {title} {Exploring dynamical phase transitions with cold atoms in
  an optical cavity},\ }\href@noop {} {\bibfield  {journal} {\bibinfo
  {journal} {Nature}\ }\textbf {\bibinfo {volume} {580}},\ \bibinfo {pages}
  {602} (\bibinfo {year} {2020})}\BibitemShut {NoStop}%
\bibitem [{\citenamefont {Kurucz}\ and\ \citenamefont
  {M{\o}lmer}(2010)}]{kurucz2010multilevel}%
  \BibitemOpen
  \bibfield  {author} {\bibinfo {author} {\bibfnamefont {Z.}~\bibnamefont
  {Kurucz}}\ and\ \bibinfo {author} {\bibfnamefont {K.}~\bibnamefont
  {M{\o}lmer}},\ }\bibfield  {title} {\bibinfo {title} {Multilevel
  holstein-primakoff approximation and its application to atomic spin squeezing
  and ensemble quantum memories},\ }\href@noop {} {\bibfield  {journal}
  {\bibinfo  {journal} {Physical Review A—Atomic, Molecular, and Optical
  Physics}\ }\textbf {\bibinfo {volume} {81}},\ \bibinfo {pages} {032314}
  (\bibinfo {year} {2010})}\BibitemShut {NoStop}%
\bibitem [{\citenamefont {Khan}\ \emph {et~al.}(2025)\citenamefont {Khan},
  \citenamefont {Chaparro}, \citenamefont {Sundar}, \citenamefont {Carter},
  \citenamefont {Bollinger}, \citenamefont {Molmer},\ and\ \citenamefont
  {Rey}}]{khan2025generating}%
  \BibitemOpen
  \bibfield  {author} {\bibinfo {author} {\bibfnamefont {M.~M.}\ \bibnamefont
  {Khan}}, \bibinfo {author} {\bibfnamefont {E.}~\bibnamefont {Chaparro}},
  \bibinfo {author} {\bibfnamefont {B.}~\bibnamefont {Sundar}}, \bibinfo
  {author} {\bibfnamefont {A.}~\bibnamefont {Carter}}, \bibinfo {author}
  {\bibfnamefont {J.}~\bibnamefont {Bollinger}}, \bibinfo {author}
  {\bibfnamefont {K.}~\bibnamefont {Molmer}},\ and\ \bibinfo {author}
  {\bibfnamefont {A.~M.}\ \bibnamefont {Rey}},\ }\bibfield  {title} {\bibinfo
  {title} {Generating einstein-podolsky-rosen correlations for teleporting
  collective spin states in a two-dimensional trapped ion crystal},\
  }\href@noop {} {\bibfield  {journal} {\bibinfo  {journal} {Physical Review
  Research}\ }\textbf {\bibinfo {volume} {7}},\ \bibinfo {pages} {L022019}
  (\bibinfo {year} {2025})}\BibitemShut {NoStop}%
\bibitem [{\citenamefont {Chase}\ and\ \citenamefont
  {Geremia}(2008)}]{chase2008collective}%
  \BibitemOpen
  \bibfield  {author} {\bibinfo {author} {\bibfnamefont {B.~A.}\ \bibnamefont
  {Chase}}\ and\ \bibinfo {author} {\bibfnamefont {J.}~\bibnamefont
  {Geremia}},\ }\bibfield  {title} {\bibinfo {title} {Collective processes of
  an ensemble of spin-1/ 2 particles},\ }\href@noop {} {\bibfield  {journal}
  {\bibinfo  {journal} {Physical Review A—Atomic, Molecular, and Optical
  Physics}\ }\textbf {\bibinfo {volume} {78}},\ \bibinfo {pages} {052101}
  (\bibinfo {year} {2008})}\BibitemShut {NoStop}%
\bibitem [{\citenamefont {Gong}\ \emph {et~al.}(2016)\citenamefont {Gong},
  \citenamefont {Xu}, \citenamefont {Foss-Feig}, \citenamefont {Thompson},
  \citenamefont {Rey}, \citenamefont {Holland},\ and\ \citenamefont
  {Gorshkov}}]{gong2016steady}%
  \BibitemOpen
  \bibfield  {author} {\bibinfo {author} {\bibfnamefont {Z.-X.}\ \bibnamefont
  {Gong}}, \bibinfo {author} {\bibfnamefont {M.}~\bibnamefont {Xu}}, \bibinfo
  {author} {\bibfnamefont {M.}~\bibnamefont {Foss-Feig}}, \bibinfo {author}
  {\bibfnamefont {J.~K.}\ \bibnamefont {Thompson}}, \bibinfo {author}
  {\bibfnamefont {A.~M.}\ \bibnamefont {Rey}}, \bibinfo {author} {\bibfnamefont
  {M.}~\bibnamefont {Holland}},\ and\ \bibinfo {author} {\bibfnamefont {A.~V.}\
  \bibnamefont {Gorshkov}},\ }\bibfield  {title} {\bibinfo {title}
  {Steady-state superradiance with rydberg polaritons},\ }\href@noop {}
  {\bibfield  {journal} {\bibinfo  {journal} {arXiv preprint arXiv:1611.00797}\
  } (\bibinfo {year} {2016})}\BibitemShut {NoStop}%
\bibitem [{\citenamefont {Shammah}\ \emph {et~al.}(2018)\citenamefont
  {Shammah}, \citenamefont {Ahmed}, \citenamefont {Lambert}, \citenamefont
  {De~Liberato},\ and\ \citenamefont {Nori}}]{shammah2018open}%
  \BibitemOpen
  \bibfield  {author} {\bibinfo {author} {\bibfnamefont {N.}~\bibnamefont
  {Shammah}}, \bibinfo {author} {\bibfnamefont {S.}~\bibnamefont {Ahmed}},
  \bibinfo {author} {\bibfnamefont {N.}~\bibnamefont {Lambert}}, \bibinfo
  {author} {\bibfnamefont {S.}~\bibnamefont {De~Liberato}},\ and\ \bibinfo
  {author} {\bibfnamefont {F.}~\bibnamefont {Nori}},\ }\bibfield  {title}
  {\bibinfo {title} {Open quantum systems with local and collective incoherent
  processes: Efficient numerical simulations using permutational invariance},\
  }\href@noop {} {\bibfield  {journal} {\bibinfo  {journal} {Physical Review
  A}\ }\textbf {\bibinfo {volume} {98}},\ \bibinfo {pages} {063815} (\bibinfo
  {year} {2018})}\BibitemShut {NoStop}%
\bibitem [{\citenamefont {Nadolny}\ \emph {et~al.}(2025)\citenamefont
  {Nadolny}, \citenamefont {Bruder},\ and\ \citenamefont
  {Brunelli}}]{nadolny2025nonreciprocal}%
  \BibitemOpen
  \bibfield  {author} {\bibinfo {author} {\bibfnamefont {T.}~\bibnamefont
  {Nadolny}}, \bibinfo {author} {\bibfnamefont {C.}~\bibnamefont {Bruder}},\
  and\ \bibinfo {author} {\bibfnamefont {M.}~\bibnamefont {Brunelli}},\
  }\bibfield  {title} {\bibinfo {title} {Nonreciprocal synchronization of
  active quantum spins},\ }\href@noop {} {\bibfield  {journal} {\bibinfo
  {journal} {Physical Review X}\ }\textbf {\bibinfo {volume} {15}},\ \bibinfo
  {pages} {011010} (\bibinfo {year} {2025})}\BibitemShut {NoStop}%
\bibitem [{\citenamefont {Zhang}\ \emph {et~al.}(2018)\citenamefont {Zhang},
  \citenamefont {Zhang},\ and\ \citenamefont {M{\o}lmer}}]{zhang2018monte}%
  \BibitemOpen
  \bibfield  {author} {\bibinfo {author} {\bibfnamefont {Y.}~\bibnamefont
  {Zhang}}, \bibinfo {author} {\bibfnamefont {Y.-X.}\ \bibnamefont {Zhang}},\
  and\ \bibinfo {author} {\bibfnamefont {K.}~\bibnamefont {M{\o}lmer}},\
  }\bibfield  {title} {\bibinfo {title} {Monte-carlo simulations of
  superradiant lasing},\ }\href@noop {} {\bibfield  {journal} {\bibinfo
  {journal} {New Journal of Physics}\ }\textbf {\bibinfo {volume} {20}},\
  \bibinfo {pages} {112001} (\bibinfo {year} {2018})}\BibitemShut {NoStop}%
\bibitem [{\citenamefont {Chu}\ \emph {et~al.}(2021)\citenamefont {Chu},
  \citenamefont {He}, \citenamefont {Thompson},\ and\ \citenamefont
  {Rey}}]{chu2021quantum}%
  \BibitemOpen
  \bibfield  {author} {\bibinfo {author} {\bibfnamefont {A.}~\bibnamefont
  {Chu}}, \bibinfo {author} {\bibfnamefont {P.}~\bibnamefont {He}}, \bibinfo
  {author} {\bibfnamefont {J.~K.}\ \bibnamefont {Thompson}},\ and\ \bibinfo
  {author} {\bibfnamefont {A.~M.}\ \bibnamefont {Rey}},\ }\bibfield  {title}
  {\bibinfo {title} {Quantum enhanced cavity qed interferometer with partially
  delocalized atoms in lattices},\ }\href@noop {} {\bibfield  {journal}
  {\bibinfo  {journal} {Physical Review Letters}\ }\textbf {\bibinfo {volume}
  {127}},\ \bibinfo {pages} {210401} (\bibinfo {year} {2021})}\BibitemShut
  {NoStop}%
\bibitem [{\citenamefont {Koppenh{\"o}fer}\ and\ \citenamefont
  {Clerk}(2023)}]{koppenhofer2023revisiting}%
  \BibitemOpen
  \bibfield  {author} {\bibinfo {author} {\bibfnamefont {M.}~\bibnamefont
  {Koppenh{\"o}fer}}\ and\ \bibinfo {author} {\bibfnamefont {A.}~\bibnamefont
  {Clerk}},\ }\bibfield  {title} {\bibinfo {title} {Revisiting the impact of
  dissipation on time-reversed one-axis-twist quantum-sensing protocols},\
  }\href@noop {} {\bibfield  {journal} {\bibinfo  {journal} {Physical Review
  Research}\ }\textbf {\bibinfo {volume} {5}},\ \bibinfo {pages} {043279}
  (\bibinfo {year} {2023})}\BibitemShut {NoStop}%
\bibitem [{\citenamefont {Masson}\ and\ \citenamefont
  {Parkins}(2019)}]{masson2019rapid}%
  \BibitemOpen
  \bibfield  {author} {\bibinfo {author} {\bibfnamefont {S.~J.}\ \bibnamefont
  {Masson}}\ and\ \bibinfo {author} {\bibfnamefont {S.}~\bibnamefont
  {Parkins}},\ }\bibfield  {title} {\bibinfo {title} {Rapid production of
  many-body entanglement in spin-1 atoms via cavity output photon counting},\
  }\href@noop {} {\bibfield  {journal} {\bibinfo  {journal} {Physical Review
  Letters}\ }\textbf {\bibinfo {volume} {122}},\ \bibinfo {pages} {103601}
  (\bibinfo {year} {2019})}\BibitemShut {NoStop}%
\bibitem [{\citenamefont {Norcia}\ \emph {et~al.}(2018)\citenamefont {Norcia},
  \citenamefont {Lewis-Swan}, \citenamefont {Cline}, \citenamefont {Zhu},
  \citenamefont {Rey},\ and\ \citenamefont {Thompson}}]{norcia2018cavity}%
  \BibitemOpen
  \bibfield  {author} {\bibinfo {author} {\bibfnamefont {M.~A.}\ \bibnamefont
  {Norcia}}, \bibinfo {author} {\bibfnamefont {R.~J.}\ \bibnamefont
  {Lewis-Swan}}, \bibinfo {author} {\bibfnamefont {J.~R.}\ \bibnamefont
  {Cline}}, \bibinfo {author} {\bibfnamefont {B.}~\bibnamefont {Zhu}}, \bibinfo
  {author} {\bibfnamefont {A.~M.}\ \bibnamefont {Rey}},\ and\ \bibinfo {author}
  {\bibfnamefont {J.~K.}\ \bibnamefont {Thompson}},\ }\bibfield  {title}
  {\bibinfo {title} {Cavity-mediated collective spin-exchange interactions in a
  strontium superradiant laser},\ }\href@noop {} {\bibfield  {journal}
  {\bibinfo  {journal} {Science}\ }\textbf {\bibinfo {volume} {361}},\ \bibinfo
  {pages} {259} (\bibinfo {year} {2018})}\BibitemShut {NoStop}%
\bibitem [{\citenamefont {Cox}\ \emph {et~al.}(2016)\citenamefont {Cox},
  \citenamefont {Greve}, \citenamefont {Weiner},\ and\ \citenamefont
  {Thompson}}]{cox2016deterministic}%
  \BibitemOpen
  \bibfield  {author} {\bibinfo {author} {\bibfnamefont {K.~C.}\ \bibnamefont
  {Cox}}, \bibinfo {author} {\bibfnamefont {G.~P.}\ \bibnamefont {Greve}},
  \bibinfo {author} {\bibfnamefont {J.~M.}\ \bibnamefont {Weiner}},\ and\
  \bibinfo {author} {\bibfnamefont {J.~K.}\ \bibnamefont {Thompson}},\
  }\bibfield  {title} {\bibinfo {title} {Deterministic squeezed states with
  collective measurements and feedback},\ }\href@noop {} {\bibfield  {journal}
  {\bibinfo  {journal} {Physical review letters}\ }\textbf {\bibinfo {volume}
  {116}},\ \bibinfo {pages} {093602} (\bibinfo {year} {2016})}\BibitemShut
  {NoStop}%
\bibitem [{\citenamefont {Hosten}\ \emph {et~al.}(2016)\citenamefont {Hosten},
  \citenamefont {Engelsen}, \citenamefont {Krishnakumar},\ and\ \citenamefont
  {Kasevich}}]{hosten2016measurement}%
  \BibitemOpen
  \bibfield  {author} {\bibinfo {author} {\bibfnamefont {O.}~\bibnamefont
  {Hosten}}, \bibinfo {author} {\bibfnamefont {N.~J.}\ \bibnamefont
  {Engelsen}}, \bibinfo {author} {\bibfnamefont {R.}~\bibnamefont
  {Krishnakumar}},\ and\ \bibinfo {author} {\bibfnamefont {M.~A.}\ \bibnamefont
  {Kasevich}},\ }\bibfield  {title} {\bibinfo {title} {Measurement noise 100
  times lower than the quantum-projection limit using entangled atoms},\
  }\href@noop {} {\bibfield  {journal} {\bibinfo  {journal} {Nature}\ }\textbf
  {\bibinfo {volume} {529}},\ \bibinfo {pages} {505} (\bibinfo {year}
  {2016})}\BibitemShut {NoStop}%
\bibitem [{\citenamefont {Milner}\ \emph {et~al.}(2025)\citenamefont {Milner},
  \citenamefont {Lannig}, \citenamefont {Mamaev}, \citenamefont {Yan},
  \citenamefont {Chu}, \citenamefont {Lewis}, \citenamefont {Frankel},
  \citenamefont {Hutson}, \citenamefont {Rey},\ and\ \citenamefont
  {Ye}}]{milner2025coherent}%
  \BibitemOpen
  \bibfield  {author} {\bibinfo {author} {\bibfnamefont {W.~R.}\ \bibnamefont
  {Milner}}, \bibinfo {author} {\bibfnamefont {S.}~\bibnamefont {Lannig}},
  \bibinfo {author} {\bibfnamefont {M.}~\bibnamefont {Mamaev}}, \bibinfo
  {author} {\bibfnamefont {L.}~\bibnamefont {Yan}}, \bibinfo {author}
  {\bibfnamefont {A.}~\bibnamefont {Chu}}, \bibinfo {author} {\bibfnamefont
  {B.}~\bibnamefont {Lewis}}, \bibinfo {author} {\bibfnamefont {M.~N.}\
  \bibnamefont {Frankel}}, \bibinfo {author} {\bibfnamefont {R.~B.}\
  \bibnamefont {Hutson}}, \bibinfo {author} {\bibfnamefont {A.~M.}\
  \bibnamefont {Rey}},\ and\ \bibinfo {author} {\bibfnamefont {J.}~\bibnamefont
  {Ye}},\ }\bibfield  {title} {\bibinfo {title} {Coherent evolution of
  superexchange interaction in seconds-long optical clock spectroscopy},\
  }\href@noop {} {\bibfield  {journal} {\bibinfo  {journal} {Science}\ }\textbf
  {\bibinfo {volume} {388}},\ \bibinfo {pages} {503} (\bibinfo {year}
  {2025})}\BibitemShut {NoStop}%
\bibitem [{\citenamefont {Franke}\ \emph {et~al.}(2023)\citenamefont {Franke},
  \citenamefont {Muleady}, \citenamefont {Kaubruegger}, \citenamefont {Kranzl},
  \citenamefont {Blatt}, \citenamefont {Rey}, \citenamefont {Joshi},\ and\
  \citenamefont {Roos}}]{franke2023quantum}%
  \BibitemOpen
  \bibfield  {author} {\bibinfo {author} {\bibfnamefont {J.}~\bibnamefont
  {Franke}}, \bibinfo {author} {\bibfnamefont {S.~R.}\ \bibnamefont {Muleady}},
  \bibinfo {author} {\bibfnamefont {R.}~\bibnamefont {Kaubruegger}}, \bibinfo
  {author} {\bibfnamefont {F.}~\bibnamefont {Kranzl}}, \bibinfo {author}
  {\bibfnamefont {R.}~\bibnamefont {Blatt}}, \bibinfo {author} {\bibfnamefont
  {A.~M.}\ \bibnamefont {Rey}}, \bibinfo {author} {\bibfnamefont {M.~K.}\
  \bibnamefont {Joshi}},\ and\ \bibinfo {author} {\bibfnamefont {C.~F.}\
  \bibnamefont {Roos}},\ }\bibfield  {title} {\bibinfo {title}
  {Quantum-enhanced sensing on optical transitions through finite-range
  interactions},\ }\href@noop {} {\bibfield  {journal} {\bibinfo  {journal}
  {Nature}\ }\textbf {\bibinfo {volume} {621}},\ \bibinfo {pages} {740}
  (\bibinfo {year} {2023})}\BibitemShut {NoStop}%
\bibitem [{\citenamefont {Bornet}\ \emph {et~al.}(2023)\citenamefont {Bornet},
  \citenamefont {Emperauger}, \citenamefont {Chen}, \citenamefont {Ye},
  \citenamefont {Block}, \citenamefont {Bintz}, \citenamefont {Boyd},
  \citenamefont {Barredo}, \citenamefont {Comparin}, \citenamefont {Mezzacapo}
  \emph {et~al.}}]{bornet2023scalable}%
  \BibitemOpen
  \bibfield  {author} {\bibinfo {author} {\bibfnamefont {G.}~\bibnamefont
  {Bornet}}, \bibinfo {author} {\bibfnamefont {G.}~\bibnamefont {Emperauger}},
  \bibinfo {author} {\bibfnamefont {C.}~\bibnamefont {Chen}}, \bibinfo {author}
  {\bibfnamefont {B.}~\bibnamefont {Ye}}, \bibinfo {author} {\bibfnamefont
  {M.}~\bibnamefont {Block}}, \bibinfo {author} {\bibfnamefont
  {M.}~\bibnamefont {Bintz}}, \bibinfo {author} {\bibfnamefont {J.~A.}\
  \bibnamefont {Boyd}}, \bibinfo {author} {\bibfnamefont {D.}~\bibnamefont
  {Barredo}}, \bibinfo {author} {\bibfnamefont {T.}~\bibnamefont {Comparin}},
  \bibinfo {author} {\bibfnamefont {F.}~\bibnamefont {Mezzacapo}}, \emph
  {et~al.},\ }\bibfield  {title} {\bibinfo {title} {Scalable spin squeezing in
  a dipolar rydberg atom array},\ }\href@noop {} {\bibfield  {journal}
  {\bibinfo  {journal} {Nature}\ }\textbf {\bibinfo {volume} {621}},\ \bibinfo
  {pages} {728} (\bibinfo {year} {2023})}\BibitemShut {NoStop}%
\bibitem [{\citenamefont {Hines}\ \emph {et~al.}(2023)\citenamefont {Hines},
  \citenamefont {Rajagopal}, \citenamefont {Moreau}, \citenamefont {Wahrman},
  \citenamefont {Lewis}, \citenamefont {Markovi{\'c}},\ and\ \citenamefont
  {Schleier-Smith}}]{hines2023spin}%
  \BibitemOpen
  \bibfield  {author} {\bibinfo {author} {\bibfnamefont {J.~A.}\ \bibnamefont
  {Hines}}, \bibinfo {author} {\bibfnamefont {S.~V.}\ \bibnamefont
  {Rajagopal}}, \bibinfo {author} {\bibfnamefont {G.~L.}\ \bibnamefont
  {Moreau}}, \bibinfo {author} {\bibfnamefont {M.~D.}\ \bibnamefont {Wahrman}},
  \bibinfo {author} {\bibfnamefont {N.~A.}\ \bibnamefont {Lewis}}, \bibinfo
  {author} {\bibfnamefont {O.}~\bibnamefont {Markovi{\'c}}},\ and\ \bibinfo
  {author} {\bibfnamefont {M.}~\bibnamefont {Schleier-Smith}},\ }\bibfield
  {title} {\bibinfo {title} {Spin squeezing by rydberg dressing in an array of
  atomic ensembles},\ }\href@noop {} {\bibfield  {journal} {\bibinfo  {journal}
  {Physical Review Letters}\ }\textbf {\bibinfo {volume} {131}},\ \bibinfo
  {pages} {063401} (\bibinfo {year} {2023})}\BibitemShut {NoStop}%
\bibitem [{\citenamefont {Kaubruegger}\ \emph {et~al.}(2019)\citenamefont
  {Kaubruegger}, \citenamefont {Silvi}, \citenamefont {Kokail}, \citenamefont
  {van Bijnen}, \citenamefont {Rey}, \citenamefont {Ye}, \citenamefont
  {Kaufman},\ and\ \citenamefont {Zoller}}]{kaubruegger2019variational}%
  \BibitemOpen
  \bibfield  {author} {\bibinfo {author} {\bibfnamefont {R.}~\bibnamefont
  {Kaubruegger}}, \bibinfo {author} {\bibfnamefont {P.}~\bibnamefont {Silvi}},
  \bibinfo {author} {\bibfnamefont {C.}~\bibnamefont {Kokail}}, \bibinfo
  {author} {\bibfnamefont {R.}~\bibnamefont {van Bijnen}}, \bibinfo {author}
  {\bibfnamefont {A.~M.}\ \bibnamefont {Rey}}, \bibinfo {author} {\bibfnamefont
  {J.}~\bibnamefont {Ye}}, \bibinfo {author} {\bibfnamefont {A.~M.}\
  \bibnamefont {Kaufman}},\ and\ \bibinfo {author} {\bibfnamefont
  {P.}~\bibnamefont {Zoller}},\ }\bibfield  {title} {\bibinfo {title}
  {Variational spin-squeezing algorithms on programmable quantum sensors},\
  }\href@noop {} {\bibfield  {journal} {\bibinfo  {journal} {Physical review
  letters}\ }\textbf {\bibinfo {volume} {123}},\ \bibinfo {pages} {260505}
  (\bibinfo {year} {2019})}\BibitemShut {NoStop}%
\bibitem [{\citenamefont {Koczor}\ \emph {et~al.}(2020)\citenamefont {Koczor},
  \citenamefont {Endo}, \citenamefont {Jones}, \citenamefont {Matsuzaki},\ and\
  \citenamefont {Benjamin}}]{koczor2020variational}%
  \BibitemOpen
  \bibfield  {author} {\bibinfo {author} {\bibfnamefont {B.}~\bibnamefont
  {Koczor}}, \bibinfo {author} {\bibfnamefont {S.}~\bibnamefont {Endo}},
  \bibinfo {author} {\bibfnamefont {T.}~\bibnamefont {Jones}}, \bibinfo
  {author} {\bibfnamefont {Y.}~\bibnamefont {Matsuzaki}},\ and\ \bibinfo
  {author} {\bibfnamefont {S.~C.}\ \bibnamefont {Benjamin}},\ }\bibfield
  {title} {\bibinfo {title} {Variational-state quantum metrology},\ }\href@noop
  {} {\bibfield  {journal} {\bibinfo  {journal} {New Journal of Physics}\
  }\textbf {\bibinfo {volume} {22}},\ \bibinfo {pages} {083038} (\bibinfo
  {year} {2020})}\BibitemShut {NoStop}%
\bibitem [{\citenamefont {Yu}\ \emph {et~al.}(2026)\citenamefont {Yu},
  \citenamefont {Muleady}, \citenamefont {Wang}, \citenamefont {Schine},
  \citenamefont {Gorshkov},\ and\ \citenamefont {Childs}}]{yu2026efficient}%
  \BibitemOpen
  \bibfield  {author} {\bibinfo {author} {\bibfnamefont {J.}~\bibnamefont
  {Yu}}, \bibinfo {author} {\bibfnamefont {S.~R.}\ \bibnamefont {Muleady}},
  \bibinfo {author} {\bibfnamefont {Y.-X.}\ \bibnamefont {Wang}}, \bibinfo
  {author} {\bibfnamefont {N.}~\bibnamefont {Schine}}, \bibinfo {author}
  {\bibfnamefont {A.~V.}\ \bibnamefont {Gorshkov}},\ and\ \bibinfo {author}
  {\bibfnamefont {A.~M.}\ \bibnamefont {Childs}},\ }\bibfield  {title}
  {\bibinfo {title} {Efficient preparation of dicke states},\ }\href@noop {}
  {\bibfield  {journal} {\bibinfo  {journal} {Physical Review Letters}\
  }\textbf {\bibinfo {volume} {136}},\ \bibinfo {pages} {030601} (\bibinfo
  {year} {2026})}\BibitemShut {NoStop}%
\bibitem [{\citenamefont {Pi{\~n}eiro~Orioli}\ and\ \citenamefont
  {Rey}(2019)}]{pineiro2019dark}%
  \BibitemOpen
  \bibfield  {author} {\bibinfo {author} {\bibfnamefont {A.}~\bibnamefont
  {Pi{\~n}eiro~Orioli}}\ and\ \bibinfo {author} {\bibfnamefont {A.~M.}\
  \bibnamefont {Rey}},\ }\bibfield  {title} {\bibinfo {title} {Dark states of
  multilevel fermionic atoms in doubly filled optical lattices},\ }\href@noop
  {} {\bibfield  {journal} {\bibinfo  {journal} {Physical Review Letters}\
  }\textbf {\bibinfo {volume} {123}},\ \bibinfo {pages} {223601} (\bibinfo
  {year} {2019})}\BibitemShut {NoStop}%
\bibitem [{\citenamefont {Gammelmark}\ and\ \citenamefont
  {M{\o}lmer}(2014)}]{gammelmark2014fisher}%
  \BibitemOpen
  \bibfield  {author} {\bibinfo {author} {\bibfnamefont {S.}~\bibnamefont
  {Gammelmark}}\ and\ \bibinfo {author} {\bibfnamefont {K.}~\bibnamefont
  {M{\o}lmer}},\ }\bibfield  {title} {\bibinfo {title} {Fisher information and
  the quantum cram{\'e}r-rao sensitivity limit of continuous measurements},\
  }\href@noop {} {\bibfield  {journal} {\bibinfo  {journal} {Physical review
  letters}\ }\textbf {\bibinfo {volume} {112}},\ \bibinfo {pages} {170401}
  (\bibinfo {year} {2014})}\BibitemShut {NoStop}%
\bibitem [{\citenamefont {Shankar}\ \emph {et~al.}(2019)\citenamefont
  {Shankar}, \citenamefont {Greve}, \citenamefont {Wu}, \citenamefont
  {Thompson},\ and\ \citenamefont {Holland}}]{shankar2019continuous}%
  \BibitemOpen
  \bibfield  {author} {\bibinfo {author} {\bibfnamefont {A.}~\bibnamefont
  {Shankar}}, \bibinfo {author} {\bibfnamefont {G.~P.}\ \bibnamefont {Greve}},
  \bibinfo {author} {\bibfnamefont {B.}~\bibnamefont {Wu}}, \bibinfo {author}
  {\bibfnamefont {J.~K.}\ \bibnamefont {Thompson}},\ and\ \bibinfo {author}
  {\bibfnamefont {M.}~\bibnamefont {Holland}},\ }\bibfield  {title} {\bibinfo
  {title} {Continuous real-time tracking of a quantum phase below the standard
  quantum limit},\ }\href@noop {} {\bibfield  {journal} {\bibinfo  {journal}
  {Physical Review Letters}\ }\textbf {\bibinfo {volume} {122}},\ \bibinfo
  {pages} {233602} (\bibinfo {year} {2019})}\BibitemShut {NoStop}%
\bibitem [{\citenamefont {Yang}\ \emph {et~al.}(2023)\citenamefont {Yang},
  \citenamefont {Huelga},\ and\ \citenamefont {Plenio}}]{yang2023efficient}%
  \BibitemOpen
  \bibfield  {author} {\bibinfo {author} {\bibfnamefont {D.}~\bibnamefont
  {Yang}}, \bibinfo {author} {\bibfnamefont {S.~F.}\ \bibnamefont {Huelga}},\
  and\ \bibinfo {author} {\bibfnamefont {M.~B.}\ \bibnamefont {Plenio}},\
  }\bibfield  {title} {\bibinfo {title} {Efficient information retrieval for
  sensing via continuous measurement},\ }\href@noop {} {\bibfield  {journal}
  {\bibinfo  {journal} {Physical Review X}\ }\textbf {\bibinfo {volume} {13}},\
  \bibinfo {pages} {031012} (\bibinfo {year} {2023})}\BibitemShut {NoStop}%
\bibitem [{\citenamefont {Duan}\ \emph {et~al.}(2025)\citenamefont {Duan},
  \citenamefont {Hu}, \citenamefont {Lu}, \citenamefont {Xiao}, \citenamefont
  {Jia}, \citenamefont {M{\o}lmer},\ and\ \citenamefont
  {Xiao}}]{duan2025concurrent}%
  \BibitemOpen
  \bibfield  {author} {\bibinfo {author} {\bibfnamefont {J.}~\bibnamefont
  {Duan}}, \bibinfo {author} {\bibfnamefont {Z.}~\bibnamefont {Hu}}, \bibinfo
  {author} {\bibfnamefont {X.}~\bibnamefont {Lu}}, \bibinfo {author}
  {\bibfnamefont {L.}~\bibnamefont {Xiao}}, \bibinfo {author} {\bibfnamefont
  {S.}~\bibnamefont {Jia}}, \bibinfo {author} {\bibfnamefont {K.}~\bibnamefont
  {M{\o}lmer}},\ and\ \bibinfo {author} {\bibfnamefont {Y.}~\bibnamefont
  {Xiao}},\ }\bibfield  {title} {\bibinfo {title} {Concurrent spin squeezing
  and field tracking with machine learning},\ }\href@noop {} {\bibfield
  {journal} {\bibinfo  {journal} {Nature Physics}\ ,\ \bibinfo {pages} {1}}
  (\bibinfo {year} {2025})}\BibitemShut {NoStop}%
\bibitem [{\citenamefont {Chu}\ \emph {et~al.}(2025)\citenamefont {Chu},
  \citenamefont {Mamaev}, \citenamefont {Koppenh{\"o}fer}, \citenamefont
  {Yuan},\ and\ \citenamefont {Clerk}}]{chu2025reconfigurable}%
  \BibitemOpen
  \bibfield  {author} {\bibinfo {author} {\bibfnamefont {A.}~\bibnamefont
  {Chu}}, \bibinfo {author} {\bibfnamefont {M.}~\bibnamefont {Mamaev}},
  \bibinfo {author} {\bibfnamefont {M.}~\bibnamefont {Koppenh{\"o}fer}},
  \bibinfo {author} {\bibfnamefont {M.}~\bibnamefont {Yuan}},\ and\ \bibinfo
  {author} {\bibfnamefont {A.~A.}\ \bibnamefont {Clerk}},\ }\bibfield  {title}
  {\bibinfo {title} {Reconfigurable dissipative entanglement between many spin
  ensembles: from robust quantum sensing to many-body state engineering},\
  }\href@noop {} {\bibfield  {journal} {\bibinfo  {journal} {arXiv preprint
  arXiv:2510.07616}\ } (\bibinfo {year} {2025})}\BibitemShut {NoStop}%
\bibitem [{\citenamefont {Bu{\v{z}}ek}\ \emph {et~al.}(1999)\citenamefont
  {Bu{\v{z}}ek}, \citenamefont {Derka},\ and\ \citenamefont
  {Massar}}]{buvzek1999optimal}%
  \BibitemOpen
  \bibfield  {author} {\bibinfo {author} {\bibfnamefont {V.}~\bibnamefont
  {Bu{\v{z}}ek}}, \bibinfo {author} {\bibfnamefont {R.}~\bibnamefont {Derka}},\
  and\ \bibinfo {author} {\bibfnamefont {S.}~\bibnamefont {Massar}},\
  }\bibfield  {title} {\bibinfo {title} {Optimal quantum clocks},\ }\href@noop
  {} {\bibfield  {journal} {\bibinfo  {journal} {Physical review letters}\
  }\textbf {\bibinfo {volume} {82}},\ \bibinfo {pages} {2207} (\bibinfo {year}
  {1999})}\BibitemShut {NoStop}%
\bibitem [{\citenamefont {Pegg}\ and\ \citenamefont
  {Barnett}(1988)}]{pegg1988unitary}%
  \BibitemOpen
  \bibfield  {author} {\bibinfo {author} {\bibfnamefont {D.}~\bibnamefont
  {Pegg}}\ and\ \bibinfo {author} {\bibfnamefont {S.}~\bibnamefont {Barnett}},\
  }\bibfield  {title} {\bibinfo {title} {Unitary phase operator in quantum
  mechanics},\ }\href@noop {} {\bibfield  {journal} {\bibinfo  {journal}
  {Europhysics Letters}\ }\textbf {\bibinfo {volume} {6}},\ \bibinfo {pages}
  {483} (\bibinfo {year} {1988})}\BibitemShut {NoStop}%
\bibitem [{\citenamefont {Direkci}\ \emph {et~al.}(2026)\citenamefont
  {Direkci}, \citenamefont {Finkelstein}, \citenamefont {Endres},\ and\
  \citenamefont {Gefen}}]{direkci2026heisenberg}%
  \BibitemOpen
  \bibfield  {author} {\bibinfo {author} {\bibfnamefont {S.}~\bibnamefont
  {Direkci}}, \bibinfo {author} {\bibfnamefont {R.}~\bibnamefont
  {Finkelstein}}, \bibinfo {author} {\bibfnamefont {M.}~\bibnamefont
  {Endres}},\ and\ \bibinfo {author} {\bibfnamefont {T.}~\bibnamefont
  {Gefen}},\ }\bibfield  {title} {\bibinfo {title} {Heisenberg-limited bayesian
  phase estimation with low-depth digital quantum circuits},\ }\href@noop {}
  {\bibfield  {journal} {\bibinfo  {journal} {npj Quantum Information}\ }
  (\bibinfo {year} {2026})}\BibitemShut {NoStop}%
\bibitem [{\citenamefont {Marciniak}\ \emph {et~al.}(2022)\citenamefont
  {Marciniak}, \citenamefont {Feldker}, \citenamefont {Pogorelov},
  \citenamefont {Kaubruegger}, \citenamefont {Vasilyev}, \citenamefont {van
  Bijnen}, \citenamefont {Schindler}, \citenamefont {Zoller}, \citenamefont
  {Blatt},\ and\ \citenamefont {Monz}}]{marciniak2022optimal}%
  \BibitemOpen
  \bibfield  {author} {\bibinfo {author} {\bibfnamefont {C.~D.}\ \bibnamefont
  {Marciniak}}, \bibinfo {author} {\bibfnamefont {T.}~\bibnamefont {Feldker}},
  \bibinfo {author} {\bibfnamefont {I.}~\bibnamefont {Pogorelov}}, \bibinfo
  {author} {\bibfnamefont {R.}~\bibnamefont {Kaubruegger}}, \bibinfo {author}
  {\bibfnamefont {D.~V.}\ \bibnamefont {Vasilyev}}, \bibinfo {author}
  {\bibfnamefont {R.}~\bibnamefont {van Bijnen}}, \bibinfo {author}
  {\bibfnamefont {P.}~\bibnamefont {Schindler}}, \bibinfo {author}
  {\bibfnamefont {P.}~\bibnamefont {Zoller}}, \bibinfo {author} {\bibfnamefont
  {R.}~\bibnamefont {Blatt}},\ and\ \bibinfo {author} {\bibfnamefont
  {T.}~\bibnamefont {Monz}},\ }\bibfield  {title} {\bibinfo {title} {Optimal
  metrology with programmable quantum sensors},\ }\href@noop {} {\bibfield
  {journal} {\bibinfo  {journal} {Nature}\ }\textbf {\bibinfo {volume} {603}},\
  \bibinfo {pages} {604} (\bibinfo {year} {2022})}\BibitemShut {NoStop}%
\bibitem [{\citenamefont {Cao}\ \emph {et~al.}(2024)\citenamefont {Cao},
  \citenamefont {Eckner}, \citenamefont {Lukin~Yelin}, \citenamefont {Young},
  \citenamefont {Jandura}, \citenamefont {Yan}, \citenamefont {Kim},
  \citenamefont {Pupillo}, \citenamefont {Ye}, \citenamefont {Darkwah~Oppong}
  \emph {et~al.}}]{cao2024multi}%
  \BibitemOpen
  \bibfield  {author} {\bibinfo {author} {\bibfnamefont {A.}~\bibnamefont
  {Cao}}, \bibinfo {author} {\bibfnamefont {W.~J.}\ \bibnamefont {Eckner}},
  \bibinfo {author} {\bibfnamefont {T.}~\bibnamefont {Lukin~Yelin}}, \bibinfo
  {author} {\bibfnamefont {A.~W.}\ \bibnamefont {Young}}, \bibinfo {author}
  {\bibfnamefont {S.}~\bibnamefont {Jandura}}, \bibinfo {author} {\bibfnamefont
  {L.}~\bibnamefont {Yan}}, \bibinfo {author} {\bibfnamefont {K.}~\bibnamefont
  {Kim}}, \bibinfo {author} {\bibfnamefont {G.}~\bibnamefont {Pupillo}},
  \bibinfo {author} {\bibfnamefont {J.}~\bibnamefont {Ye}}, \bibinfo {author}
  {\bibfnamefont {N.}~\bibnamefont {Darkwah~Oppong}}, \emph {et~al.},\
  }\bibfield  {title} {\bibinfo {title} {Multi-qubit gates and schr{\"o}dinger
  cat states in an optical clock},\ }\href@noop {} {\bibfield  {journal}
  {\bibinfo  {journal} {Nature}\ }\textbf {\bibinfo {volume} {634}},\ \bibinfo
  {pages} {315} (\bibinfo {year} {2024})}\BibitemShut {NoStop}%
\bibitem [{\citenamefont {Finkelstein}\ \emph {et~al.}(2024)\citenamefont
  {Finkelstein}, \citenamefont {Tsai}, \citenamefont {Sun}, \citenamefont
  {Scholl}, \citenamefont {Direkci}, \citenamefont {Gefen}, \citenamefont
  {Choi}, \citenamefont {Shaw},\ and\ \citenamefont
  {Endres}}]{finkelstein2024universal}%
  \BibitemOpen
  \bibfield  {author} {\bibinfo {author} {\bibfnamefont {R.}~\bibnamefont
  {Finkelstein}}, \bibinfo {author} {\bibfnamefont {R.~B.-S.}\ \bibnamefont
  {Tsai}}, \bibinfo {author} {\bibfnamefont {X.}~\bibnamefont {Sun}}, \bibinfo
  {author} {\bibfnamefont {P.}~\bibnamefont {Scholl}}, \bibinfo {author}
  {\bibfnamefont {S.}~\bibnamefont {Direkci}}, \bibinfo {author} {\bibfnamefont
  {T.}~\bibnamefont {Gefen}}, \bibinfo {author} {\bibfnamefont
  {J.}~\bibnamefont {Choi}}, \bibinfo {author} {\bibfnamefont {A.~L.}\
  \bibnamefont {Shaw}},\ and\ \bibinfo {author} {\bibfnamefont
  {M.}~\bibnamefont {Endres}},\ }\bibfield  {title} {\bibinfo {title}
  {Universal quantum operations and ancilla-based read-out for tweezer
  clocks},\ }\href@noop {} {\bibfield  {journal} {\bibinfo  {journal} {Nature}\
  }\textbf {\bibinfo {volume} {634}},\ \bibinfo {pages} {321} (\bibinfo {year}
  {2024})}\BibitemShut {NoStop}%
\bibitem [{\citenamefont {Chen}\ \emph {et~al.}(2011)\citenamefont {Chen},
  \citenamefont {Bohnet}, \citenamefont {Sankar}, \citenamefont {Dai},\ and\
  \citenamefont {Thompson}}]{chen2011conditional}%
  \BibitemOpen
  \bibfield  {author} {\bibinfo {author} {\bibfnamefont {Z.}~\bibnamefont
  {Chen}}, \bibinfo {author} {\bibfnamefont {J.~G.}\ \bibnamefont {Bohnet}},
  \bibinfo {author} {\bibfnamefont {S.~R.}\ \bibnamefont {Sankar}}, \bibinfo
  {author} {\bibfnamefont {J.}~\bibnamefont {Dai}},\ and\ \bibinfo {author}
  {\bibfnamefont {J.~K.}\ \bibnamefont {Thompson}},\ }\bibfield  {title}
  {\bibinfo {title} {Conditional spin squeezing of a large ensemble via the
  vacuum rabi splitting},\ }\href@noop {} {\bibfield  {journal} {\bibinfo
  {journal} {Physical Review Letters}\ }\textbf {\bibinfo {volume} {106}},\
  \bibinfo {pages} {133601} (\bibinfo {year} {2011})}\BibitemShut {NoStop}%
\bibitem [{\citenamefont {Chen}\ \emph {et~al.}(2014)\citenamefont {Chen},
  \citenamefont {Bohnet}, \citenamefont {Weiner}, \citenamefont {Cox},\ and\
  \citenamefont {Thompson}}]{chen2014cavity}%
  \BibitemOpen
  \bibfield  {author} {\bibinfo {author} {\bibfnamefont {Z.}~\bibnamefont
  {Chen}}, \bibinfo {author} {\bibfnamefont {J.~G.}\ \bibnamefont {Bohnet}},
  \bibinfo {author} {\bibfnamefont {J.~M.}\ \bibnamefont {Weiner}}, \bibinfo
  {author} {\bibfnamefont {K.~C.}\ \bibnamefont {Cox}},\ and\ \bibinfo {author}
  {\bibfnamefont {J.~K.}\ \bibnamefont {Thompson}},\ }\bibfield  {title}
  {\bibinfo {title} {Cavity-aided nondemolition measurements for atom counting
  and spin squeezing},\ }\href@noop {} {\bibfield  {journal} {\bibinfo
  {journal} {Physical Review A}\ }\textbf {\bibinfo {volume} {89}},\ \bibinfo
  {pages} {043837} (\bibinfo {year} {2014})}\BibitemShut {NoStop}%
\bibitem [{\citenamefont {Reddy}\ \emph {et~al.}(2020)\citenamefont {Reddy},
  \citenamefont {Nerem}, \citenamefont {Nam}, \citenamefont {Mirin},\ and\
  \citenamefont {Verma}}]{reddy2020superconducting}%
  \BibitemOpen
  \bibfield  {author} {\bibinfo {author} {\bibfnamefont {D.~V.}\ \bibnamefont
  {Reddy}}, \bibinfo {author} {\bibfnamefont {R.~R.}\ \bibnamefont {Nerem}},
  \bibinfo {author} {\bibfnamefont {S.~W.}\ \bibnamefont {Nam}}, \bibinfo
  {author} {\bibfnamefont {R.~P.}\ \bibnamefont {Mirin}},\ and\ \bibinfo
  {author} {\bibfnamefont {V.~B.}\ \bibnamefont {Verma}},\ }\bibfield  {title}
  {\bibinfo {title} {Superconducting nanowire single-photon detectors with 98\%
  system detection efficiency at 1550 nm},\ }\href@noop {} {\bibfield
  {journal} {\bibinfo  {journal} {Optica}\ }\textbf {\bibinfo {volume} {7}},\
  \bibinfo {pages} {1649} (\bibinfo {year} {2020})}\BibitemShut {NoStop}%
\bibitem [{\citenamefont {Ding}\ \emph {et~al.}(2025)\citenamefont {Ding},
  \citenamefont {Zhang}, \citenamefont {Xiong}, \citenamefont {Xiao},
  \citenamefont {Zhang}, \citenamefont {Huang}, \citenamefont {Xu},
  \citenamefont {Liu}, \citenamefont {You}, \citenamefont {Wang} \emph
  {et~al.}}]{ding2025photon}%
  \BibitemOpen
  \bibfield  {author} {\bibinfo {author} {\bibfnamefont {C.}~\bibnamefont
  {Ding}}, \bibinfo {author} {\bibfnamefont {X.}~\bibnamefont {Zhang}},
  \bibinfo {author} {\bibfnamefont {J.}~\bibnamefont {Xiong}}, \bibinfo
  {author} {\bibfnamefont {Y.}~\bibnamefont {Xiao}}, \bibinfo {author}
  {\bibfnamefont {T.}~\bibnamefont {Zhang}}, \bibinfo {author} {\bibfnamefont
  {J.}~\bibnamefont {Huang}}, \bibinfo {author} {\bibfnamefont
  {H.}~\bibnamefont {Xu}}, \bibinfo {author} {\bibfnamefont {X.}~\bibnamefont
  {Liu}}, \bibinfo {author} {\bibfnamefont {L.}~\bibnamefont {You}}, \bibinfo
  {author} {\bibfnamefont {Z.}~\bibnamefont {Wang}}, \emph {et~al.},\
  }\bibfield  {title} {\bibinfo {title} {Photon-number-resolving single-photon
  detector with a system detection efficiency of 98\% and photon-number
  resolution of 32},\ }\href@noop {} {\bibfield  {journal} {\bibinfo  {journal}
  {ACS Photonics}\ } (\bibinfo {year} {2025})}\BibitemShut {NoStop}%
\bibitem [{\citenamefont {Beekman}\ \emph {et~al.}(2019)\citenamefont
  {Beekman}, \citenamefont {Rademaker},\ and\ \citenamefont
  {Van~Wezel}}]{beekman2019introduction}%
  \BibitemOpen
  \bibfield  {author} {\bibinfo {author} {\bibfnamefont {A.}~\bibnamefont
  {Beekman}}, \bibinfo {author} {\bibfnamefont {L.}~\bibnamefont {Rademaker}},\
  and\ \bibinfo {author} {\bibfnamefont {J.}~\bibnamefont {Van~Wezel}},\
  }\bibfield  {title} {\bibinfo {title} {An introduction to spontaneous
  symmetry breaking},\ }\href@noop {} {\bibfield  {journal} {\bibinfo
  {journal} {SciPost Physics Lecture Notes}\ ,\ \bibinfo {pages} {011}}
  (\bibinfo {year} {2019})}\BibitemShut {NoStop}%
\bibitem [{\citenamefont {Rademaker}(2019)}]{rademaker2019exact}%
  \BibitemOpen
  \bibfield  {author} {\bibinfo {author} {\bibfnamefont {L.}~\bibnamefont
  {Rademaker}},\ }\bibfield  {title} {\bibinfo {title} {Exact ground state of
  the lieb-mattis hamiltonian as a superposition of n{\'e}el states},\
  }\href@noop {} {\bibfield  {journal} {\bibinfo  {journal} {Physical Review
  Research}\ }\textbf {\bibinfo {volume} {1}},\ \bibinfo {pages} {032018}
  (\bibinfo {year} {2019})}\BibitemShut {NoStop}%
\bibitem [{\citenamefont {Liu}\ \emph {et~al.}(2014)\citenamefont {Liu},
  \citenamefont {Jing}, \citenamefont {Zhong},\ and\ \citenamefont
  {Wang}}]{liu2014quantum}%
  \BibitemOpen
  \bibfield  {author} {\bibinfo {author} {\bibfnamefont {J.}~\bibnamefont
  {Liu}}, \bibinfo {author} {\bibfnamefont {X.-X.}\ \bibnamefont {Jing}},
  \bibinfo {author} {\bibfnamefont {W.}~\bibnamefont {Zhong}},\ and\ \bibinfo
  {author} {\bibfnamefont {X.-G.}\ \bibnamefont {Wang}},\ }\bibfield  {title}
  {\bibinfo {title} {Quantum fisher information for density matrices with
  arbitrary ranks},\ }\href@noop {} {\bibfield  {journal} {\bibinfo  {journal}
  {Communications in Theoretical Physics}\ }\textbf {\bibinfo {volume} {61}},\
  \bibinfo {pages} {45} (\bibinfo {year} {2014})}\BibitemShut {NoStop}%
\bibitem [{\citenamefont {Pezze}\ and\ \citenamefont
  {Smerzi}(2014)}]{pezze2014quantum}%
  \BibitemOpen
  \bibfield  {author} {\bibinfo {author} {\bibfnamefont {L.}~\bibnamefont
  {Pezze}}\ and\ \bibinfo {author} {\bibfnamefont {A.}~\bibnamefont {Smerzi}},\
  }\bibfield  {title} {\bibinfo {title} {Quantum theory of phase estimation},\
  }in\ \href@noop {} {\emph {\bibinfo {booktitle} {Atom interferometry}}}\
  (\bibinfo  {publisher} {IOS Press},\ \bibinfo {year} {2014})\ pp.\ \bibinfo
  {pages} {691--741}\BibitemShut {NoStop}%
\end{thebibliography}
\end{document}